\documentclass[nofootinbib,preprint,preprintnumbers,a4paper,10pt]{revtex4}
\usepackage{multirow}
\usepackage{textcomp}
\usepackage{amsmath,graphicx,color,epsfig}

\usepackage{footnote}
\usepackage{ulem}
\usepackage{booktabs}
\usepackage{array}
\usepackage{amssymb}
\usepackage{subfigure}
\usepackage{longtable}
\usepackage{verbatim}
\usepackage{amsfonts}
\usepackage{hyperref}
\usepackage{cancel}

\begin{document}

\title{Topological amplitudes of charmed baryon decays in the $SU(3)_F$ limit}

\author{Di Wang$^{1}$}\email{wangdi@hunnu.edu.cn}
\author{Jin-Feng Luo$^{1}$}
\address{%
$^1$Department of Physics, Hunan Normal University, Changsha 410081, China
}

\begin{abstract}
Charmed baryon decay plays an important role in studying the weak and strong interactions.
Topological diagram is an intuitive tool for analyzing the dynamics of heavy hadron decays.
In this work, we investigate the topological diagrams of charmed baryon antitriplet ($\mathcal{B}_{c\overline 3}$) decays into a light baryon octet ($\mathcal{B}_8$) and a light meson ($M$).
A one-to-one mapping between the topological diagram and the invariant tensor is established.
The topological diagrams of the $\mathcal{B}_{c\overline 3}\to \mathcal{B}_8^S M$ modes (where $\mathcal{B}_8^S$ and $\mathcal{B}_8^A$ are the $q_1\leftrightarrow q_2$ symmetric and antisymmetric octets) and the diagrams with a quark loop are presented for the first time.
The completeness of topologies is confirmed by permutation.
The linear relations of topologies are obtained by deriving the relation between the topological amplitudes constructed by the third- and second-rank octet tensors.
It is found the topologies contributing to the $\mathcal{B}_{c\overline 3}\to \mathcal{B}_8^S M$ modes can be determined by the topologies contributing to the $\mathcal{B}_{c\overline 3}\to \mathcal{B}_8^A M$ modes, and vice versa.
The equations of $SU(3)$ irreducible amplitudes decomposed by topologies are derived through two different intermediate amplitudes.
However, the inverse solution does not exist since the number of topologies exceeds number of $SU(3)$ irreducible amplitudes.
Applying this framework to the Standard Model, it is found there are thirteen independent $SU(3)$ irreducible amplitudes contributing to the $\mathcal{B}_{c\overline 3}\to \mathcal{B}_8 M$ decays. Among these, four amplitudes associated with three-dimensional operators are significant for $CP$ asymmetries.
Considering the suppressions due to small Cabibbo-Kobayashi-Maskawa matrix elements $V_{cb}^*V_{ub}$ and the K\"orner-Pati-Woo theorem, the branching fractions of charmed baryon decays are dominated by five $SU(3)$ irreducible amplitudes in the $SU(3)_F$ limit.
Quark-loop diagrams could enhance the $U$-spin breaking effects and increase the branching fraction difference of two decay channels.
Systematic measurements of branching fractions of the singly Cabibbo-suppressed modes could help identify promising channels for searching for $CP$ asymmetries in the charmed baryon sector.
\end{abstract}

\maketitle

\section{Introduction}

Charmed baryon decay plays an important role in studying the nonperturbative baryonic transitions.
Among charmed baryon decays, the decays of the charmed baryon antitriplet ($\ \mathcal{B}_{c\overline 3}$) into a light baryon octet ($\mathcal{B}_{8}$) and a light meson ($M$) are the most widely studied in the literature.
Many data on the $\mathcal{B}_{c\overline 3}\to \mathcal{B}_{8} M$ decays have been collected by
BESIII \cite{BESIII:2023uvs,BESIII:2023vfi,BESIII:2023ooh,BESIII:2023rky,BESIII:2023iwu,
BESIII:2022udq,BESIII:2022izy,BESIII:2022vrr,BESIII:2022xne,BESIII:2022onh,
BESIII:2022aok,BESIII:2022bkj,BESIII:2021fqx,BESIII:2020kap,BESIII:2020cpu,
BESIII:2019odb,BESIII:2018qyg,
Ablikim:2018jfs,Ablikim:2018woi,Ablikim:2018bir,Ablikim:2015prg,
Ablikim:2015flg,Ablikim:2016tze,Ablikim:2016mcr,
Ablikim:2016vqd,Ablikim:2017ors,Ablikim:2017iqd},
Belle (II) \cite{Belle:2022uod,Belle:2022bsi,Belle-II:2024jql,Belle:2021crz,Belle:2021dgc,Belle:2021btl,Berger:2018pli,
Zupanc:2013iki,Yang:2015ytm,Pal:2017ypp}
and LHCb \cite{Aaij:2017xva,Aaij:2017nsd} over the past decades.
In theoretical aspect, the quantum chromodynamics (QCD)-inspired approaches do not work well in the charmed hadron decays because the expansion parameters $\alpha_s(m_c)$ and $\Lambda_{\rm QCD}/m_c$ are larger than those in the $b$-hadron decays.
Thereby, it is significant to study charmed baryon decays based on the flavor $SU(3)$ symmetry.
Topological diagram is an intuitive and helpful tool for analyzing the dynamics of heavy hadron decays.
It provides a theoretical framework that allows not only for model-dependent data analysis but also model calculations.

Topological diagrams have been used in studies of charmed baryon decays \cite{Groote:2021pxt,Hsiao:2020iwc,Zhao:2018mov,He:2018joe,Zhong:2024qqs,Hsiao:2021nsc,Kohara:1991ug,Chau:1995gk,Zhong:2024zme}.
However, some theoretical issues remain unresolved.
According to the Pauli exclusion principle, the combination of flavor and spin wave functions of octet baryons, $\phi_{\rm flavor}\chi_{\rm spin}$, should be symmetric under the interchange of any two quarks.
However, neither combination $\phi_S\chi_S$ nor $\phi_A\chi_A$ is symmetric under the interchange of $q_1\leftrightarrow q_3$ or $q_2\leftrightarrow q_3$.
Only the equal-weight combination, $\psi = (\phi_S\chi_S+\phi_A\chi_A)/\sqrt{2}$, is symmetric under the interchange of any two quarks \cite{Fayyazuddin:1994wh}.
Due to the different spin wave functions, the topological diagrams of charmed baryon decays into octet baryons have two distinct sets.
The decay amplitude of a $\mathcal{B}_{c\overline 3}\to \mathcal{B}_{8} M$ mode is the sum of the $\mathcal{B}_{c\overline 3}\to \mathcal{B}_{8}^S M$ and $\mathcal{B}_{c\overline 3}\to \mathcal{B}_{8}^A M$ transitions.
However, the topological diagrams of the $\mathcal{B}_{c\overline 3}\to \mathcal{B}_{8}^S M$ modes are usually missing in the literature \cite{Groote:2021pxt,Hsiao:2021nsc,Zhao:2018mov,He:2018joe}.
Papers such as \cite{Zhong:2024qqs,Zhong:2024zme} consider some topological amplitudes of the $\mathcal{B}_{c\overline 3}\to \mathcal{B}_{8}^S M$ modes, but many topologies are overlooked, and the relations between the topologies of the $\mathcal{B}_{c\overline 3}\to \mathcal{B}_{8}^S M$ and $\mathcal{B}_{c\overline 3}\to \mathcal{B}_{8}^A M$ transitions are ambiguous.

On the other hand, the penguin diagrams are regarded as the dominant contributions to $CP$ asymmetries in heavy hadron weak decays.
Unfortunately, the penguin diagrams of charmed baryon decays are missing in the literature.
This omission affects the mathematical completeness of topological amplitudes and limits theoretical studies of $CP$ asymmetries in charmed baryon decays.
Recent LHCb measurements have revealed that the penguin diagrams are comparable to the tree amplitudes in the $D$ meson decays \cite{Aaij:2019kcg,LHCb:2022lry}.
However, there is no evidence of $CP$ violation in charmed baryon decays so far.
Studies of penguin diagrams in the charmed baryon decays could be valuable for searching for $CP$ asymmetries in the baryon sector.

Topological diagrams in the heavy hadron decays can be formalized as invariant tensors constructed by the four-fermion operators and initial/final states \cite{He:2018php}.
This allows us to study topologies using mathematical tools such as tensor contraction, permutation, and Lie group decomposition.
A systematic analysis of the topological amplitudes for $D$ meson decays was performed in Ref.~\cite{Wang:2020gmn}, in which the universality, completeness, and linear correlation of topological amplitudes in the $SU(3)_F$ limit are clarified.
However, it is not straightforward to extend the theoretical framework of $D$ meson decays to charmed baryon decays.
The reason is that the topological amplitudes of $\mathcal{B}_{c\overline 3}\to \mathcal{B}_8M$ decays are constructed by the third-rank octet tensors, whereas the $SU(3)$ irreducible amplitudes are constructed by the second-rank octet tensors.
The relations between the topological diagrams and $SU(3)$ irreducible amplitudes of the $\mathcal{B}_{c\overline 3}\to \mathcal{B}_8M$ decays are more complicated than those of the two-body $D$ meson decays.

In this work, we study the topological diagrams of $\mathcal{B}_{c\overline 3}\to \mathcal{B}_8M$ decays within the flavor $SU(3)$ symmetry.
A one-to-one mapping between the topological diagrams and the invariant tensors is established.
The topological diagrams of the $\mathcal{B}_{c\overline 3}\to \mathcal{B}_8^S M$ modes and the quark-loop diagrams are presented for the first time.
The completeness of topologies is confirmed by permutation.
To study linear correlation of topologies, we derive the relations between the topological amplitudes constructed by the third-rank and second-rank octet tensors.
It is found the topologies contributing to the $\mathcal{B}_{c\overline 3}\to \mathcal{B}_8^S M$ modes can be determined by the topologies contributing to the $\mathcal{B}_{c\overline 3}\to \mathcal{B}_8^A M$ modes, and vice versa.
The equations of $SU(3)$ irreducible amplitudes decomposed by topologies are derived through two different approaches.
However, the inverse solution does not exist since the number of topologies exceeds the number of $SU(3)$ irreducible amplitudes.

As a phenomenological application, we analyze the $\mathcal{B}_{c\overline 3}\to \mathcal{B}_8M$ decays in the Standard Model.
It is found there are thirteen independent tree induced amplitudes contributing to the $\mathcal{B}_{c\overline 3}\to \mathcal{B}_8 M$ decays, with four amplitudes associated with three-dimensional operators are non-negligible in $CP$ asymmetries.
The penguin induced amplitudes can be determined once the tree induced amplitudes are known.
Considering that some topological amplitudes are suppressed by the small CKM matrix elements and the K\"orner-Pati-Woo (KPW) theorem \cite{Pati:1970fg,Korner:1970xq}, the dominant $SU(3)$ irreducible amplitudes in the branching fractions are reduced to five.
Similar to the charm meson decays, the quark-loop diagrams in charmed baryon decays could enhance the $U$-spin breaking effects.
We suggest measuring of branching fractions of the singly Cabibbo-suppressed charmed baryon decays systematically before searching for $CP$ asymmetries.
It could help us to identify potential "golden modes" for experimental studies.

The rest of this paper is organized as follows.
In Sec.~\ref{tensor}, we show the general discussions about the topological amplitudes of charmed baryon decays in the $SU(3)_F$ limit.
The topological amplitudes in the Standard Model and the phenomenological discussions are presented in Sec.~\ref{sm}.
Sec.~\ref{summary} is a brief summary.
The analysis of the third-rank $SU(3)$ irreducible amplitudes are shown in Appendix.~\ref{su3}.

\section{Tensor analysis of topologies}\label{tensor}
The charm quark decays are categorized into three types, Cabibbo-favored (CF), singly Cabibbo-suppressed (SCS), and doubly Cabibbo-suppressed (DCS) decays, with flavor structures of
\begin{align}
  c\to s\bar d u, \qquad   c\to d\bar du/s\bar su,\qquad c\to d\bar s u,
\end{align}
respectively.
In the $SU(3)$ picture, the weak Hamiltonian of charm decay can be written as \cite{Wang:2020gmn}
 \begin{equation}\label{h}
 \mathcal H_{\rm eff}= \sum_p \sum_{i,j,k=1}^3 (H^{(p)})_{ij}^{k}\mathcal{O}_{ij}^{(p)k},
 \end{equation}
in which
\begin{equation}\label{a5}
\mathcal{O}_{ij}^{(p)k} = \frac{G_F}{\sqrt{2}} \sum_{\rm color} \sum_{\rm current}C_p(\overline q_iq_k)(\overline q_jc).
\end{equation}
$\mathcal{O}_{ij}^{(p)k}$ denotes the four-quark operator with the Fermi coupling constant $G_F$ and the Wilson coefficient $C_p$.
Superscript $p$ in $\mathcal{O}_{ij}^{(p)k}$ is the order of perturbation in the effective theory.
In the Standard Model, $p=0,1$ denote the tree and penguin operators.
For each value of $p$, there are $27$ different $ \mathcal{O}^{(p)k}_{ij}$. The matrix $H^{(p)}$ is the $3\times 3\times 3$ coefficient matrix.
The color indices and current structures of four quark operators are summed because, once the flavor structure of the operator is determined, operators with different color indices and current structures always appear simultaneously.

The charmed antitriplet baryon is expressed as
\begin{eqnarray}
 \mathcal{B}_{c\overline 3}=  \left( \begin{array}{ccc}
   0   & \Lambda_c^+  & \Xi_c^+ \\
    -\Lambda_c^+ &   0   & \Xi_c^0 \\
    -\Xi_c^+ & -\Xi_c^0 & 0 \\
  \end{array}\right).
\end{eqnarray}
The light pseudoscalar nonet meson is
\begin{eqnarray}
 M=  \left( \begin{array}{ccc}
   \frac{1}{\sqrt 2} \pi^0+  \frac{1}{\sqrt 6} \eta_8    & \pi^+  & K^+ \\
    \pi^- &   - \frac{1}{\sqrt 2} \pi^0+ \frac{1}{\sqrt 6} \eta_8   & K^0 \\
    K^- & \overline K^0 & -\sqrt{2/3}\eta_8 \\
  \end{array}\right) +  \frac{1}{\sqrt 3} \left( \begin{array}{ccc}
   \eta_1    & 0  & 0 \\
    0 &  \eta_1   & 0 \\
   0 & 0 & \eta_1 \\
  \end{array}\right).
\end{eqnarray}
The pseudoscalar mesons $\eta_8$ and $\eta_1$ are not mass eigenstates.
The mass eigenstates $\eta$ and $\eta^\prime$ are mixtures of $\eta_8$ and $\eta_1$,
\begin{eqnarray}
\left( \begin{array}{ccc}
\eta\\
\eta^\prime
\end{array}
\right)
=
\left(
\begin{array}{cc}
\cos\xi  &  -\sin\xi\\
\sin\xi  &  \cos\xi
\end{array}
\right)\left(
\begin{array}{c}
\eta_8\\
\eta_1
\end{array}\right).
\end{eqnarray}
There are two different octets with symmetric and antisymmetric flavor wave functions under $q_1\leftrightarrow q_2$.
They are labeled by $\mathcal{B}_8^S$ and $\mathcal{B}_8^A$ and their flavor and spin wave functions are $\phi_S\chi_S$ and $\phi_A\chi_A$, respectively.
The light baryon octets $\mathcal{B}_8^S$ and $\mathcal{B}_8^A$ can be written as tensors with three covariant indices.
$\mathcal{B}_8^S$ is given by
\begin{align}\label{BS}
  \Sigma^+ & = \frac{1}{\sqrt{6}}(2\mathcal{B}_8^{113}-\mathcal{B}_8^{311}-\mathcal{B}_8^{131}),
  \qquad   p =\frac{1}{\sqrt{6}}(-2\mathcal{B}_8^{112}+\mathcal{B}_8^{211}+\mathcal{B}_8^{121}), \qquad   \Sigma^- = \frac{1}{\sqrt{6}}(-2\mathcal{B}_8^{223}+\mathcal{B}_8^{322}+\mathcal{B}_8^{232}), \nonumber\\
    n &=\frac{1}{\sqrt{6}}(2\mathcal{B}_8^{221}-\mathcal{B}_8^{122}-\mathcal{B}_8^{212}), \qquad\Xi^- = \frac{1}{\sqrt{6}}(2\mathcal{B}_8^{332}-\mathcal{B}_8^{233}-\mathcal{B}_8^{323}), \qquad   \Xi^0 =\frac{1}{\sqrt{6}}(-2\mathcal{B}_8^{331}+\mathcal{B}_8^{133}+\mathcal{B}_8^{313}),\nonumber\\
 \Sigma^0 &=\frac{1}{\sqrt{12}}(\mathcal{B}_8^{321}+\mathcal{B}_8^{231}+\mathcal{B}_8^{312}
 +\mathcal{B}_8^{132}-2\mathcal{B}_8^{123}-2\mathcal{B}_8^{213}), \qquad \Lambda^0 = \frac{1}{2}(\mathcal{B}_8^{231}-\mathcal{B}_8^{132}+\mathcal{B}_8^{321}
 -\mathcal{B}_8^{312}),
\end{align}
and $\mathcal{B}_8^A$ is given by
\begin{align}\label{BA}
  \Sigma^+ & = \frac{1}{\sqrt{2}}(\mathcal{B}_8^{311}-\mathcal{B}_8^{131}),\qquad   p =\frac{1}{\sqrt{2}}(\mathcal{B}_8^{121}-\mathcal{B}_8^{211}), \qquad \Sigma^- = \frac{1}{\sqrt{2}}(\mathcal{B}_8^{232}-\mathcal{B}_8^{322})  \nonumber\\
  n& =\frac{1}{\sqrt{2}}(\mathcal{B}_8^{122}-\mathcal{B}_8^{212}), \qquad  \Xi^- = \frac{1}{\sqrt{2}}(\mathcal{B}_8^{233}-\mathcal{B}_8^{323}),\qquad   \Xi^0 =\frac{1}{\sqrt{2}}(\mathcal{B}_8^{313}-\mathcal{B}_8^{133}),  \nonumber\\
\Sigma^0 &=\frac{1}{2}(\mathcal{B}_8^{132}-\mathcal{B}_8^{312}+\mathcal{B}_8^{231}
-\mathcal{B}_8^{321}),   \qquad   \Lambda^0  = \frac{1}{\sqrt{12}}(2\mathcal{B}_8^{213}-2\mathcal{B}_8^{123}+\mathcal{B}_8^{231}
-\mathcal{B}_8^{321}+\mathcal{B}_8^{312}-\mathcal{B}_8^{132}).
\end{align}
Similarly to the effective Hamiltonian, the initial and final states, for example light meson $M$, can be written as
\begin{align}
  |M^\alpha\rangle = (M^\alpha)^{i}_{j}|M^{i}_{j} \rangle,
\end{align}
where $|M^{i}_{j} \rangle$ is the quark composition of meson state, $|M^{i}_{j} \rangle = |q_i\bar q_j\rangle$, and $(M^\alpha)$ is the coefficient matrix.

Decay amplitude of the $\mathcal{B}^\gamma_{c\overline 3}\to \mathcal{B}^\alpha_{8} M^\beta$ mode is constructed to be
\begin{align}\label{amp}
\mathcal{A}(\mathcal{B}^\gamma_{c\overline 3}\to \mathcal{B}_{8}^\alpha M^\beta)& = \langle \mathcal{B}_{8}^\alpha M^\beta |\mathcal{H}_{\rm eff}| \mathcal{B}^\gamma_{c\overline 3}\rangle\nonumber\\&~=\sum_p \sum_{\rm Per.}\,(\mathcal{B}_{8}^\alpha)^{ijk}\langle \mathcal{B}_{8}^{ijk}|(M^\beta)^l_m\langle M^l_m||H_{np}^q\mathcal{O}_{np}^q||(\mathcal{B}^\gamma_{c\overline 3})_{rs}|[\mathcal{B}_{c\overline 3}]_{rs}\rangle\nonumber\\& ~~=\sum_p \sum_{\rm Per.}\,\langle \mathcal{B}_{8}^{ijk} M^l_m |\mathcal{O}_{np}^q|[\mathcal{B}_{c\overline 3}]_{rs}\rangle \times (\mathcal{B}_{8}^\alpha)^{ijk}(M^\beta)^l_m H_{np}^q(\mathcal{B}^\gamma_{c\overline 3})_{rs}.
\end{align}
In above formula, $\sum_{\rm Per.}$ presents summing over all possible full contractions, i.e., all the flavor indices of $\langle \mathcal{B}_{8}^{ijk} M^l_m |\mathcal{O}_{np}^q|[\mathcal{B}_{c\overline 3}]_{rs}\rangle$ are contracted with each other.
For simplicity, the decay amplitude of $\mathcal{B}^\gamma_{c\overline 3}\to \mathcal{B}^\alpha_{8} M^\beta$ mode is expressed as
\begin{align}\label{ampx}
\mathcal{A}(\mathcal{B}^\gamma_{c\overline 3}\to \mathcal{B}_{8}^\alpha M^\beta)= \sum_\omega X_{\omega}(C_\omega)_{\alpha\beta\gamma},
\end{align}
where $\omega$ labels the different contractions of the $SU(3)$ indices, $X_\omega = \langle \mathcal{B}_{8}^{ijk} M^l_m |\mathcal{O}_{np}^q|[\mathcal{B}_{c\overline 3}]_{rs}\rangle$ is the reduced matrix element, and $(C_\omega)_{\alpha\beta\gamma}=(\mathcal{B}_{8}^\alpha)^{ijk}(M^\beta)^l_m H_{np}^q(\mathcal{B}^\gamma_{c\overline 3})_{rs}$ is the Clebsch-Gordan (CG) coefficient.
According to the Wigner-Eckhart theorem \cite{Eckart30,Wigner59}, $X_\omega$ is independent of indices $\alpha$, $\beta$ and $\gamma$.
All the information of initial/final states is absorbed into the Clebsch-Gordan coefficient $(C_\omega)_{\alpha\beta\gamma}$.

\begin{figure}
  \centering
  \includegraphics[width=14cm]{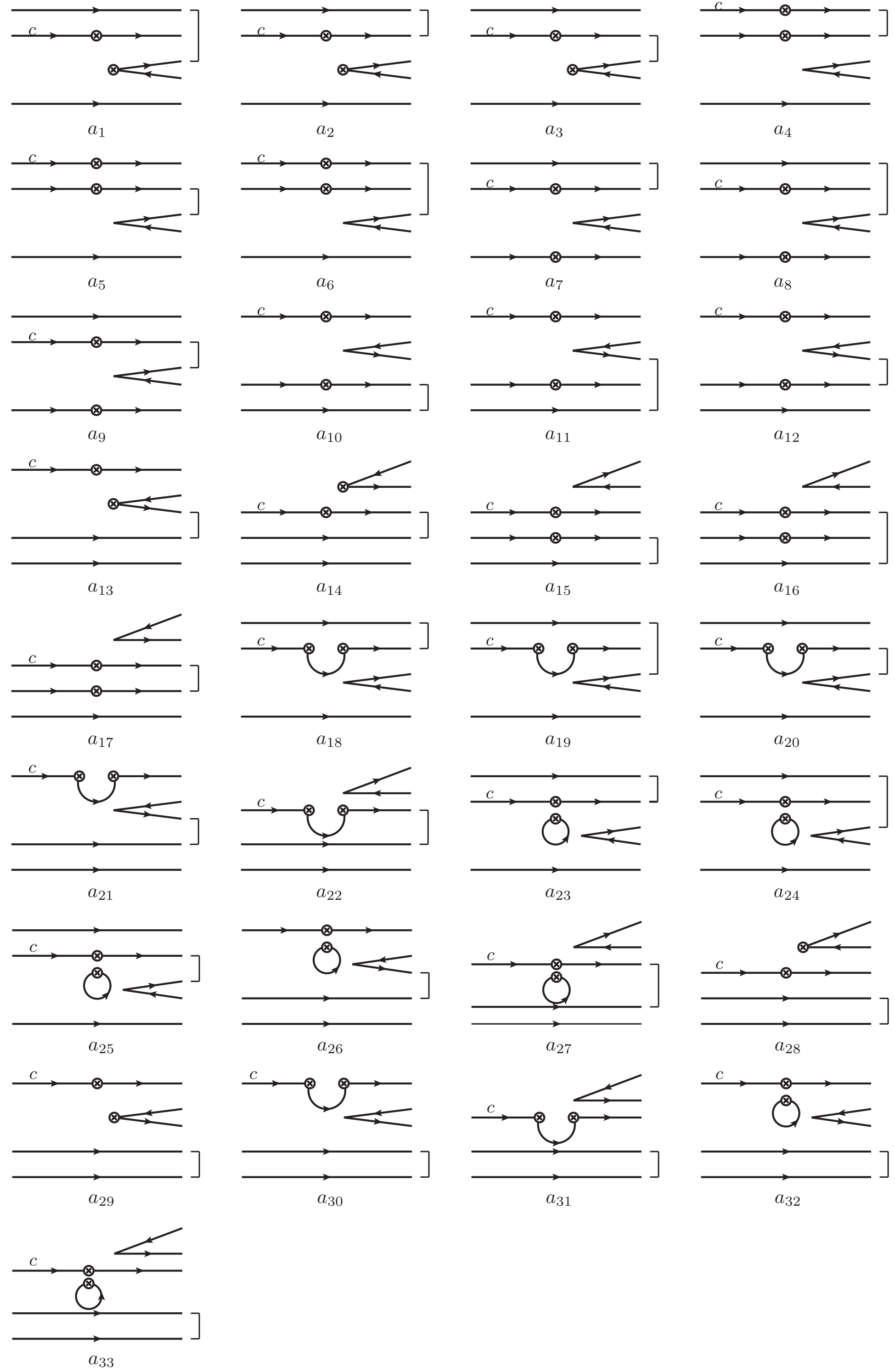}
  \caption{Topological diagrams of the charmed baryon antitriplet $\mathcal{B}_{c\overline 3}$ decays into a light baryon octet $\mathcal{B}^A_8$ and a light meson $M$, in which $"]"$ means two antisymmetric light quarks in baryon.}\label{top7}
\end{figure}

The overall wave function for a bound $qqq$ state, accounting for all degrees of freedom, is written as
\begin{align}
  \Psi = \phi_{\rm flavor}\chi_{\rm spin}\xi_{\rm color}\eta_{\rm space},
\end{align}
where $\phi_{\rm flavor}$, $\chi_{\rm spin}$, $\xi_{\rm color}$ and $\eta_{\rm space}$ are the flavor, spin, color, and space wave functions, respectively.
The wave function $\Psi$ must be
antisymmetric under the interchange of any two quarks because of the Pauli exclusion principle.
The color wave function $\xi_{\rm color}$ is necessarily totally antisymmetric since all hadrons are $SU(3)_C$ singlets.
For the $L=0$ ground state baryon, there is no orbital angular momentum.
The space wave function $\eta_{\rm space}$ is symmetric under the interchange of any two quarks.
Consequently, the combined flavour and spin wave function $\phi_{\rm flavor}\chi_{\rm spin}$ must be symmetric under the interchange of any two quarks.
However, neither $\phi_S\chi_S$ or $\phi_A\chi_A$ has a definite symmetry under the interchange of $q_1\leftrightarrow q_3$ or $q_2\leftrightarrow q_3$.
Only the equal-weight combination of $\phi_S\chi_S$ and $\phi_A\chi_A$ is symmetric under the interchange of any two quarks \cite{Fayyazuddin:1994wh},
\begin{align}\label{a7}
 \Psi = \frac{1}{\sqrt{2}}(\phi_S\chi_S+\phi_A\chi_A)\xi_{\rm color}\eta_{\rm space}.
\end{align}
Because of the different spin wave functions of $\mathcal{B}_{8}^S$ and $\mathcal{B}_{8}^A$, the topological diagrams of charmed baryon decays into octet baryons have two different sets.
For distinction, we label the topological diagrams of charmed baryon decays into $\mathcal{B}^S_8$ and $\mathcal{B}^A_8$ with superscripts $S$ and $A$, respectively.
The total amplitude for the $\mathcal{B}_{c\overline 3}\to \mathcal{B}_8 M$ decay is obtained by summing the amplitudes of the $\mathcal{B}_{c\overline 3}\to \mathcal{B}_8^S M$ and $\mathcal{B}_{c\overline 3}\to \mathcal{B}_8^A M$ transitions,
\begin{align}\label{a6}
  \mathcal{A}(\mathcal{B}_{c\overline 3}\to \mathcal{B}_8 M) = \frac{1}{\sqrt{2}}\big[\mathcal{A}^S(\mathcal{B}_{c\overline 3}\to \mathcal{B}_8^S M) + \mathcal{A}^A(\mathcal{B}_{c\overline 3}\to \mathcal{B}_8^A M)\big].
\end{align}

The number of topologies contributing to the $\mathcal{B}_{c\overline 3}\to \mathcal{B}_8 M$ decay can be counted by permutation.
There are five covariant/contravariant indices in tensor $\langle \mathcal{B}_{8}^{ijk} M^l_m |\mathcal{O}_{np}^q|[\mathcal{B}_{c\overline 3}]_{rs}\rangle$. The number of full contractions is $N=A^5_5=120$.
Considering that the two flavor indices in the charmed baryon antitriplet are antisymmetric, the number of term contributing to the $\mathcal{B}_{c\overline 3}\to \mathcal{B}_8 M$ decay, including both $\mathcal{B}_{c\overline 3}\to\mathcal{B}_8^S M$ and $\mathcal{B}_{c\overline 3}\to\mathcal{B}_8^A M$ modes, is $N_S+N_A=N/2=60$.
The other $60$ terms appear in the charmed baryon sextet decays.
Because of the symmetric light quarks in $\mathcal{B}^S_8$ conflict with the antisymmetric light quarks in $\mathcal{B}_{c\overline 3}$,  the number of topological diagram in charmed antitriplet baryon decays into $\mathcal{B}^S_8$ is less than the one in decays into $\mathcal{B}^A_8$.
The difference between them, $N_A-N_S$, is computed to be $A_{3}^3 = 6$.
Then $N_S$ and $N_A$ are solved to be $N_S=27$, $N_A=33$.
Furthermore, we can use permutation to count the amount of different types of topologies.
For example, the number of diagrams with a quark loop is $2\times A^4_4/2 =24$, where the factor $2$ arises from two types of quark loops, and the factor $1/2$ arises from the antisymmetric indices in charmed baryons.
The number of diagrams involving a singlet meson is $A^4_4/2 =12$.
The number of diagram involving $q\overline q$ produced from the vacuum is $3\times A^4_4/2 = 36$, where the factor $3$ arises from the three choices for a meson tensor contracting with a baryon tensor.

The topological amplitude of $\mathcal{B}_{c\overline 3}\to \mathcal{B}_8^SM$ decay is constructed as
\begin{align}\label{amp1}
 \mathcal{A}^S(\mathcal{B}_{c\overline 3}\to \mathcal{B}_8^SM) = & a^S_1(\mathcal{B}_{c\overline 3})_{ij} H^m_{kl}M^i_m ( \mathcal{B}_8^S)^{jkl} + a^S_2(\mathcal{B}_{c\overline 3})_{ij}H^m_{kl}M^i_m ( \mathcal{B}_8^S)^{jlk}+ a^S_3(\mathcal{B}_{c\overline 3})_{ij}H^m_{kl}M^i_m ( \mathcal{B}_8^S)^{klj} \nonumber\\
   & + a^S_4(\mathcal{B}_{c\overline 3})_{ij} H^i_{kl}M^j_m ( \mathcal{B}_8^S)^{klm}  + a^S_5(\mathcal{B}_{c\overline 3})_{ij} H^i_{kl}M^j_m ( \mathcal{B}_8^S)^{kml}  + a^S_6(\mathcal{B}_{c\overline 3})_{ij} H^i_{kl}M^j_m ( \mathcal{B}_8^S)^{lmk} \nonumber\\
   &+ a^S_7(\mathcal{B}_{c\overline 3})_{ij} H^i_{kl}M^k_m ( \mathcal{B}_8^S)^{jlm}+ a^S_8(\mathcal{B}_{c\overline 3})_{ij} H^i_{kl}M^k_m ( \mathcal{B}_8^S)^{jml} + a^S_9(\mathcal{B}_{c\overline 3})_{ij} H^i_{kl}M^k_m ( \mathcal{B}_8^S)^{lmj} \nonumber\\
 &+ a^S_{10}(\mathcal{B}_{c\overline 3})_{ij} H^i_{kl}M^l_m ( \mathcal{B}_8^S)^{jkm}  + a^S_{11}(\mathcal{B}_{c\overline 3})_{ij} H^i_{kl}M^l_m ( \mathcal{B}_8^S)^{jmk} + a^S_{12}(\mathcal{B}_{c\overline 3})_{ij} H^i_{kl}M^l_m ( \mathcal{B}_8^S)^{kmj} \nonumber\\
&+ a^S_{13}(\mathcal{B}_{c\overline 3})_{ij} H^m_{kl}M^l_m (\mathcal{B}_8^S)^{ikj}+ a^S_{14}(\mathcal{B}_{c\overline 3})_{ij} H^m_{kl}M^k_m ( \mathcal{B}_8^S)^{ilj} + a^S_{15}(\mathcal{B}_{c\overline 3})_{ij} H^i_{kl}M^m_m ( \mathcal{B}_8^S)^{jkl} \nonumber\\
 &+ a^S_{16}(\mathcal{B}_{c\overline 3})_{ij} H^i_{kl}M^m_m (\mathcal{B}_8^S)^{jlk} + a^S_{17}(\mathcal{B}_{c\overline 3})_{ij} H^i_{kl}M^m_m ( \mathcal{B}_8^S)^{klj}  + a^S_{18}(\mathcal{B}_{c\overline 3})_{ij} H^l_{kl}M^i_m ( \mathcal{B}_8^S)^{jkm}\nonumber\\
 & + a^S_{19}(\mathcal{B}_{c\overline 3})_{ij} H^l_{kl}M^i_m (\mathcal{B}_8^S)^{jmk} + a^S_{20}(\mathcal{B}_{c\overline 3})_{ij} H^l_{kl}M^i_m (\mathcal{B}_8^S)^{kmj} + a^S_{21}(\mathcal{B}_{c\overline 3})_{ij} H^l_{kl}M^k_m ( \mathcal{B}_8^S)^{imj}\nonumber\\
& + a^S_{22}(\mathcal{B}_{c\overline 3})_{ij} H^l_{kl}M^m_m (\mathcal{B}_8^S)^{ikj} + a^S_{23}(\mathcal{B}_{c\overline 3})_{ij} H^l_{lk}M^i_m ( \mathcal{B}_8^S)^{jkm}+ a^S_{24}(\mathcal{B}_{c\overline 3})_{ij} H^l_{lk}M^i_m ( \mathcal{B}_8^S)^{jmk}\nonumber\\
& + a^S_{25}(\mathcal{B}_{c\overline 3})_{ij} H^l_{lk}M^i_m (\mathcal{B}_8^S)^{kmj} + a^S_{26}(\mathcal{B}_{c\overline 3})_{ij} H^l_{lk}M^k_m ( \mathcal{B}_8^S)^{imj}+ a^S_{27}(\mathcal{B}_{c\overline 3})_{ij} H^l_{lk}M^m_m ( \mathcal{B}_8^S)^{ikj},
\end{align}
and the topological amplitude of $\mathcal{B}_{c\overline 3}\to \mathcal{B}_8^AM$ decay is constructed as
\begin{align}\label{amp2}
 \mathcal{A}^A(\mathcal{B}_{c\overline 3}\to \mathcal{B}_8^AM) = & a^A_1(\mathcal{B}_{c\overline 3})_{ij} H^m_{kl}M^i_m ( \mathcal{B}_8^A)^{jkl} + a^A_2(\mathcal{B}_{c\overline 3})_{ij}H^m_{kl}M^i_m ( \mathcal{B}_8^A)^{jlk}+ a^A_3(\mathcal{B}_{c\overline 3})_{ij}H^m_{kl}M^i_m ( \mathcal{B}_8^A)^{klj} \nonumber\\
   & + a^A_4(\mathcal{B}_{c\overline 3})_{ij} H^i_{kl}M^j_m (\mathcal{B}_8^A)^{klm}  + a^A_5(\mathcal{B}_{c\overline 3})_{ij} H^i_{kl}M^j_m (\mathcal{B}_8^A)^{kml}  + a^A_6(\mathcal{B}_{c\overline 3})_{ij} H^i_{kl}M^j_m ( \mathcal{B}_8^A)^{lmk} \nonumber\\
   &+ a^A_7(\mathcal{B}_{c\overline 3})_{ij} H^i_{kl}M^k_m (\mathcal{B}_8^A)^{jlm}+ a^A_8(\mathcal{B}_{c\overline 3})_{ij} H^i_{kl}M^k_m (\mathcal{B}_8^A)^{jml} + a^A_9(\mathcal{B}_{c\overline 3})_{ij} H^i_{kl}M^k_m ( \mathcal{B}_8^A)^{lmj} \nonumber\\
 &+ a^A_{10}(\mathcal{B}_{c\overline 3})_{ij} H^i_{kl}M^l_m ( \mathcal{B}_8^A)^{jkm}  + a^A_{11}(\mathcal{B}_{c\overline 3})_{ij} H^i_{kl}M^l_m (\mathcal{B}_8^A)^{jmk} + a^A_{12}(\mathcal{B}_{c\overline 3})_{ij} H^i_{kl}M^l_m ( \mathcal{B}_8^A)^{kmj} \nonumber\\
&+ a^A_{13}(\mathcal{B}_{c\overline 3})_{ij} H^m_{kl}M^l_m ( \mathcal{B}_8^A)^{ikj}+ a^A_{14}(\mathcal{B}_{c\overline 3})_{ij} H^m_{kl}M^k_m (\mathcal{B}_8^A)^{ilj} + a^A_{15}(\mathcal{B}_{c\overline 3})_{ij} H^i_{kl}M^m_m ( \mathcal{B}_8^A)^{jkl} \nonumber\\
 &+ a^A_{16}(\mathcal{B}_{c\overline 3})_{ij} H^i_{kl}M^m_m ( \mathcal{B}_8^A)^{jlk} + a^A_{17}(\mathcal{B}_{c\overline 3})_{ij} H^i_{kl}M^m_m (\mathcal{B}_8^A)^{klj}  + a^A_{18}(\mathcal{B}_{c\overline 3})_{ij} H^l_{kl}M^i_m ( \mathcal{B}_8^A)^{jkm}\nonumber\\
 & + a^A_{19}(\mathcal{B}_{c\overline 3})_{ij} H^l_{kl}M^i_m ( \mathcal{B}_8^A)^{jmk} + a^A_{20}(\mathcal{B}_{c\overline 3})_{ij} H^l_{kl}M^i_m ( \mathcal{B}_8^A)^{kmj} + a^A_{21}(\mathcal{B}_{c\overline 3})_{ij} H^l_{kl}M^k_m ( \mathcal{B}_8^A)^{imj}\nonumber\\
& + a^A_{22}(\mathcal{B}_{c\overline 3})_{ij} H^l_{kl}M^m_m ( \mathcal{B}_8^A)^{ikj} + a^A_{23}(\mathcal{B}_{c\overline 3})_{ij} H^l_{lk}M^i_m ( \mathcal{B}_8^A)^{jkm}+ a^A_{24}(\mathcal{B}_{c\overline 3})_{ij} H^l_{lk}M^i_m ( \mathcal{B}_8^A)^{jmk}\nonumber\\
& + a^A_{25}(\mathcal{B}_{c\overline 3})_{ij} H^l_{lk}M^i_m ( \mathcal{B}_8^A)^{kmj} + a^A_{26}(\mathcal{B}_{c\overline 3})_{ij} H^l_{lk}M^k_m ( \mathcal{B}_8^A)^{imj}+ a^A_{27}(\mathcal{B}_{c\overline 3})_{ij} H^l_{lk}M^m_m ( \mathcal{B}_8^A)^{ikj}\nonumber\\
&+ a^A_{28}(\mathcal{B}_{c\overline 3})_{ij} H^m_{kl}M^k_m (\mathcal{B}_8^A)^{ijl} + a^A_{29}(\mathcal{B}_{c\overline 3})_{ij} H^m_{kl}M^l_m (\mathcal{B}_8^A)^{ijk} + a^A_{30}(\mathcal{B}_{c\overline 3})_{ij} H^l_{kl}M^k_m ( \mathcal{B}_8^A)^{ijm} \nonumber\\
&+ a^A_{31}(\mathcal{B}_{c\overline 3})_{ij} H^l_{kl}M^m_m ( \mathcal{B}_8^A)^{ijk}+ a^A_{32}(\mathcal{B}_{c\overline 3})_{ij} H^l_{lk}M^k_m ( \mathcal{B}_8^A)^{ijm}+ a^A_{33}(\mathcal{B}_{c\overline 3})_{ij} H^l_{lk}M^m_m ( \mathcal{B}_8^A)^{ijk}.
\end{align}
Each term in Eq.~\eqref{amp1} and Eq.~\eqref{amp2} can be interpreted as a topological diagram with following rules.
\begin{itemize}
  \item In tensor $H_{ij}^k$, the index $j$ presents the quark $q_j$ produced in the vertex that connects to the $c$ quark line, while indices $i$ and $k$ denote quark $q_i$ and antiquark $\overline q_k$ produced in the other vertex.
  \item The contraction of the final state meson or baryon with the four-quark Hamiltonian denotes that the quark produced in one vertex enters the final state meson or baryon.
  \item  The contraction between initial state and final state denotes the light quark in charmed baryon enters the final state as a spectator quark.
\item  The contraction between two indices in $H_{ij}^k$ denotes a quark loop. There are two different ways to contract two indices in $H_{ij}^k$. $H_{il}^l$ presents the quark loop connecting two vertices in the topological diagram, and $H_{lj}^l$ presents the quark loop induced by one vertexes in the topological diagram.
\end{itemize}
To distinguish other decay amplitudes defined in this paper, we also refer to the topological amplitudes defined in Eqs.~\eqref{amp1} and \eqref{amp2} as "3-rank topological amplitudes" since the octet baryons are written as third-rank tensors.

The topological diagrams contributing to the $\mathcal{B}_{c\overline 3}\to \mathcal{B}^A_8 M$ decay are shown in Fig.~\ref{top7}.
The topological diagrams contributing to the $\mathcal{B}_{c\overline 3}\to \mathcal{B}^S_8 M$ decay can be obtained by replacing the anti-symmetric quarks of $a_1$ $\sim$ $a_{27}$ in Fig.~\ref{top7} to symmetric ones.
In Eqs.~\eqref{amp1} and \eqref{amp2}, we do not write the order of perturbation of four-quark operators $p$ explicitly.
It should be emphasized that the same contractions with different $p$ are different topological amplitudes.
In the SM, $p=0$ corresponds to the diagrams induced by tree operators, and $p=1$ corresponds to the diagrams induced by penguin operators.
By excluding diagrams with quark loops, the topological amplitudes contributing to $\mathcal{B}_{c\overline 3}\to \mathcal{B}^A_8 M$ mode are consistent with Ref.~\cite{He:2018joe}.

The octet baryon can also be written as a tensor with one covariant index and one contravariant index, $(\mathcal{B}_8)^i_j$, where $i\neq j$.
In the literature, the third-rank octet tensor is usually written as $(\mathcal{B}_8)^{ijk}= \epsilon^{ijl}(\mathcal{B}_8)^k_l$, where $\epsilon^{ijk}$ is the Levi-Civita tensor.
This expression is valid only for the $q_1\leftrightarrow q_2$ anti-symmetric octet $\mathcal{B}_8^A$ because the indices $i$ and $j$ are antisymmetric.
For the $q_1 \leftrightarrow q_2$ symmetric octet, the third-rank tensor $(\mathcal{B}^S_8)^{ijk}$ can be constructed by summing two antisymmetric tensors $\epsilon^{ikl}(\mathcal{B}_8)^j_l$ and $\epsilon^{jkl}(\mathcal{B}_8)^i_l$.
Then third-rank tensors $(\mathcal{B}^S_8)^{ijk}$ and $(\mathcal{B}^A_8)^{ijk}$ are expressed in terms of the second-rank tensors as follows
\begin{eqnarray}\label{sy1}
(\mathcal{B}^S_8)^{ijk} = \epsilon^{kil}(\mathcal{B}_8)^j_l
+\epsilon^{kjl}(\mathcal{B}_8)^i_l,\qquad\quad  (\mathcal{B}^A_8)^{ijk}= \epsilon^{ijl}(\mathcal{B}_8)^k_l.
\end{eqnarray}
The indices $i$ and $j$ are $i \leftrightarrow j$ symmetric in $(\mathcal{B}^S_8)^{ijk}$ and antisymmetric in $(\mathcal{B}^A_8)^{ijk}$, i.e., $(\mathcal{B}^S_8)^{ijk} = (\mathcal{B}^S_8)^{jik}$ and  $(\mathcal{B}^A_8)^{ijk}=-(\mathcal{B}^A_8)^{jik}$.
We take proton $p$ as an example to illustrate Eq.~\eqref{sy1}.
According to the first formula of Eq.~\eqref{sy1}, we have $(\mathcal{B}^S_8)^{112}= -2(\mathcal{B}_8)^1_3$ and $(\mathcal{B}^S_8)^{211}=(\mathcal{B}^S_8)^{121}= (\mathcal{B}_8)^1_3$.
So the third-rank tensors $(\mathcal{B}^S_8)^{112}$, $(\mathcal{B}^S_8)^{211}$, and $(\mathcal{B}^S_8)^{121}$ contribute to proton as $-2(\mathcal{B}^S_8)^{112}+(\mathcal{B}^S_8)^{211}+(\mathcal{B}^S_8)^{121}$. Taking into account the normalization factor $1/\sqrt{6}$, the first formula of Eq.~\eqref{sy1} is consistent with Eq.~\eqref{BS}.
Similarly, we conclude that $(\mathcal{B}^A_8)^{121}=-(\mathcal{B}^A_8)^{211}= (\mathcal{B}_8)^1_3$, according to the second formula of Eq.~\eqref{sy1}.
Taking into account the normalization factor $1/\sqrt{2}$, the second formula of Eq.~\eqref{sy1} is consistent with Eq.~\eqref{BA}.

The second-rank octet is given by
\begin{eqnarray}\label{B8}
 \mathcal{B}_8=  \left( \begin{array}{ccc}
   \frac{1}{\sqrt 2} \Sigma^0+  \frac{1}{\sqrt 6} \Lambda^0    & \Sigma^+  & p \\
    \Sigma^- &   - \frac{1}{\sqrt 2} \Sigma^0+ \frac{1}{\sqrt 6} \Lambda^0   & n \\
    \Xi^- & \Xi^0 & -\sqrt{2/3}\Lambda^0 \\
  \end{array}\right).
\end{eqnarray}
The charmed baryon antitriplet can be written as
\begin{eqnarray}\label{sy2}
(\mathcal{B}_{c\overline 3})_{ij}=\epsilon_{ijk}(\mathcal{B}_{c\overline 3})^{k},
\end{eqnarray}
in which
\begin{eqnarray}
 (\mathcal{B}_{c\overline 3})^{k}=\left( \begin{array}{ccc}
     \Xi_c^0 \\
    -\Xi_c^+  \\
    \Lambda_c^+ \\
  \end{array}\right).
\end{eqnarray}
The decay amplitude of $\mathcal{B}_{c\overline 3}\to \mathcal{B}_8 M$ mode could be constructed by second-rank baryon octet and first-rank charmed baryon antitriplet.
There are four covariant/contravariant indices in tensor $\langle (\mathcal{B}_{8})^{i}_j M^k_l |\mathcal{O}_{mn}^p|\mathcal{B}_{c\overline 3}^{q}\rangle$.
The number of full contractions is $N=A^4_4=24$.
Considering that $i\neq j$ in $(\mathcal{B}_{8})^{i}_j$, six terms are excluded.
Then the number of terms contributing to the $\mathcal{B}_{c\overline 3}\to\mathcal{B}_8 M$ mode is $24-6=18$.
The decay amplitude of the $\mathcal{B}_{c\overline 3}\to \mathcal{B}_8 M$ mode constructed by the second-rank octet tensors is
\begin{align}\label{amp3}
 \mathcal{A}(\mathcal{B}_{c\overline 3}\to \mathcal{B}_8M) = & A_{1}(\mathcal{B}_{c\overline 3})^{i} H^j_{kl}M^l_i( \mathcal{B}_8)^k_j
 +A_{2}(\mathcal{B}_{c\overline 3})^{i} H^j_{lk}M^l_j( \mathcal{B}_8)^k_i+ A_{3}(\mathcal{B}_{c\overline 3})^{i} H^j_{lk}M^l_i( \mathcal{B}_8)^k_j\nonumber\\
&+A_{4}(\mathcal{B}_{c\overline 3})^{i} H^j_{kl}M^l_j( \mathcal{B}_8)^k_i+ A_5(\mathcal{B}_{c\overline 3})^{i} H^j_{ik}M^l_j( \mathcal{B}_8)^k_l+ A_6(\mathcal{B}_{c\overline 3})^{i} H^j_{il}M^l_k( \mathcal{B}_8)^k_j\nonumber\\
&+A_7(\mathcal{B}_{c\overline 3})^{i} H^j_{ki}M^l_j( \mathcal{B}_8)^k_l  + A_8(\mathcal{B}_{c\overline 3})^{i} H^j_{li}M^l_k( \mathcal{B}_8)^k_j+
+ A_9(\mathcal{B}_{c\overline 3})^{i} H^j_{ik}M^l_l( \mathcal{B}_8)^k_j\nonumber\\
&+A_{10}(\mathcal{B}_{c\overline 3})^{i} H^j_{ki}M^l_l( \mathcal{B}_8)^k_j
+A_{11}(\mathcal{B}_{c\overline 3})^{i} H^j_{kj}M^l_i( \mathcal{B}_8)^k_l
+ A_{12}(\mathcal{B}_{c\overline 3})^{i} H^j_{lj}M^l_k( \mathcal{B}_8)^k_i\nonumber\\
& + A_{13}(\mathcal{B}_{c\overline 3})^{i} H^j_{kj}M^l_l( \mathcal{B}_8)^k_i
+A_{14}(\mathcal{B}_{c\overline 3})^{i} H^j_{ij}M^l_k( \mathcal{B}_8)^k_l
+ A_{15}(\mathcal{B}_{c\overline 3})^{i} H^j_{jl} M^l_k
(\mathcal{B}_8)^k_i\nonumber\\
&+A_{16}(\mathcal{B}_{c\overline 3})^{i} H^j_{jk}M^l_l( \mathcal{B}_8)^k_i
 + A_{17}(\mathcal{B}_{c\overline 3})^{i} H^j_{jk}M^l_i( \mathcal{B}_8)^k_l
 + A_{18}(\mathcal{B}_{c\overline 3})^{i} H^j_{ji}M^l_k( \mathcal{B}_8)^k_l.
\end{align}
In this work, we refer to the decay amplitudes defined in Eq.~\eqref{amp3} as "2-rank topological amplitudes".

By inserting Eqs.~\eqref{sy1} and \eqref{sy2} into each term of Eqs.~\eqref{amp1} and \eqref{amp2}, we can derive the relations between decay amplitudes constructed by second- and third-rank octets.
For example, the first term of Eq.~\eqref{amp1} is simplified to
\begin{align}
a^S_1(\mathcal{B}_{c\overline 3})_{ij} H^m_{kl}M^i_m ( \mathcal{B}_8^S)^{jkl} & = a^S_1\epsilon_{ijp}(\mathcal{B}_{c\overline 3})^{p} H^m_{kl}M^i_m \epsilon^{ljq}(\mathcal{B}_8)^k_q+a^S_1\epsilon_{ijp}(\mathcal{B}_{c\overline 3})^{p} H^m_{kl}M^i_m\epsilon^{lkq}(\mathcal{B}_8)^j_q\nonumber\\
& = -a^S_1\,(\mathcal{B}_{c\overline 3})^{i} H^j_{lk}M^l_j( \mathcal{B}_8)^k_i+2a^S_1\,(\mathcal{B}_{c\overline 3})^{i} H^j_{kl}M^l_j( \mathcal{B}_8)^k_i\nonumber\\&~~~~~+a^S_1\,(\mathcal{B}_{c\overline 3})^{i} H^j_{ik}M^l_j( \mathcal{B}_8)^k_l-2a^S_1\,(\mathcal{B}_{c\overline 3})^{i} H^j_{ki}M^l_j( \mathcal{B}_8)^k_l.
\end{align}
Thus, the diagram $a^S_1$ contributes to Eq.~\eqref{amp3} as
\begin{align}
&A_2 = -2\sqrt{3}\,a^S_1+..., \qquad A_4 = 4\sqrt{3}\,a^S_1+...,\qquad A_5 = 2\sqrt{3}\,a^S_1+...,\qquad A_7 = -4\sqrt{3}\,a^S_1+...,
\end{align}
where factor $2\sqrt{3}$ arises from the inverse of the normalization factor in Eq.~\eqref{BS} times $\sqrt{2}$ introduced in Eq.~\eqref{a6}, $\sqrt{6}\times \sqrt{2} = 2\sqrt{3}$.
The relations between the third- and second-rank topological amplitudes for the $\mathcal{B}_{c\overline 3}\to \mathcal{B}_8 M$ modes are therefore derived to be
\begin{align}\label{sol}
 A_1  = &2(a^A_6+a^A_{11}+a^A_{12})-2\sqrt{3}(a^S_4-a^S_{5}+2a^S_{10}-a^S_{11}-a^S_{12}), \nonumber\\
 A_2  = & 2(-a^A_1+a^A_{3}-a^A_{5}-a^A_8-a^A_{9}+a^A_{14}+2a^A_{28})-2\sqrt{3}(a^S_1-2a^S_{2}+a^S_{3}-a^S_{4}
 +a^S_{6}-2a^S_7+a^S_{8}+a^S_{9}+3a^S_{14}), \nonumber\\
 A_3  = &2(a^A_5+a^A_{8}+a^A_{9})-2\sqrt{3}(a^S_4-a^S_{6}+2a^S_{7}-a^S_{8}
 -a^S_{9}), \nonumber\\
 A_4  = &2(-a^A_2-a^A_{3}-a^A_{6}-a^A_{11}-a^A_{12}+a^A_{13}+2a^A_{29})
 -2\sqrt{3}(-2a^S_1+a^S_{2}+a^S_{3}-a^S_{4}
 +a^S_{5}-2a^S_{10}+a^S_{11}+a^S_{12}
 +3a^S_{13}), \nonumber\\
 A_5  = &2(a^A_1-a^A_{3}+a^A_{5})-2\sqrt{3}(-a^S_1+2a^S_{2}-a^S_{3}+a^S_{4}
 -a^S_{6}), \nonumber\\
 A_6  = &2(-a^A_4+a^A_{10}-a^A_{12})-2\sqrt{3}(-a^S_5+a^S_{6}-a^S_{10}+2a^S_{11}
 -a^S_{12}), \nonumber\\
 A_7  = &2(a^A_2+a^A_{3}+a^A_{6})-2\sqrt{3}(2a^S_1-a^S_{2}-a^S_{3}+a^S_{4}
 -a^S_{5}), \nonumber\\
 A_8  = &2(a^A_4+a^A_{7}-a^A_{9})-2\sqrt{3}(a^S_5-a^S_{6}-a^S_{7}+2a^S_{8}
 -a^S_{9}), \nonumber\\
 A_9  = &2(-a^A_5+a^A_{15}-a^A_{17})-2\sqrt{3}(-a^S_4+a^S_{6}-a^S_{15}+2a^S_{16}
 -a^S_{17}), \nonumber\\
 A_{10}  = &2(-a^A_6+a^A_{16}+a^A_{17})-2\sqrt{3}(-a^S_4+a^S_{5}+2a^S_{15}-a^S_{16}
 -a^S_{17}), \nonumber\\
 A_{11}  = &2(-a^A_6+a^A_{19}+a^A_{20})-2\sqrt{3}(-a^S_4+a^S_{5}+2a^S_{18}-a^S_{19}
 -a^S_{20}), \nonumber\\
 A_{12}  = &2(-a^A_4-a^A_{7}+a^A_{9}-a^A_{18}+a^A_{20}+a^A_{21}+2a^A_{30})-2\sqrt{3}(-a^S_5+a^S_{6}+a^S_{7}-2a^S_{8}
 +a^S_{9}+a^S_{18}-2a^S_{19}
 +a^S_{20}+3a^S_{21}), \nonumber\\
 A_{13}  = &2(a^A_6-a^A_{16}-a^A_{17}-a^A_{19}-a^A_{20}+a^A_{22}+2a^A_{31})-2\sqrt{3}(a^S_4-a^S_{5}-2a^S_{15}+a^S_{16}
 +a^S_{17}-2a^S_{18}+a^S_{19}
 +a^S_{20}+3a^S_{22}), \nonumber\\
 A_{14}  = &2(a^A_4+a^A_{18}-a^A_{20})-2\sqrt{3}(a^S_5-a^S_{6}-a^S_{18}+2a^S_{19}
 -a^S_{20}), \nonumber\\
 A_{15}  = &2(a^A_4-a^A_{10}+a^A_{12}-a^A_{23}+a^A_{25}+a^A_{26}+2a^A_{32})-2\sqrt{3}(a^S_5-a^S_{6}+a^S_{10}-2a^S_{11}
 +a^S_{12}+a^S_{23}-2a^S_{24}
 +a^S_{25}+3a^S_{26}), \nonumber\\
 A_{16}  = &2(a^A_5-a^A_{15}+a^A_{17}-a^A_{24}-a^A_{25}+a^A_{27}+2a^A_{33})-2\sqrt{3}(a^S_4-a^S_{6}+a^S_{15}-2a^S_{16}
 +a^S_{17}-2a^S_{23}+a^S_{24}
 +a^S_{25}+3a^S_{27}), \nonumber\\
 A_{17}  = &2(-a^A_5+a^A_{24}+a^A_{25})-2\sqrt{3}(-a^S_4+a^S_{6}+2a^S_{23}-a^S_{24}
 -a^S_{25}), \nonumber\\
 A_{18}  = &2(-a^A_4+a^A_{23}-a^A_{25})-2\sqrt{3}(-a^S_5+a^S_{6}-a^S_{23}+2a^S_{24}
 -a^S_{25}).
\end{align}
According to Eq.~\eqref{sol}, the topological diagrams of $\mathcal{B}_{c\overline 3}\to \mathcal{B}^S_{8}M$ and $\mathcal{B}_{c\overline 3}\to \mathcal{B}^A_{8}M$ transitions are correlated.
If the topological diagrams in a $\mathcal{B}_{c\overline 3}\to \mathcal{B}^S_{8}M$ mode are known, the topological diagrams contributing to the corresponding $\mathcal{B}_{c\overline 3}\to \mathcal{B}^A_{8}M$ mode are determined, and vice versa.

Operator $\mathcal{O}_{ij}^k$ is a reducible representation of the $SU(3)$ group.
It can be decomposed as $3 \otimes\overline 3 \otimes 3 =  3_p\oplus3_t\oplus \overline 6 \oplus 15$.
Then the coefficient matrix $H$ can be written as
\begin{align}\label{c8}
  H^k_{ij}= &\frac{1}{8}H(15)^k_{ij}+\frac{1}{4}\epsilon_{ijl}H(\overline 6)^{lk}+\delta_j^k\Big(\frac{3}{8}H( 3_t)_i-\frac{1}{8}H(3_p)_i\Big)+
  \delta_i^k\Big(\frac{3}{8}H( 3_p)_j-\frac{1}{8}H( 3_t)_j\Big),
\end{align}
in which $H( 3_p)_i = H_{li}^l$ and $H( 3_t)_i = H_{il}^l$.
The $SU(3)$ irreducible amplitude of $\mathcal{B}_{c\overline 3}\to \mathcal{B}_8M$ decay is constructed as
\begin{align}\label{amp4}
{\cal A}^{\rm IR}(\mathcal{B}_{c\overline 3}\to \mathcal{B}_8M)& =
b_1(\mathcal{B}_{c\overline 3})^i H(\overline 6)_{ij}^kM^j_l( \mathcal{B}_8)^l_k +b_2(\mathcal{B}_{c\overline 3})^i H(\overline 6)_{ij}^kM^l_k(\mathcal{B}_8)^j_l+ b_3(\mathcal{B}_{c\overline 3})^i H(\overline 6)_{ij}^kM^l_l(\mathcal{B}_8)^j_k\nonumber\\ &  +b_4(\mathcal{B}_{c\overline 3})^i H(\overline 6)_{jl}^kM^j_i(\mathcal{B}_8)^l_k +b_5(\mathcal{B}_{c\overline 3})^i H(\overline 6)_{jl}^kM^l_k(\mathcal{B}_8)^j_i + b_{6}(\mathcal{B}_{c\overline 3})^i H({15})_{ij}^kM^j_l(\mathcal{B}_8)^l_k\nonumber\\
  & + b_{7}(\mathcal{B}_{c\overline 3})^i H({15})_{ij}^kM^l_k(\mathcal{B}_8)^j_l
  +b_{8}(\mathcal{B}_{c\overline 3})^i H({15})_{ij}^kM^l_l(\mathcal{B}_8)^j_k  + b_{9}(\mathcal{B}_{c\overline 3})^i H({15})_{jl}^kM^j_i(\mathcal{B}_8)^l_k \nonumber\\&+ b_{10}(\mathcal{B}_{c\overline 3})^i H({15})_{jl}^kM^l_k(\mathcal{B}_8)^j_i
+b_{11} (\mathcal{B}_{c\overline 3})^i H(3_p)_i M_k^j(\mathcal{B}_8)_j^k
  +b_{12} (\mathcal{B}_{c\overline 3})^i H(3_p)_k M_j^j(\mathcal{B}_8)_i^k\nonumber\\ &+b_{13} (\mathcal{B}_{c\overline 3})^i H(3_p)_k M_i^j(\mathcal{B}_8)_j^k +b_{14} (\mathcal{B}_{c\overline 3})^i H(3_p)_k M_j^k(\mathcal{B}_8)_i^j
+b_{15} (\mathcal{B}_{c\overline 3})^i H(3_t)_i M_k^j(\mathcal{B}_8)_j^k
\nonumber\\  & +b_{16} (\mathcal{B}_{c\overline 3})^i H(3_t)_k M_j^j(\mathcal{B}_8)_i^k +b_{17} (\mathcal{B}_{c\overline 3})^i H(3_t)_k M_i^j(\mathcal{B}_8)_j^k +b_{18} (\mathcal{B}_{c\overline 3})^i H(3_t)_k M_j^k(\mathcal{B}_8)_i^j.
\end{align}
In this work, we refer to the decay amplitudes defined in Eq.~\eqref{amp4} as "2-rank $SU(3)$ irreducible amplitudes".
By substituting Eq.~\eqref{c8} into each term of Eq.~\eqref{amp3},
the relations between Eq.~\eqref{amp3} and Eq.~\eqref{amp4} are derived to be
\begin{align}\label{sol2}
 &b_1  =\frac{A_6-A_8}{4}, \qquad b_2  =\frac{A_5-A_7}{4},  \qquad b_3 = \frac{A_9-A_{10}}{4}, \qquad b_4 = \frac{-A_1+A_3}{4}, \qquad b_5 = \frac{-A_2+A_4}{4}, \nonumber\\ &b_{6} =\frac{A_6+A_8}{8}, \qquad
  b_{7}  =\frac{A_5+A_7}{8},  \qquad b_{8} = \frac{A_9+A_{10}}{8}, \qquad b_{9} = \frac{A_1+A_3}{8},\qquad  b_{10} = \frac{A_2+A_4}{8},\nonumber\\
& b_{11}= -\frac{1}{8}A_5+\frac{3}{8}A_7-\frac{1}{8}A_6+\frac{3}{8}A_8+ A_{18},\qquad b_{15} =\frac{3}{8}A_5-\frac{1}{8}A_7+\frac{3}{8}A_6-\frac{1}{8}A_8+A_{14},
\nonumber\\
 & b_{12} = \frac{3}{8}A_2-\frac{1}{8}A_4+\frac{3}{8}A_9 - \frac{1}{8}A_{10}+A_{16}, \qquad  b_{16} = -\frac{1}{8}A_2+\frac{3}{8}A_4-\frac{1}{8}A_9 + \frac{3}{8}A_{10}+A_{13},
   \nonumber\\
& b_{13} = -\frac{1}{8}A_1+\frac{3}{8}A_3+\frac{3}{8}A_5 - \frac{1}{8}A_7+A_{17},\qquad   b_{17} = \frac{3}{8}A_1-\frac{1}{8}A_3-\frac{1}{8}A_5 + \frac{3}{8}A_7+A_{11},
 \nonumber\\
& b_{14} = -\frac{1}{8}A_2+\frac{3}{8}A_4+\frac{3}{8}A_6 - \frac{1}{8}A_8+A_{15},\qquad   b_{18}  = \frac{3}{8}A_2-\frac{1}{8}A_4-\frac{1}{8}A_6 + \frac{3}{8}A_8+A_{12}.
\end{align}
The second-rank topological amplitudes and the $SU(3)$ irreducible amplitude are equivalent since the only difference between them is whether the effective Hamiltonian is decomposed into $SU(3)$ irreducible representation or not.
The inverse solution of Eq.~\eqref{sol2} is
\begin{align}\label{sol3}
 &A_1  =-2b_4+4b_9, \qquad A_2  =-2b_5+4b_{10},  \qquad A_3  =2b_4+4b_{9}, \qquad A_4  =2b_5+4b_{10}, \qquad A_5  =2b_2+4b_{7}, \nonumber\\  &A_6  =2b_1+4b_6, \qquad A_7  =-2b_2+4b_{7},  \qquad A_8  =-2b_1+4b_{6}, \qquad A_9  =2b_3+4b_{8}, \qquad A_{10}  =-2b_3+4b_{8}, \nonumber\\
&A_{11}  =b_{17}+b_2+b_4-b_7-b_9, \qquad A_{12}  =b_{18}+b_1+b_5-b_6-b_{10}, \qquad A_{13}  =b_{16}+b_3-b_5-b_8-b_{10}, \nonumber\\
&A_{14}  =b_{15}-b_1-b_2-b_6-b_7, \qquad A_{15}  =b_{14}-b_1-b_5-b_6-b_{10}, \qquad A_{16}  =b_{12}-b_3+b_5-b_8-b_{10}, \nonumber\\
&A_{17}  =b_{13}-b_2-b_4-b_7-b_9, \qquad A_{18}  =b_{11}+b_1+b_2-b_6-b_7.
\end{align}
By inserting Eq.~\eqref{sol} into Eq.~\eqref{sol2}, we obtain the equations for $SU(3)$ irreducible amplitudes decomposed in terms of topological amplitudes,
\begin{align}\label{sol11}
b_1& =-(2 a^A_4+a^A_7-a^A_9-a^A_{10}+a^A_{12})/2+\sqrt{3}(2 a^S_{5}-2 a^S_6-a^S_7+2 a^S_8-a^S_9+a^S_{10}-2 a^S_{11}+a^S_{12})/2,\nonumber\\
b_2&=(a^A_1-a^A_2-2 a^A_3+a^A_5-a^A_6)/2+\sqrt{3}(3 a^S_1-3 a^S_2-a^S_5+a^S_6)/2,\nonumber\\
b_3&=-(a^A_5-a^A_6-a^A_{15}+a^A_{16}+2 a^A_{17})/2+\sqrt{3}(a^S_5-a^S_6+3 a^S_{15}-3 a^S_{16})/2,\nonumber\\
b_4&=(a^A_5-a^A_6+a^A_8+a^A_9-a^A_{11}-a^A_{12})/2
-\sqrt{3}(a^S_5-a^S_6+2 a^S_7-a^S_8-a^S_9-2 a^S_{10}+a^S_{11}+a^S_{12})/2,\nonumber\\
b_5&=(a^A_1-a^A_2-2 a^A_3+a^A_5-a^A_6+a^A_8+a^A_9-a^A_{11}-a^A_{12}+a^A_{13}-a^A_{14}-2 a^A_{28}+2 a^A_{29})/2\nonumber\\&~~~~~~+\sqrt{3}(3 a^S_1-3 a^S_2-a^S_5+a^S_6-2 a^S_7+a^S_8+a^S_9+2 a^S_{10}-a^S_{11}-a^S_{12}-3 a^S_{13}+3 a^S_{14})/2,\nonumber\\
b_6&=(a^A_7-a^A_9+a^A_{10}-a^A_{12})/4+\sqrt{3}(a^S_7-2 a^S_8+a^S_9+a^S_{10}-2 a^S_{11}+a^S_{12})/4,\nonumber\\
b_7&=(a^A_1+a^A_2+a^A_5+a^A_6)/4-\sqrt{3}(a^S_1+a^S_2-2 a^S_3+2 a^S_4-a^S_5-a^S_6)/4, \nonumber\\
b_8&=-(a^A_5+a^A_6-a^A_{15}-a^A_{16})/4+\sqrt{3}(2 a^S_4-a^S_5-a^S_6-a^S_{15}-a^S_{16}+2 a^S_{17})/4, \nonumber\\
b_9&=(a^A_5+a^A_6+a^A_8+a^A_9+a^A_{11}+a^A_{12})/4+\sqrt{3}(-2 a^S_4+a^S_5+a^S_6-2 a^S_7+a^S_8+a^S_9-2 a^S_{10}+a^S_{11}+a^S_{12})/4,\nonumber\\
b_{10}&=-(a^A_1+a^A_2+a^A_5+a^A_6+a^A_8+a^A_9+a^A_{11}+a^A_{12}
-a^A_{13}-a^A_{14}-2 a^A_{28}-2 a^A_{29})/4\nonumber\\&~~~~~~+\sqrt{3}(a^S_1+a^S_2-2 a^S_3+2 a^S_4-a^S_5-a^S_6+2 a^S_7-a^S_8-a^S_9+2 a^S_{10}-a^S_{11}-a^S_{12}-3 a^S_{13}-3 a^S_{14})/4,\nonumber\\
b_{11}&=-(a^A_1-3 a^A_2-4 a^A_3+4 a^A_4+a^A_5-3 a^A_6-3a^A_7+3a^A_9+a^A_{10}-a^A_{12}-8 a^A_{23}+8 a^A_{25})/4\nonumber\\&~~~~~~-\sqrt{3}(7 a^S_1-5 a^S_2-2 a^S_3+2 a^S_4-7 a^S_5+5 a^S_6-3a^S_7+6 a^S_8-3a^S_9+a^S_{10}-2 a^S_{11}+a^S_{12}-8 a^S_{23}+16 a^S_{24}-8 a^S_{25})/4,\nonumber\\
b_{12}&=-(3 a^A_1-a^A_2-4 a^A_3-2 a^A_5-2 a^A_6+3 a^A_8+3a^A_9-a^A_{11}-a^A_{12}+a^A_{13}-3 a^A_{14}+5 a^A_{15}+a^A_{16}-4 a^A_{17}\nonumber\\&~~~~~~+8 a^A_{24}+8 a^A_{25}-8 a^A_{27}-6 a^A_{28}+2 a^A_{29}-16 a^A_{33}) /4-\sqrt{3}(5 a^S_1-7 a^S_2+2 a^S_3+4 a^S_4-2 a^S_5-2 a^S_6-6 a^S_7+3 a^S_8\nonumber\\&~~~~~~~~+3a^S_9+2 a^S_{10}-a^S_{11}-a^S_{12}-3 a^S_{13}+9 a^S_{14}+3 a^S_{15}-9 a^S_{16}+6 a^S_{17}-16 a^S_{23}+8 a^S_{24}+8 a^S_{25}+24 a^S_{27})/4,\nonumber\\
b_{13}&=(3 a^A_1-a^A_2-4 a^A_3-2 a^A_5-2 a^A_6+3 a^A_8+3a^A_9-a^A_{11}-a^A_{12}+8 a^A_{24}+8 a^A_{25})/4\nonumber\\&~~~~~~+\sqrt{3}(5 a^S_1-7 a^S_2+2 a^S_3+4 a^S_4-2 a^S_5-2 a^S_6-6 a^S_7+3a^S_8+3a^S_9+2 a^S_{10}-a^S_{11}-a^S_{12}-16 a^S_{23}+8 a^S_{24}+8 a^S_{25})/4,\nonumber\\
b_{14}&=(a^A_1-3 a^A_2-4 a^A_3+4 a^A_4+a^A_5-3 a^A_6-a^A_7+a^A_8+2 a^A_9-5 a^A_{10}-3 a^A_{11}+2 a^A_{12}+3 a^A_{13}-a^A_{14}-8 a^A_{23}+8 a^A_{25}\nonumber\\&~~~~~~+8 a^A_{26}-2 a^A_{28}+6 a^A_{29}+16 a^A_{32})/4+\sqrt{3}(7 a^S_1-5 a^S_2-2 a^S_3+2 a^S_4-7 a^S_5+5 a^S_6-3 a^S_7+3 a^S_8+a^S_{10}+7 a^S_{11}\nonumber\\&~~~~~~~~-8 a^S_{12}-9 a^S_{13}+3 a^S_{14}-8 a^S_{23}+16 a^S_{24}-8 a^S_{25}-24 a^S_{26})/4,\nonumber\\
b_{15}&=(3 a^A_1-a^A_2-4 a^A_3+4 a^A_4+3 a^A_5-a^A_6-a^A_7+a^A_9+3 a^A_{10}-3 a^A_{12}+8 a^A_{18}-8 a^A_{20})/4\nonumber\\&~~~~~~+\sqrt{3}(5 a^S_1-7 a^S_2+2 a^S_3-2 a^S_4-5 a^S_5+7 a^S_6-a^S_7+2 a^S_8-a^S_9+3 a^S_{10}-6 a^S_{11}+3 a^S_{12}+8 a^S_{18}-16 a^S_{19}+8 a^S_{20})/4,\nonumber\\
b_{16}&=(a^A_1-3 a^A_2-4 a^A_3+2 a^A_5+2 a^A_6+a^A_8+a^A_9-3 a^A_{11}-3 a^A_{12}+3 a^A_{13}-a^A_{14}-a^A_{15}-5 a^A_{16}-4 a^A_{17}-8 a^A_{19}-8 a^A_{20}\nonumber\\&~~~~~~+8 a^A_{22}-2 a^A_{28}+6 a^A_{29}+16 a^A_{31})/4+\sqrt{3}(7 a^S_1-5 a^S_2-2 a^S_3-4 a^S_4+2 a^S_5+2 a^S_6-2 a^S_7+a^S_8+a^S_9+6 a^S_{10}\nonumber\\&~~~~~~~~-3 a^S_{11}-3 a^S_{12}-9 a^S_{13}+3 a^S_{14}+9 a^S_{15}-3 a^S_{16}-6 a^S_{17}+16 a^S_{18}-8 a^S_{19}-8 a^S_{20}-24 a^S_{22})/4,\nonumber\\
b_{17}&=-(a^A_1-3 a^A_2-4 a^A_3+2 a^A_5+2 a^A_6+a^A_8+a^A_9-3 a^A_{11}-3 a^A_{12}-8 a^A_{19}-8 a^A_{20})/4\nonumber\\&~~~~~~-\sqrt{3}(7 a^S_1-5 a^S_2-2 a^S_3-4 a^S_4+2 a^S_5+2 a^S_6-2 a^S_7+a^S_8+a^S_9+6 a^S_{10}-3 a^S_{11}-3 a^S_{12}+16 a^S_{18}-8 a^S_{19}-8 a^S_{20})/4,\nonumber\\
b_{18}&=-(3 a^A_1-a^A_2-4 a^A_3+4 a^A_4+3 a^A_5-a^A_6+5 a^A_7+3 a^A_8-2 a^A_9+a^A_{10}-a^A_{11}-2 a^A_{12}+a^A_{13}-3 a^A_{14}+8 a^A_{18}-8 a^A_{20}\nonumber\\&~~~~~~-8 a^A_{21}-6 a^A_{28}+2 a^A_{29}-16 a^A_{30})/4-\sqrt{3}(5 a^S_1-7 a^S_2+2 a^S_3-2 a^S_4-5 a^S_5+7 a^S_6-a^S_7-7 a^S_8+8 a^S_9+3 a^S_{10}\nonumber\\&~~~~~~~~-3 a^S_{11}-3 a^S_{13}+9 a^S_{14}+8 a^S_{18}-16 a^S_{19}+8 a^S_{20}+24 a^S_{21})/4.
\end{align}
Note the inverse solution does not exist because the number of terms in the topological amplitudes is larger than the number of terms in the $SU(3)$ irreducible amplitudes.

The three indices of $H(\overline 6)_{ij}^k$ can be simplified to two upper indices through Levi-Civita tensor, $H(\overline 6)_{ij}^k =\epsilon_{ijl}H(\overline 6)^{lk}$.
The terms constructed by $H(\overline 6)_{ij}^k$ in Eq.~\eqref{amp4} can be written as
\begin{align}\label{c1}
 b_1(\mathcal{B}_{c\overline 3})^i H(\overline 6)_{ij}^kM^j_l( \mathcal{B}_8)^l_k &= b_1(\mathcal{B}_{c\overline 3})_{ji} H(\overline 6)^{ki}M^j_l( \mathcal{B}_8)^l_k,\nonumber\\
b_2(\mathcal{B}_{c\overline 3})^i H(\overline 6)_{ij}^kM^l_k(\mathcal{B}_8)^j_l &= b_2(\mathcal{B}_{c\overline 3})_{ji} H(\overline 6)^{ki}M^l_k( \mathcal{B}_8)^j_l,\nonumber\\
b_3(\mathcal{B}_{c\overline 3})^i H(\overline 6)_{ij}^kM^l_l(\mathcal{B}_8)^j_k&=b_3(\mathcal{B}_{c\overline 3})_{ji} H(\overline 6)^{ki}M^l_l( \mathcal{B}_8)^j_k,\nonumber\\
b_4(\mathcal{B}_{c\overline 3})^i H(\overline 6)_{jl}^kM^j_i(\mathcal{B}_8)^l_k&=b_4\big[(\mathcal{B}_{c\overline 3})_{jl} H(\overline 6)^{ki}M^j_i( \mathcal{B}_8)^l_k-(\mathcal{B}_{c\overline 3})_{ji} H(\overline 6)^{ki}M^j_l( \mathcal{B}_8)^l_k\nonumber\\&~~~~~+(\mathcal{B}_{c\overline 3})_{ji} H(\overline 6)^{ki}M^l_l( \mathcal{B}_8)^j_k\big],\nonumber\\
 b_5(\mathcal{B}_{c\overline 3})^i H(\overline 6)_{jl}^kM^l_k(\mathcal{B}_8)^j_i&=-b_5\big[(\mathcal{B}_{c\overline 3})_{ji} H(\overline 6)^{ki}M^l_k( \mathcal{B}_8)^j_l+(\mathcal{B}_{c\overline 3})_{jl} H(\overline 6)^{ki}M^j_i( \mathcal{B}_8)^l_k\big],
\end{align}
in which $(\mathcal{B}_{c\overline 3})^i=\epsilon^{ijk}(\mathcal{B}_{c\overline 3})_{jk}/2$
is used.
There are four invariant tensors in the RHS of Eq.~\eqref{c1}.
It indicates the degree of freedom of the $SU(3)$ irreducible amplitudes associated with $\overline 6$ representation is four, which is consistent with Refs.~\cite{He:2018joe,Groote:2021pxt}.
We can define four new parameters to replace $b_1\sim b_5$,
\begin{align}\label{c9}
  b_1^\prime = b_1 - b_4,\qquad b_2^\prime = b_2 - b_5,\qquad b_3^\prime = b_3 +b_4,\qquad b_4^\prime = b_4 - b_5.
\end{align}
Since Eq.~\eqref{amp3} and Eq.~\eqref{amp4} are equivalent, one of the amplitudes in Eq.~\eqref{amp3} is not independent.
According to Eq.~\eqref{sol2} and Eq.~\eqref{sol}, the $SU(3)$ irreducible amplitudes associated with $\overline 6$ representation are independent of the topologies involving quark-loops.
It means one of the amplitudes $A_1\sim A_{10}$ is not independent under the $SU(3)_F$ symmetry, regardless of whether the amplitudes $A_{11}\sim A_{18}$ are neglected or not.

\begin{figure}
  \centering
  \includegraphics[width=4cm]{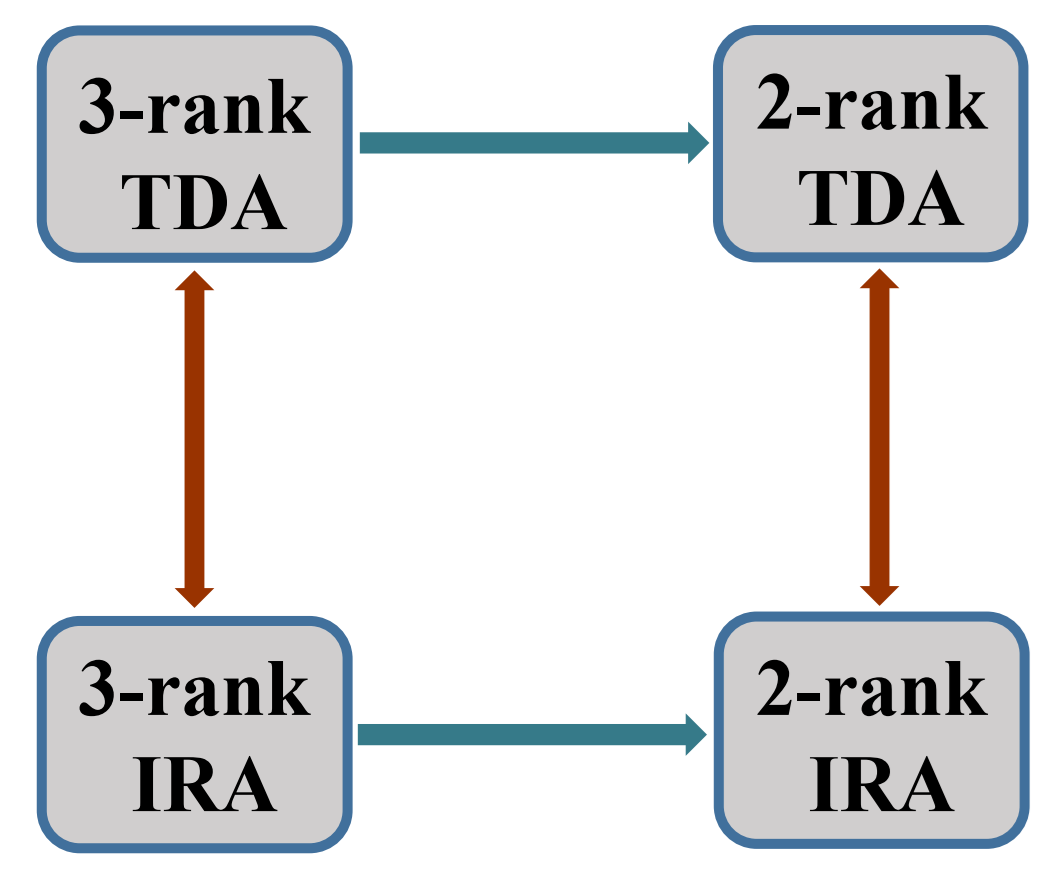}
  \caption{Relations between 3-rank topological amplitudes, 2-rank topological amplitudes, 3-rank $SU(3)$ irreducible amplitudes and 2-rank $SU(3)$ irreducible amplitudes.}\label{tt1}
\end{figure}
The amplitude of $\mathcal{B}_{c\overline 3}\to \mathcal{B}_8M$ decay can also be constructed by third-rank octet tensors and $SU(3)$ irreducible effective operators.
The third-rank $SU(3)$ irreducible amplitudes of $\mathcal{B}_{c\overline 3}\to \mathcal{B}_8^SM$ and $\mathcal{B}_{c\overline 3}\to \mathcal{B}_8^AM$ modes and the equivalence relations between the third-rank $SU(3)$ irreducible amplitudes and topological amplitudes are shown in Appendix.~\ref{su3}.
Moreover, the equations for the second-rank $SU(3)$ irreducible amplitudes decomposed by the third-rank $SU(3)$ irreducible amplitudes are derived.
The cross-check of Eq.~\eqref{sol11} is then undertaken via the approach of 3-rank topological amplitudes $\Rightarrow$ 3-rank $SU(3)$ irreducible amplitudes $\Rightarrow$  2-rank $SU(3)$ irreducible amplitudes.

In summary, we show the relations between the 3-rank topological amplitudes, the 2-rank topological amplitudes, the 3-rank $SU(3)$ irreducible amplitudes and the 2-rank $SU(3)$ irreducible amplitudes in Fig.~\ref{tt1}.
The double-headed arrow presents the 3(2)-rank amplitudes can be decomposed by the other 3(2)-rank amplitudes and the inverse solution exist.
The single-headed arrow denotes the two 2-rank amplitudes can be decomposed by corresponding 3-rank amplitudes, but inverse solutions do not exist.

\begin{table*}
\caption{Topological amplitudes of the Cabibbo-favored and doubly Cabibbo-suppressed $\mathcal{B}_{c\overline 3}\to \mathcal{B}_8^SM$ decays.}\label{ta1}
 \scriptsize
\begin{tabular}{|c|c|}
\hline\hline
 Channel & Topological amplitude \\\hline
$\Lambda^+_c\to \Lambda^0 \pi^+$ & $\frac{1}{2}\lambda_1(a^S_2-a^S_3+a^S_4-a^S_6+a^S_7-a^S_9-2a^S_{14})$ \\\hline
$\Lambda^+_c\to \Sigma^0 \pi^+$ & $-\frac{1}{2\sqrt{3}}\lambda_1(2a^S_1-a^S_2-a^S_3+a^S_4-2a^S_5+a^S_6+a^S_{7}
-2a^S_8+a^S_9)$ \\\hline
$\Lambda^+_c\to \Sigma^+ \pi^0$ & $\frac{1}{2\sqrt{3}}\lambda_1(2a^S_1-a^S_2-a^S_3+a^S_4-2a^S_5+a^S_6+a^S_{7}
-2a^S_8+a^S_9)$ \\\hline
$\Lambda^+_c\to \Sigma^+ \eta_8$ &$\frac{1}{6}\lambda_1(-2a^S_1+a^S_2+a^S_3+a^S_4-2a^S_5+a^S_6+a^S_{7}-2a^S_8
+a^S_9+4a^S_{10}-2a^S_{11}-2a^S_{12})$  \\\hline
$\Lambda^+_c\to \Sigma^+ \eta_1$ & $\frac{1}{3\sqrt{2}}\lambda_1(-2a^S_1+a^S_2+a^S_3+a^S_4-2a^S_5+a^S_6+a^S_{7}-2a^S_8
+a^S_9-2a^S_{10}+a^S_{11}+a^S_{12}-6a^S_{15}+3a^S_{16}+3a^S_{17})$ \\\hline
$\Lambda^+_c\to p \overline K^0$ & $\frac{1}{\sqrt{6}}\lambda_1(2a^S_{10}-a^S_{11}-a^S_{12}-3a^S_{13})$ \\\hline
$\Lambda^+_c\to \Xi^0 K^+$ & $\frac{1}{\sqrt{6}}\lambda_1(-a^S_4-a^S_5+2a^S_6-a^S_{7}-a^S_8+2a^S_9)$ \\\hline
$\Xi^0_c\to \Xi^- \pi^+$ & $\frac{1}{\sqrt{6}}\lambda_1(2a^S_7-a^S_8-a^S_9-3a^S_{14})$ \\\hline
$\Xi^0_c\to \Xi^0 \pi^0$ & $\frac{1}{2\sqrt{3}}\lambda_1(-a^S_1+2a^S_2-a^S_3-2a^S_{7}+a^S_8+a^S_9)$ \\\hline
$\Xi^0_c\to \Xi^0 \eta_8$ & $\frac{1}{6}\lambda_1(a^S_1-2a^S_2+a^S_3-2a^S_{4}-2a^S_5+4a^S_6-
2a^S_7+a^S_8+a^S_9-2a^S_{10}+4a^S_{11}
-2J^S_{12})$ \\\hline
$\Xi^0_c\to \Xi^0 \eta_1$ & $\frac{1}{3\sqrt{2}}\lambda_1(a^S_1-2a^S_2+a^S_3+a^S_{4}+a^S_5-2a^S_6-2a^S_7+a^S_8
+a^S_9+a^S_{10}-2a^S_{11}
+a^S_{12}+3a^S_{15}-6a^S_{16}+3a^S_{17})$ \\\hline
$\Xi^0_c\to \Lambda^0 \overline K^0$ & $\frac{1}{2}\lambda_1(-a^S_2+a^S_3-a^S_4+a^S_6-a^S_{10}+a^S_{11}+a^S_{13})$ \\\hline
$\Xi^0_c\to \Sigma^+ K^-$ & $\frac{1}{\sqrt{6}}\lambda_1(-a^S_{4}+2a^S_5-a^S_6-a^S_{10}-a^S_{11}+2a^S_{12})$ \\\hline
$\Xi^0_c\to \Sigma^0 \overline K^0$ &$-\frac{1}{2\sqrt{3}}\lambda_1(-2a^S_{1}+a^S_2+a^S_3-a^S_{4}+2a^S_{5}-a^S_{6}
-a^S_{10}-a^S_{11}+2a^S_{12}+3a^S_{13})$  \\\hline
$\Xi^+_c\to \Xi^0 \pi^+$ & $\frac{1}{\sqrt{6}}\lambda_1(a^S_{1}-2a^S_2+a^S_3+3a^S_{14})$ \\\hline
$\Xi^+_c\to \Sigma^+ \overline K^0$ & $\frac{1}{\sqrt{6}}\lambda_1(-2a^S_{1}+a^S_2+a^S_3+3a^S_{13})$ \\\toprule[1.2pt]
$\Lambda^+_c\to pK^0$ & $\frac{1}{\sqrt{6}}\lambda_2(2a^S_1-a^S_2-a^S_3-3a^S_{13})$ \\\hline
$\Lambda^+_c\to nK^+$ & $\frac{1}{\sqrt{6}}\lambda_2(-a^S_1+2a^S_2-a^S_3-3a^S_{14})$ \\\hline
$\Xi^0_c\to \Sigma^-K^+$ & $\frac{1}{\sqrt{6}}\lambda_2(2a^S_7-a^S_8-a^S_9-3a^S_{14})$ \\\hline
$\Xi^0_c\to \Lambda^0K^0$ & $\frac{1}{2}\lambda_2(-a^S_1+a^S_2+a^S_5-a^S_6-a^S_{11}+a^S_{12}+a^S_{13})$ \\\hline
$\Xi^0_c\to \Sigma^0K^0$ & $-\frac{1}{2\sqrt{3}}\lambda_1(-a^S_1-a^S_2+2a^S_3-2a^S_4+a^S_{5}+a^S_{6}-2a^S_{10}
+a^S_{11}+a^S_{12}+3a^S_{13})$ \\\hline
$\Xi^0_c\to p\pi^-$ & $\frac{1}{\sqrt{6}}\lambda_2(-a^S_4+2a^S_5-a^S_6-a^S_{10}-a^S_{11}+2a^S_{12})$ \\\hline
$\Xi^0_c\to n\pi^0$ & $\frac{1}{2\sqrt{3}}\lambda_2(-a^S_4-a^S_5+2a^S_6-2a^S_7+a^S_{8}+a^S_{9}-a^S_{10}
+2a^S_{11}-a^S_{12})$ \\\hline
$\Xi^0_c\to n\eta_8$ & $\frac{1}{6}\lambda_2(-2a^S_1+4a^S_2-2a^S_3+a^S_4+a^S_{5}-2a^S_{6}-2a^S_{7}+a^S_{8}
+a^S_{9}+a^S_{10}-2a^S_{11}+a^S_{12})$ \\\hline
$\Xi^0_c\to n\eta_1$ & $\frac{1}{3\sqrt{2}}\lambda_2(a^S_1-2a^S_2+a^S_3+a^S_4+a^S_{5}-2a^S_{6}-2a^S_{7}
+a^S_{8}+a^S_{9}+a^S_{10}-2a^S_{11}+a^S_{12}-3a^S_{15}+6a^S_{16}-3a^S_{17})$ \\\hline
$\Xi^+_c\to \Lambda^0K^+$ &  $\frac{1}{2}\lambda_2(-a^S_1+a^S_2+a^S_5-a^S_{6}+a^S_{8}-a^S_{9}-a^S_{14} )$ \\\hline
$\Xi^+_c\to \Sigma^0K^+$ & $-\frac{1}{2\sqrt{3}}\lambda_2(-a^S_1-a^S_2+2a^S_3-2a^S_4+a^S_{5}+a^S_{6}-2a^S_{7}
+a^S_{8}+a^S_{9}+3a^S_{14})$ \\\hline
$\Xi^+_c\to \Sigma^+K^0$ & $\frac{1}{\sqrt{6}}\lambda_2(-2a^S_{10}+a^S_{11}+a^S_{12}+3a^S_{13})$ \\\hline
$\Xi^+_c\to p\pi^0$ & $\frac{1}{2\sqrt{3}}\lambda_2(-a^S_4+2a^S_5-a^S_6-a^S_7+2a^S_{8}-a^S_{9}-2a^S_{10}
+a^S_{11}+a^S_{12})$ \\\hline
$\Xi^+_c\to p\eta_8$ & $\frac{1}{6}\lambda_2(-4a^S_1+2a^S_2+2a^S_3-a^S_4+2a^S_{5}-a^S_{6}-a^S_{7}+2a^S_{8}
-a^S_{9}+2a^S_{10}-a^S_{11}-a^S_{12})$ \\\hline
$\Xi^+_c\to p\eta_1$ & $\frac{1}{3\sqrt{2}}\lambda_2(2a^S_1-a^S_2-a^S_3-a^S_4+2a^S_{5}-a^S_{6}-a^S_{7}
+2a^S_{8}-a^S_{9}+2a^S_{10}-a^S_{11}-a^S_{12}+6a^S_{15}-3a^S_{16}-3a^S_{17})$ \\\hline
$\Xi^+_c\to n\pi^+$ & $\frac{1}{\sqrt{6}}\lambda_2(a^S_{4}+a^S_{5}-2a^S_{6}+a^S_{7}+a^S_{8}-2a^S_{9})$ \\\hline
\end{tabular}
\end{table*}
\begin{table*}
\caption{Topological amplitudes of the singly Cabibbo-favored $\mathcal{B}_{c\overline 3}\to \mathcal{B}_8^SM$ decays $(1)$.}\label{ta2}
 \scriptsize
\begin{tabular}{|c|c|}
\hline\hline
 Channel & Topological amplitude \\\hline
$\Lambda^+_c\to \Lambda^0K^+$ & $\frac{1}{2}\lambda_d(a^S_5-a^S_6+a^S_8-a^S_9+a^S_{19}-a^S_{20}-2a^S_{21})
+\frac{1}{2}\lambda_s(a^S_2-a^S_3-2a^S_{14}+a^S_{19}-a^S_{20}-2a^S_{21})$ \\ & $-\frac{1}{2}\lambda_b(a^{SP}_1-a^{SP}_5+a^{SP}_8+a^{SP}_{19}+3a^{SP}_{24}-a^{SP}_3
+a^{SP}_6-a^{SP}_9-a^{SP}_{20}-3a^{SP}_{25}
+2a^{SP}_{11}-2a^{SP}_{12}-2a^{SP}_{13}-2a^{SP}_{21}-6a^{PS}_{26})$\\\hline
$\Lambda^+_c\to \Sigma^0K^+$ &  $\frac{1}{2\sqrt{3}}\lambda_d(2a^S_4-a^S_5-a^S_6+2a^S_7-a^S_{8}-a^S_{9}-2a^S_{18}
+a^S_{19}+a^S_{20})
+\frac{1}{2\sqrt{3}}\lambda_s(-2a^S_1+a^S_2+a^S_{3}-2a^S_{18}+a^S_{19}+a^S_{20})$ \\ &
$-\frac{1}{2\sqrt{3}}\lambda_b(-2a^{SP}_2+2a^{SP}_4-2a^{SP}_7-2a^{SP}_{18}-6a^{SP}_{23}
+a^{SP}_1-a^{SP}_5+a^{SP}_8+a^{SP}_{19}+3a^{SP}_{24}
+a^{SP}_3-a^{SP}_6+a^{SP}_9+a^{SP}_{20}+3a^{SP}_{25})$\\\hline
$\Lambda^+_c\to \Sigma^+K^0$ & $\frac{1}{\sqrt{6}}\lambda_d(-2a^S_{10}+a^S_{11}+a^S_{12}-2a^S_{18}+a^S_{19}+a^S_{20})
+\frac{1}{\sqrt{6}}\lambda_s(-2a^S_1+a^S_2+a^S_{3}-2a^S_{18}+a^S_{19}+a^S_{20})$ \\
 &$-\frac{1}{\sqrt{6}}\lambda_b(-2a^{SP}_2+2a^{SP}_4-2a^{SP}_7-2a^{SP}_{18}-6a^{SP}_{23}
 +a^{SP}_1-a^{SP}_5+a^{SP}_8+a^{SP}_{19}+3a^{SP}_{24}
 +a^{SP}_3-a^{SP}_6+a^{SP}_9+a^{SP}_{20}+3a^{SP}_{25})$\\\hline
$\Lambda^+_c\to p\pi^0$ & $\frac{1}{2\sqrt{3}}\lambda_d(-2a^S_{1}+a^S_{2}+a^S_{3}-a^S_{4}+2a^S_{5}-a^S_{6}
-a^S_{7}+2a^S_{8}-a^S_{9}-2a^S_{10}+a^S_{11}
+a^S_{12}+3a^S_{13}-a^S_{18}+2a^S_{19}-a^S_{20}-3a^S_{21})$\\&$
+\frac{1}{2\sqrt{3}}\lambda_s(-a^S_{18}+2a^S_{19}-a^S_{20}-3a^S_{21})
-\frac{1}{2\sqrt{3}}\lambda_b(-a^{SP}_2+a^{SP}_4-a^{SP}_7-a^{SP}_{18}-3a^{SP}_{23} $ \\
 & $+ 2a^{SP}_1-2a^{SP}_5+2a^{SP}_8+2a^{SP}_{19}+6a^{SP}_{24}-a^{SP}_3+a^{SP}_6-a^{SP}_9
 -a^{SP}_{20}-3a^{SP}_{25}+3a^{SP}_{11}-3a^{SP}_{12}
 -3a^{SP}_{13}-3a^{SP}_{21}-9a^{SP}_{26})$\\ \hline
$\Lambda^+_c\to p\eta_8$ & $\frac{1}{6}\lambda_d(2a^S_{1}-a^S_{2}-a^S_{3}-a^S_{4}+2a^S_{5}-a^S_{6}-a^S_{7}
+2a^S_{8}-a^S_{9}+2a^S_{10}-a^S_{11}
-a^S_{12}-3a^S_{13}+3a^S_{18}-3a^S_{20}-3a^S_{21})$\\&$
+\frac{1}{6}\lambda_s(6a^S_{13}+3a^S_{18}-3a^S_{20}-3a^S_{21})
-\frac{1}{6}\lambda_b(3a^{SP}_2-3a^{SP}_4+3a^{SP}_7$ \\&
$+3a^{SP}_{18}+9a^{SP}_{23}   -3a^{SP}_3+3a^{SP}_6-3a^{SP}_9-3a^{SP}_{20}-9a^{SP}_{25}+3a^{SP}_{11}-3a^{SP}_{12}
-3a^{SP}_{13}-3a^{SP}_{21}-9a^{SP}_{26})$ \\\hline
$\Lambda^+_c\to p\eta_1$ & $\frac{1}{3\sqrt{2}}\lambda_d(2a^S_{1}-a^S_{2}-a^S_{3}-a^S_{4}+2a^S_{5}-a^S_{6}
-a^S_{7}+2a^S_{8}-a^S_{9}+2a^S_{10}-a^S_{11}
-a^S_{12}-3a^S_{13}+6a^S_{15}-3a^S_{16}-3a^S_{17} $ \\& $  +3a^S_{18}-3a^S_{20}-3a^S_{21}-9a^S_{22})
+\frac{1}{3\sqrt{2}}\lambda_s(-3a^S_{13}+3a^S_{18}-3a^S_{20}-3a^S_{21}-9a^S_{22})$ \\& $-\frac{1}{3\sqrt{2}}\lambda_b(3a^{SP}_2-3a^{SP}_4+3a^{SP}_7+3a^{SP}_{18}+9a^{SP}_{23} -3a^{SP}_3+3a^{SP}_6-3a^{SP}_9-3a^{SP}_{20}-9a^{SP}_{25}$\\
&$+3a^{SP}_{11}-3a^{SP}_{12}-3a^{SP}_{13}-3a^{SP}_{21}-9a^{SP}_{26}-9a^{SP}_{14}
+9a^{SP}_{16}-9a^{SP}_{17}-9a^{SP}_{22}-27a^{SP}_{27})$
 \\\hline
$\Lambda^+_c\to n\pi^+$ & $\frac{1}{\sqrt{6}}\lambda_d(-a^S_{1}+2a^S_{2}-a^S_{3}+a^S_{4}+a^S_{5}-2a^S_{6}
+a^S_{7}+a^S_{8}-2a^S_{9}-3a^S_{14}
-a^S_{18}+2a^S_{19}-a^S_{20}-3a^S_{21})$\\&$
+\frac{1}{\sqrt{6}}\lambda_s(-a^S_{18}+2a^S_{19}-a^S_{20}-3a^S_{21})$ \\
 & $-\frac{1}{\sqrt{6}}\lambda_b(-a^{SP}_2+a^{SP}_4-a^{SP}_7-a^{SP}_{18}-3a^{SP}_{23}
 +2a^{SP}_1-2a^{SP}_5+2a^{SP}_8+2a^{SP}_{19}+6a^{SP}_{24}
 -a^{SP}_3+a^{SP}_6-a^{SP}_9$\\& $-a^{SP}_{20}-3a^{SP}_{25}+3a^{SP}_{11}-3a^{SP}_{12}-3a^{SP}_{13}
 -3a^{SP}_{21}-9a^{SP}_{26})$ \\\hline
$\Xi^0_c\to \Sigma^-\pi^+$ & $\frac{1}{\sqrt{6}}\lambda_d(a^S_{7}+a^S_{8}-2a^S_{9}-3a^S_{14}-3a^S_{21})
+\frac{1}{\sqrt{6}}\lambda_s(-a^S_{7}+2a^S_{8}-a^S_{9}-3a^S_{21})
-\frac{1}{\sqrt{6}}\lambda_b(3a^{SP}_{11}-3a^{SP}_{12}-3a^{SP}_{13}-3a^{SP}_{21}
-9a^{SP}_{26})$\\\hline
$\Xi^0_c\to \Lambda^0\pi^0$ & $\frac{1}{2\sqrt{2}}\lambda_d(a^S_{1}-a^S_{2}+a^S_{7}-a^S_{8}+a^S_{10}-a^S_{11}
-a^S_{13}+a^S_{18}-a^S_{19}+a^S_{21})
+\frac{1}{2\sqrt{2}}\lambda_s(-a^S_{4}+a^S_{6}-a^S_{7}+a^S_{9}+a^S_{18}
-a^S_{19}+a^S_{21})$ \\
 & $-\frac{1}{2\sqrt{2}}\lambda_b(a^{SP}_2-a^{SP}_4+a^{SP}_7+a^{SP}_{18}+3a^{SP}_{23}
 -a^{SP}_1+a^{SP}_5-a^{SP}_8-a^{SP}_{19}-3a^{SP}_{24}
 -a^{SP}_{11}+a^{SP}_{12}+a^{SP}_{13}+a^{SP}_{21}+3a^{SP}_{26})$ \\\hline
$\Xi^0_c\to \Lambda^0\eta_8$ & $\frac{1}{2\sqrt{6}}\lambda_d(-a^S_{1}+a^S_{2}+ 2a^S_{5}-2a^S_{6} +a^S_{7}-a^S_{8}-a^S_{10}+a^S_{11}+a^S_{13}-a^S_{18}+3a^S_{19}-2a^S_{20}+a^S_{21})$\\&$
+\frac{1}{2\sqrt{6}}\lambda_s(2a^S_{2}-2a^S_{3}+a^S_{4}-a^S_{6}-a^S_{7}+a^S_{9}
+2a^S_{11}-2a^S_{12}+2a^S_{13}-a^S_{18}+3a^S_{19}-2a^S_{20}+a^S_{21})$ \\
 & $-\frac{1}{2\sqrt{6}}\lambda_b( -a^{SP}_2+a^{SP}_4-a^{SP}_7-a^{SP}_{18}-3a^{SP}_{23} + 3a^{SP}_1-3a^{SP}_5+3a^{SP}_8+3a^{SP}_{19}+9a^{SP}_{24} -2 a^{SP}_3+2a^{SP}_6-2a^{SP}_9$
\\&$ -2a^{SP}_{20}-6a^{SP}_{25}-a^{SP}_{11}+a^{SP}_{12}+a^{SP}_{13}+a^{SP}_{21}+3a^{SP}_{26})$ \\\hline
$\Xi^0_c\to \Lambda^0\eta_1$ & $\frac{1}{2\sqrt{3}}\lambda_d(-a^S_{1}+a^S_{2}-a^S_{5}+a^S_{6} +a^S_{7}-a^S_{8}-a^S_{10}+a^S_{11}+a^S_{13}-3a^S_{15}+3a^S_{16} -a^S_{18}+a^S_{20}+a^S_{21}+3a^S_{22})$\\&$
+\frac{1}{2\sqrt{3}}\lambda_s(-a^S_{2}+a^S_{3}+a^S_{4}-a^S_{6}-a^S_{7}+a^S_{9}
-a^S_{11}+a^S_{12}+a^S_{13} -3a^S_{16}+3a^S_{17} -a^S_{18}+a^S_{20}+a^S_{21}+3a^S_{22})$ \\
 & $-\frac{1}{2\sqrt{3}}\lambda_b(-a^{SP}_2+a^{SP}_4-a^{SP}_7-a^{SP}_{18}-3a^{SP}_{23}
 +a^{SP}_3-a^{SP}_6+a^{SP}_9+a^{SP}_{20}+3a^{SP}_{25}
 -a^{SP}_{11}+a^{SP}_{12}+a^{SP}_{13}$\\&$
 +a^{SP}_{21}+3a^{SP}_{26}+3a^{SP}_{14}-3a^{SP}_{16}+3a^{SP}_{17}+3a^{SP}_{22}+9a^{SP}_{27})$ \\\hline
$\Xi^0_c\to \Sigma^0\pi^0$ & $\frac{1}{2\sqrt{6}}\lambda_d(-a^S_{1}-a^S_{2}+2a^S_{3}+a^S_{7}+a^S_{8}-2a^S_{9} -a^S_{10}-a^S_{11}+2a^S_{12}+3a^S_{13}-a^S_{18}-a^S_{19}+2a^S_{20}-3a^S_{21})$\\&$
+\frac{1}{2\sqrt{6}}\lambda_s(a^S_{4}-2a^S_{5}+a^S_{6}-a^S_{7}+2a^S_{8}-a^S_{9}
-a^S_{18}-a^S_{19}+2a^S_{20}-3a^S_{21})$\\&
$-\frac{1}{2\sqrt{6}}\lambda_b(-a^{SP}_2+a^{SP}_4-a^{SP}_7-a^{SP}_{18}-3a^{SP}_{23}
-a^{SP}_1+a^{SP}_5-a^{SP}_8-a^{SP}_{19}-3a^{SP}_{24}
+2a^{SP}_3-2a^{SP}_6+2a^{SP}_9$\\&$
+2a^{SP}_{20}+6a^{SP}_{25}+3a^{SP}_{11}-3a^{SP}_{12}-3a^{SP}_{13}-3a^{SP}_{21}-9a^{SP}_{26})$\\\hline
$\Xi^0_c\to \Sigma^0\eta_8$ &
$\frac{1}{6\sqrt{2}}\lambda_d(a^S_{1}+a^S_{2}-2a^S_{3}+4a^S_{4}-2a^S_{5}-2a^S_{6} +a^S_{7}+a^S_{8}-2a^S_{9}+a^S_{10}+a^S_{11}-2a^S_{12}-3a^S_{13}-3a^S_{18}+3a^S_{19}
-3a^S_{21})$\\&$
+\frac{1}{6\sqrt{2}}\lambda_s(-4a^S_{1}+2a^S_{2}+2a^S_{3}-a^S_{4}+2a^S_{5}-a^S_{6}
-a^S_{7}+2a^S_{8}-a^S_{9}-4a^S_{10}+2a^S_{11}+2a^S_{12}+6a^S_{13}-3a^S_{18}
+3a^S_{19}-3a^S_{21})$\\&
$-\frac{1}{2\sqrt{2}}\lambda_b(-a^{SP}_2+a^{SP}_4-a^{SP}_7-a^{SP}_{18}-3a^{SP}_{23}
+a^{SP}_1-a^{SP}_5+a^{SP}_8+a^{SP}_{19}+3a^{SP}_{24}
+a^{SP}_{11}-a^{SP}_{12}-a^{SP}_{13}-a^{SP}_{21}-3a^{SP}_{26})$\\\hline
$\Xi^0_c\to \Sigma^0\eta_1$ & $\frac{1}{6}\lambda_d(a^S_{1}+a^S_{2}-2a^S_{3}-2a^S_{4}+a^S_{5}+a^S_{6} +a^S_{7}+a^S_{8}-2a^S_{9}+a^S_{10}+a^S_{11}-2a^S_{12}-3a^S_{13}$\\&$ +3a^S_{15}+3a^S_{16}-6a^S_{17}+3a^S_{18}-3a^S_{20}-3a^S_{21}-9a^S_{22})$\\&$
+\frac{1}{6}\lambda_s(2a^S_{1}-a^S_{2}-a^S_{3}-a^S_{4}+2a^S_{5}-a^S_{6}
-a^S_{7}+2a^S_{8}-a^S_{9}+2a^S_{10}-a^S_{11}-a^S_{12}-3a^S_{13}$\\&$
+6a^S_{15}-3a^S_{16}-3a^S_{17}   +3a^S_{18}-3a^S_{20}-3a^S_{21}-9a^S_{22})$\\&
$-\frac{1}{2}\lambda_b(a^{SP}_2-a^{SP}_4+a^{SP}_7+a^{SP}_{18}+3a^{SP}_{23} - a^{SP}_3+a^{SP}_6-a^{SP}_9-a^{SP}_{20}-3a^{SP}_{25}+a^{SP}_{11}-a^{SP}_{12}
-a^{SP}_{13}-a^{SP}_{21}-3a^{SP}_{26}$\\
&$-3a^{SP}_{14}+3a^{SP}_{16}-3a^{SP}_{17}-3a^{SP}_{22}-9a^{SP}_{27})$\\\hline
$\Xi^0_c\to n\overline K^0$ & $\frac{1}{\sqrt{6}}\lambda_d(a^S_{1}-2a^S_{2}+a^S_{3}-a^S_{4}-a^S_{5}+2a^S_{6}
+a^S_{18}-2a^S_{19}+a^S_{20})
+\frac{1}{\sqrt{6}}\lambda_s(a^S_{10}-2a^S_{11}+a^S_{12}+a^S_{18}-2a^S_{19}+a^S_{20})$\\&
$-\frac{1}{\sqrt{6}}\lambda_b(a^{SP}_2-a^{SP}_4+a^{SP}_7+a^{SP}_{18}+3a^{SP}_{23}
-2a^{SP}_1+2a^{SP}_5-2a^{SP}_8-2a^{SP}_{19}-6a^{SP}_{24}
+a^{SP}_3-a^{SP}_6+a^{SP}_9+a^{SP}_{20}+3a^{SP}_{25})$\\\hline
$\Xi^0_c\to \Xi^-K^+$ &
$\frac{1}{\sqrt{6}}\lambda_d(-a^S_{7}+2a^S_{8}-a^S_{9}-3a^S_{21})
+\frac{1}{\sqrt{6}}\lambda_s(a^S_{7}+a^S_{8}-2a^S_{9}-3a^S_{14}-3a^S_{21})
-\frac{1}{\sqrt{6}}\lambda_b(-3a^{SP}_{11}+3a^{SP}_{12}+3a^{SP}_{13}+3a^{SP}_{21}
+9a^{SP}_{26})$\\\hline
$\Xi^0_c\to \Xi^0K^0$ &
$\frac{1}{\sqrt{6}}\lambda_d(a^S_{10}-2a^S_{11}+a^S_{12}+a^S_{18}-2a^S_{19}+a^S_{20})
+\frac{1}{\sqrt{6}}\lambda_s(a^S_{1}-2a^S_{2}+a^S_{3}-a^S_{4}-a^S_{5}+2a^S_{6}
+a^S_{18}-2a^S_{19}+a^S_{20})$\\&
$-\frac{1}{\sqrt{6}}\lambda_b(a^{SP}_2-a^{SP}_4+a^{SP}_7+a^{SP}_{18}+3a^{SP}_{23}
-2a^{SP}_1+2a^{SP}_5-2a^{SP}_8-2a^{SP}_{19}-6a^{SP}_{24}
+a^{SP}_3-a^{SP}_6+a^{SP}_9+a^{SP}_{20}+3a^{SP}_{25})$\\\hline
$\Xi^0_c\to \Sigma^+\pi^-$ &
$\frac{1}{\sqrt{6}}\lambda_d(-a^S_{10}-a^S_{11}+2a^S_{12}-a^S_{18}-a^S_{19}+2a^S_{20})
+\frac{1}{\sqrt{6}}\lambda_s(a^S_{4}-2a^S_{5}+a^S_{6}-a^S_{18}-a^S_{19}+2a^S_{20})$\\&
$-\frac{1}{\sqrt{6}}\lambda_b(-a^{SP}_2+a^{SP}_4-a^{SP}_7-a^{SP}_{18}-3a^{SP}_{23}
-a^{SP}_1+a^{SP}_5-a^{SP}_8-a^{SP}_{19}-3a^{SP}_{24}
+2a^{SP}_3-2a^{SP}_6+2a^{SP}_9+2a^{SP}_{20}+6a^{SP}_{25})$\\\hline
$\Xi^0_c\to pK^-$ &  $\frac{1}{\sqrt{6}}\lambda_d(a^S_{4}-2a^S_{5}+a^S_{6}-a^S_{18}-a^S_{19}+2a^S_{20})
+\frac{1}{\sqrt{6}}\lambda_s(-a^S_{10}-a^S_{11}+2a^S_{12}-a^S_{18}-a^S_{19}+2a^S_{20})$\\&
$-\frac{1}{\sqrt{6}}\lambda_b(-a^{SP}_2+a^{SP}_4-a^{SP}_7-a^{SP}_{18}-3a^{SP}_{23}
-a^{SP}_1+a^{SP}_5-a^{SP}_8-a^{SP}_{19}-3a^{SP}_{24}
+2a^{SP}_3-2a^{SP}_6+2a^{SP}_9+2a^{SP}_{20}+6a^{SP}_{25})$\\\hline
\end{tabular}
\end{table*}
\begin{table*}
\caption{Topological amplitudes of the singly Cabibbo-favored $\mathcal{B}_{c\overline 3}\to \mathcal{B}_8^SM$ decays $(2)$.}\label{ta3}
 \scriptsize
\begin{tabular}{|c|c|}
\hline\hline
 Channel & Topological amplitude \\\hline
$\Xi^+_c\to \Lambda^0\pi^+$ &
$\frac{1}{2}\lambda_d(-a^S_{1}+a^S_{2}-a^S_{14}-a^S_{18}+a^S_{19}-a^S_{21})
+\frac{1}{2}\lambda_s(a^S_{4}-a^S_{6}+a^S_{7}-a^S_{9}-a^S_{18}+a^S_{19}-a^S_{21})$\\&
$-\frac{1}{2}\lambda_b(-a^{SP}_2+a^{SP}_4-a^{SP}_7-a^{SP}_{18}-3a^{SP}_{23}
+a^{SP}_1-a^{SP}_5+a^{SP}_8+a^{SP}_{19}+3a^{SP}_{24}
+a^{SP}_{11}-a^{SP}_{12}-a^{SP}_{13}-a^{SP}_{21}-3a^{SP}_{26})$\\\hline
$\Xi^+_c\to \Sigma^0\pi^+$ &
$\frac{1}{2\sqrt{3}}\lambda_d(a^S_{1}+a^S_{2}-2a^S_{3}-3a^S_{14}+a^S_{18}+a^S_{19}
-2a^S_{20}-3a^S_{21})$\\&$
+\frac{1}{2\sqrt{3}}\lambda_s(-a^S_{4}+2a^S_{5}-a^S_{6}-a^S_{7}+2a^S_{8}-a^S_{9}
+a^S_{18}+a^S_{19}-2a^S_{20}-3a^S_{21})$\\&
$-\frac{1}{2\sqrt{3}}\lambda_b(a^{SP}_2-a^{SP}_4+a^{SP}_7+a^{SP}_{18}+3a^{SP}_{23}
+a^{SP}_1-a^{SP}_5+a^{SP}_8+a^{SP}_{19}+3a^{SP}_{24}  -2a^{SP}_3+2a^{SP}_6-2a^{SP}_9$\\&$
-2a^{SP}_{20}-6a^{SP}_{25}+3a^{SP}_{11}-3a^{SP}_{12}-3a^{SP}_{13}-3a^{SP}_{21}-9a^{SP}_{26})$\\\hline
$\Xi^+_c\to \Sigma^+\pi^0$ &  $\frac{1}{2\sqrt{3}}\lambda_d(-3a^S_{13}-a^S_{18}-a^S_{19}+2a^S_{20}+3a^S_{21})
+\frac{1}{2\sqrt{3}}\lambda_s(a^S_{4}-2a^S_{5}+a^S_{6}+a^S_{7}-2a^S_{8}+a^S_{9}
-a^S_{18}-a^S_{19}+2a^S_{20}+3a^S_{21})$\\&
$-\frac{1}{2\sqrt{3}}\lambda_b(-a^{SP}_2+a^{SP}_4-a^{SP}_7-a^{SP}_{18}-3a^{SP}_{23}  -a^{PS}_1+a^{SP}_5-a^{SP}_8-a^{SP}_{19}-3a^{SP}_{24}  +2a^{SP}_3-2a^{SP}_6+2a^{SP}_9$\\&$
+2a^{SP}_{20}+6a^{SP}_{25}-3a^{SP}_{11}+3a^{SP}_{12}+3a^{SP}_{13}+3a^{SP}_{21}
+9a^{SP}_{26})$\\\hline
$\Xi^+_c\to \Sigma^+\eta_8$ &  $\frac{1}{6}\lambda_d(3a^S_{13}+3a^S_{18}-3a^S_{19}+3a^S_{21})
+\frac{1}{6}\lambda_s(4a^S_{1}-2a^S_{2}-2a^S_{3}+a^S_{4}-2a^S_{5}+a^S_{6}$\\&$
+a^S_{7}
-2a^S_{8}+a^S_{9}+4a^S_{10}  -2a^S_{11}-2a^S_{12}-6a^S_{13}+3a^S_{18}-3a^S_{19}+3a^S_{21})$\\&
$-\frac{1}{2}\lambda_b(a^{SP}_2-a^{SP}_4+a^{SP}_7+a^{SP}_{18}+3a^{SP}_{23}  -a^{SP}_1+a^{SP}_5-a^{SP}_8-a^{SP}_{19}-3a^{SP}_{24}-a^{SP}_{11}+a^{SP}_{12}
+a^{SP}_{13}+a^{SP}_{21}+3a^{SP}_{26})$\\\hline
$\Xi^+_c\to \Sigma^+\eta_1$ &  $\frac{1}{3\sqrt{2}}\lambda_d(3a^S_{13}-3a^S_{18}+3a^S_{20}+3a^S_{21}+9a^S_{22})
+\frac{1}{3\sqrt{2}}\lambda_s(-2a^S_{1}+a^S_{2}+a^S_{3}+a^S_{4}-2a^S_{5}+a^S_{6}$\\&$
+a^S_{7}-2a^S_{8}+a^S_{9}-2a^S_{10}  +a^S_{11}+a^S_{12}+3a^S_{13}-6a^S_{15}+3a^S_{16}+3a^S_{17}-3a^S_{18}+3a^S_{20}
+3a^S_{21}+9a^S_{22})$
\\ & $-\frac{1}{\sqrt{2}}\lambda_b(-a^{SP}_2+a^{SP}_4-a^{SP}_7-a^{SP}_{18}-3a^{SP}_{23}
+a^{SP}_3-a^{SP}_6+a^{SP}_9+a^{SP}_{20}+3a^{SP}_{25}
-a^{SP}_{11}+a^{SP}_{12}+a^{SP}_{13}$\\&$+a^{SP}_{21}+3a^{SP}_{26}+3a^{SP}_{14}
-3a^{SP}_{16}+3a^{SP}_{17}+3a^{SP}_{22}+9a^{SP}_{27})$\\\hline
$\Xi^+_c\to p\overline K^0$ & $\frac{1}{\sqrt{6}}\lambda_d(2a^S_{1}-a^S_{2}-a^S_{3}+2a^S_{18}-a^S_{19}-a^S_{20})
+\frac{1}{\sqrt{6}}\lambda_s(2a^S_{10}-a^S_{11}-a^S_{12}+2a^S_{18}-a^S_{19}
-a^S_{20})-\frac{1}{\sqrt{6}}\lambda_b(2a^{SP}_2-2a^{SP}_4$\\&
$+2a^{SP}_7+2a^{SP}_{18}
+6a^{SP}_{23}  -a^{SP}_1+a^{SP}_5-a^{SP}_8-a^{SP}_{19}-3a^{SP}_{24}  -a^{SP}_3+a^{SP}_6-a^{SP}_9-a^{SP}_{20}-3a^{SP}_{25})$\\\hline
$\Xi^+_c\to \Xi^0K^+$ & $\frac{1}{\sqrt{6}}\lambda_d(a^S_{18}-2a^S_{19}+a^S_{20}+3a^S_{21})
+\frac{1}{\sqrt{6}}\lambda_s(a^S_{1}-2a^S_{2}+a^S_{3}-a^S_{4}-a^S_{5}+2a^S_{6}-a^S_{7}
-a^S_{8}+2a^S_{9}+3a^S_{14}+a^S_{18}$\\&$-2a^S_{19}+a^S_{20}+3a^S_{21} )$
$-\frac{1}{\sqrt{6}}\lambda_b(a^{SP}_2-a^{SP}_4+a^{SP}_7+a^{SP}_{18}+3a^{SP}_{23}  -2a^{SP}_1+2a^{SP}_5-2a^{SP}_8-2a^{SP}_{19}
$\\&
$-6a^{SP}_{24}+a^{SP}_3-a^{SP}_6+a^{SP}_9
+a^{SP}_{20}+3a^{SP}_{25}-3a^{SP}_{11}+3a^{SP}_{12}+3a^{SP}_{13}+3a^{SP}_{21}
+9a^{SP}_{26})$\\\hline
\end{tabular}
\end{table*}

\begin{table*}
\caption{Topological amplitudes of the Cabibbo-favored and  doubly Cabibbo-favored $\mathcal{B}_{c\overline 3}\to \mathcal{B}_8^AM$ decays.}\label{ta4}
 \scriptsize
\begin{tabular}{|c|c|}
\hline\hline
 Channel & Topological amplitude \\\hline
$\Lambda^+_c\to \Lambda^0 \pi^+$ & $\frac{1}{2\sqrt{3}}\lambda_1(2a^A_1+a^A_2-a^A_3+a^A_4+2a^A_5+a^A_6+a^A_7
+2a^A_8+a^A_9-2a^A_{14}-4a^A_{28})$ \\\hline
$\Lambda^+_c\to \Sigma^0 \pi^+$ & $\frac{1}{2}\lambda_1(a^A_2+a^A_3-a^A_4+a^A_6-a^A_7+a^A_9)$ \\\hline
$\Lambda^+_c\to \Sigma^+ \pi^0$ & $\frac{1}{2}\lambda_1(-a^A_2-a^A_3+a^A_4-a^A_6+a^A_7-a^A_9)$ \\\hline
$\Lambda^+_c\to \Sigma^+ \eta_8$ & $\frac{1}{2\sqrt{3}}\lambda_1(a^A_2+a^A_3+a^A_4-a^A_6+a^A_7-a^A_9-2a^A_{11}-2a^A_{12})$ \\\hline
$\Lambda^+_c\to \Sigma^+ \eta_1$ & $\frac{1}{\sqrt{6}}\lambda_1(a^A_2+a^A_3+a^A_4-a^A_6+a^A_7-a^A_9+a^A_{11}+a^A_{12}
+3a^A_{16}+3a^A_{17})$ \\\hline
$\Lambda^+_c\to p \overline K^0$ & $\frac{1}{\sqrt{2}}\lambda_1(-a^A_{11}-a^A_{12}+a^A_{13}+2a^A_{29})$ \\\hline
$\Lambda^+_c\to \Xi^0 K^+$ & $\frac{1}{\sqrt{2}}\lambda_1(a^A_4+a^A_5+a^A_7+a^A_8)$ \\\hline
$\Xi^0_c\to \Xi^- \pi^+$ & $\frac{1}{\sqrt{2}}\lambda_1(-a^A_8-a^A_9+a^A_{14}+2a^A_{28})$ \\\hline
$\Xi^0_c\to \Xi^0 \pi^0$ & $\frac{1}{2}\lambda_1(-a^A_1+a^A_3+a^A_8+a^A_9)$ \\\hline
$\Xi^0_c\to \Xi^0 \eta_8$ & $\frac{1}{2\sqrt{3}}\lambda_1(a^A_1-a^A_3+2a^A_4+2a^A_5+a^A_8+a^A_9-2a^A_{10}
+2a^A_{12})$ \\\hline
$\Xi^0_c\to \Xi^0 \eta_1$ & $\frac{1}{\sqrt{6}}\lambda_1(a^A_1-a^A_3-a^A_4-a^A_5+a^A_8+a^A_9+a^A_{10}-a^A_{12}
+3a^A_{15}-3a^A_{17})$ \\\hline
$\Xi^0_c\to \Lambda^0 \overline K^0$ &$-\frac{1}{2\sqrt{3}}\lambda_1(2a^A_1+a^A_2-a^A_3+a^A_4+2a^A_5+a^A_6-a^A_{10}+a^A_{11}
+2a^A_{12}-a^A_{13}-3a^A_{29})$  \\\hline
$\Xi^0_c\to \Sigma^+ K^-$ & $\frac{1}{\sqrt{2}}\lambda_1(-a^A_4+a^A_6+a^A_{10}+a^A_{11})$ \\\hline
$\Xi^0_c\to \Sigma^0 \overline K^0$ & $-\frac{1}{\sqrt{2}}\lambda_1(a^A_2+a^A_3-a^A_{4}+a^A_{6}+a^A_{10}+a^A_{11}-a^A_{13}
-2a^A_{29})$ \\\hline
$\Xi^+_c\to \Xi^0 \pi^+$ & $\frac{1}{\sqrt{2}}\lambda_1(a^A_1-a^A_3-a^A_{14}-2a^A_{28})$ \\\hline
$\Xi^+_c\to \Sigma^+ \overline K^0$ & $\frac{1}{\sqrt{2}}\lambda_1(a^A_2+a^A_3-a^A_{13}-2a^A_{29})$ \\\toprule[1.2pt]
$\Lambda^+_c\to pK^0$ & $\frac{1}{\sqrt{2}}\lambda_2(-a^A_2-a^A_3+a^A_{13}+2a^A_{29})$ \\\hline
$\Lambda^+_c\to nK^+$ & $\frac{1}{\sqrt{2}}\lambda_2(-a^A_1+a^A_3+a^A_{14}+2a^A_{28})$ \\\hline
$\Xi^0_c\to \Sigma^-K^+$ & $\frac{1}{\sqrt{2}}\lambda_2(-a^A_8-a^A_9+a^A_{14}+2a^A_{28})$ \\\hline
$\Xi^0_c\to \Lambda^0K^0$ & $\frac{1}{2\sqrt{3}}\lambda_2(a^A_1-a^A_2-2a^A_{3}+2a^A_{4}+a^A_5-a^A_6-2a^A_{10}
-a^A_{11}+a^A_{12}+a^A_{13}+2a^A_{29})$ \\\hline
$\Xi^0_c\to \Sigma^0K^0$ & $-\frac{1}{2}\lambda_2(a^A_1+a^A_2+a^A_{5}+a^A_{6}+a^A_{11}+a^A_{12}-a^A_{13}-2a^A_{29})$ \\\hline
$\Xi^0_c\to p\pi^-$ & $\frac{1}{\sqrt{2}}\lambda_2(-a^A_4+a^A_6+a^A_{10}+a^A_{11})$ \\\hline
$\Xi^0_c\to n\pi^0$ & $\frac{1}{2}\lambda_2(a^A_4+a^A_5+a^A_{8}+a^A_{9}-a^A_{10}+a^A_{12})$ \\\hline
$\Xi^0_c\to n\eta_8$ & $\frac{1}{2\sqrt{3}}\lambda_2(-2a^A_1+2a^A_3-a^A_{4}-a^A_{5}+a^A_8+a^A_9+a^A_{10}
-a^A_{12})$ \\\hline
$\Xi^0_c\to n\eta_1$ &$\frac{1}{\sqrt{6}}\lambda_2(a^A_1-a^A_3-a^A_{4}-a^A_{5}+a^A_8+a^A_9+a^A_{10}
-a^A_{12}+3a^A_{15}-3a^A_{17})$  \\\hline
$\Xi^+_c\to \Lambda^0K^+$ & $\frac{1}{2\sqrt{3}}\lambda_2(a^A_1-a^A_2-2a^A_{3}+2a^A_{4}+a^A_5-a^A_6+2a^A_{7}
+a^A_{8}-a^A_{9}-a^A_{14}-2a^A_{28})$ \\\hline
$\Xi^+_c\to \Sigma^0K^+$ & $-\frac{1}{2}\lambda_2(a^A_1+a^A_2+a^A_{5}+a^A_{6}+a^A_{8}+a^A_{9}-a^A_{14}-2a^A_{28})$  \\\hline
$\Xi^+_c\to \Sigma^+K^0$ & $\frac{1}{\sqrt{2}}\lambda_2(a^A_{11}+a^A_{12}-a^A_{13}-2a^A_{29})$ \\\hline
$\Xi^+_c\to p\pi^0$ & $\frac{1}{2}\lambda_2(-a^A_4+a^A_6-a^A_{7}+a^A_{9}+a^A_{11}+a^A_{12})$ \\\hline
$\Xi^+_c\to p\eta_8$ & $\frac{1}{2\sqrt{3}}\lambda_2(2a^A_2+2a^A_3-a^A_{4}+a^A_{6}-a^A_{7}+a^A_{9}-a^A_{11}
-a^A_{12})$ \\\hline
$\Xi^+_c\to p\eta_1$ & $\frac{1}{\sqrt{6}}\lambda_2(-a^A_2-a^A_3-a^A_{4}+a^A_{6}-a^A_{7}+a^A_{9}
-a^A_{11}-a^A_{12}-3a^A_{16}-3a^A_{17})$ \\\hline
$\Xi^+_c\to n\pi^+$ & $\frac{1}{\sqrt{2}}\lambda_2(-a^A_4-a^A_5-a^A_{7}-a^A_{8})$ \\
  \hline
\end{tabular}
\end{table*}
\begin{table*}
\caption{Topological amplitudes of the singly Cabibbo-favored $\mathcal{B}_{c\overline 3}\to \mathcal{B}_8^AM$ decays (1).}\label{ta5}
 \scriptsize
\begin{tabular}{|c|c|}
\hline\hline
 Channel & Topological amplitude \\\hline
$\Lambda^+_c\to \Lambda^0K^+$ &  $-\frac{1}{2\sqrt{3}}\lambda_d(-2a^A_4-a^A_5+a^A_6-2a^A_7-a^A_8+a^A_9-2a^A_{18}
-a^A_{19}+a^A_{20}+2a^A_{21}+4a^A_{30})$\\&$
-\frac{1}{2\sqrt{3}}\lambda_s(-2a^A_1-a^A_2+a^A_3+2a^A_{14}+4a^A_{28}-2a^A_{18}
-a^A_{19}+a^A_{20}+2a^A_{21}+4a^A_{30})$ \\ & $+\frac{1}{2\sqrt{3}}\lambda_b(-2a^{AP}_2+2a^{AP}_4-2a^{AP}_7-2a^{AP}_{18}-6a^{AP}_{23}  -a^{AP}_1+a^{AP}_5-a^{AP}_8-a^{AP}_{19}-3a^{AP}_{24}  -a^{AP}_3-a^{AP}_6+a^{AP}_9$\\&$
+a^{AP}_{20}+3a^{AP}_{25}-2a^{AP}_{11}+2a^{AP}_{12}+2a^{AP}_{13}+2a^{AP}_{21}+6a^{AP}_{26}  -4a^{AP}_{10}+4a^{AP}_{29}+4a^{AP}_{30}+12a^{AP}_{32})$\\\hline
$\Lambda^+_c\to \Sigma^0K^+$ &  $\frac{1}{2}\lambda_d(-a^A_5-a^A_6-a^A_8-a^A_{9}+a^A_{19}+a^A_{20})
+\frac{1}{2}\lambda_s(a^A_2+a^A_3+a^A_{19}+a^A_{20})$\\ & $-\frac{1}{2}\lambda_b(a^{AP}_1-a^{AP}_5+a^{AP}_8+a^{AP}_{19}+3a^{AP}_{24}  -a^{AP}_3-a^{AP}_6+a^{AP}_9+a^{AP}_{20}+3a^{AP}_{25})$ \\\hline
$\Lambda^+_c\to \Sigma^+K^0$ & $\frac{1}{\sqrt{6}}\lambda_d(a^A_{11}+a^A_{12}+a^A_{19}+a^A_{20})
+\frac{1}{\sqrt{6}}\lambda_s(a^A_2+a^A_3+a^A_{19}+a^A_{20})$\\ & $-\frac{1}{\sqrt{6}}\lambda_b(a^{AP}_2-a^{AP}_4+a^{AP}_7+a^{AP}_{18}+3a^{AP}_{23}
+a^{AP}_1-a^{AP}_5+a^{AP}_8+a^{AP}_{19}+3a^{AP}_{24})$ \\\hline
$\Lambda^+_c\to p\pi^0$ &  $\frac{1}{2}\lambda_d(a^A_{2}+a^A_{3}-a^A_{4}+a^A_{6}-a^A_{7}+a^A_{9}+a^A_{11}+a^A_{12}
-a^A_{13}-2a^A_{29}-a^A_{18}+a^A_{20}+a^A_{21}+2a^A_{30})$\\ & $+\frac{1}{2}\lambda_s(-a^A_{18}+a^A_{20}+a^A_{21}+2a^A_{30})
-\frac{1}{2}\lambda_b(-a^{AP}_2+a^{AP}_4-a^{AP}_7-a^{AP}_{18}-3a^{AP}_{23}-a^{AP}_3
-a^{AP}_6+a^{AP}_9+a^{AP}_{20}$\\&$
+3a^{AP}_{25}
-a^{AP}_{11}+a^{AP}_{12}+a^{AP}_{13}+a^{AP}_{21}+3a^{AP}_{26}-2a^{AP}_{10}+2a^{AP}_{29}+2a^{AP}_{30}+6a^{AP}_{32})$ \\\hline
$\Lambda^+_c\to p\eta_8$ & $\frac{1}{2\sqrt{3}}\lambda_d(-a^A_{2}-a^A_{3}-a^A_{4}+a^A_{6}-a^A_{7}+a^A_{9}-a^A_{11}
-a^A_{12}
+a^A_{13}+2a^A_{29}-a^A_{18}-2a^A_{19}-a^A_{20}+a^A_{21}+2a^A_{30})$\\&$
+\frac{1}{2\sqrt{3}}\lambda_s(-2a^A_{13}-4a^A_{29}-a^A_{18}-2a^A_{19}-a^A_{20}+a^A_{21}
+2a^A_{30})$\\ & $-\frac{1}{2\sqrt{3}}\lambda_b(-a^{AP}_2+a^{AP}_4-a^{AP}_7-a^{AP}_{18}-3a^{AP}_{23} -2a^{AP}_1+2a^{AP}_5-2a^{AP}_8-2a^{AP}_{19}-6a^{AP}_{24} +a^{AP}_3+a^{AP}_6
$\\&$-a^{AP}_9-a^{AP}_{20}-3a^{AP}_{25}-a^{AP}_{11}+a^{AP}_{12}+a^{AP}_{13}+a^{AP}_{21}+3a^{AP}_{26}
-2a^{AP}_{10}+2a^{AP}_{29}+2a^{AP}_{30}+6a^{AP}_{32})$ \\\hline
$\Lambda^+_c\to p\eta_1$ &  $\frac{1}{\sqrt{6}}\lambda_d(-a^A_{2}-a^A_{3}-a^A_{4}+a^A_{6}-a^A_{7}+a^A_{9}-a^A_{11}
-a^A_{12}
+a^A_{13}+2a^A_{29}-3a^A_{16}-3a^A_{17}-a^A_{18}-2a^A_{19}-a^A_{20}$\\&$
+a^A_{21}
+2a^A_{30}+3a^A_{22}+6a^A_{31})+\frac{1}{\sqrt{6}}\lambda_s(a^A_{13}+2a^A_{29}-a^A_{18}-2a^A_{19}-a^A_{20}+a^A_{21}
+2a^A_{30}+3a^A_{22}+6a^A_{31})$\\ & $-\frac{1}{\sqrt{6}}\lambda_b(-a^{AP}_2+a^{AP}_4-a^{AP}_7-a^{AP}_{18}-3a^{AP}_{23} -2a^{AP}_1+2a^{AP}_5-2a^{AP}_8-2a^{AP}_{19}-6a^{AP}_{24} +a^{AP}_3+a^{AP}_6
$\\&$-a^{AP}_9-a^{AP}_{20}-3a^{AP}_{25}-a^{AP}_{11}+a^{AP}_{12}+a^{AP}_{13}+a^{AP}_{21}+3a^{AP}_{26}+3a^{AP}_{14}
-3a^{AP}_{16}+3a^{AP}_{17}$
\\&$+3a^{AP}_{22}+9a^{AP}_{27}
-2a^{AP}_{10}+2a^{AP}_{29}+2a^{AP}_{30}+6a^{AP}_{32}-6a^{AP}_{15}+6a^{AP}_{28}
+6a^{AP}_{31}+18a^{AP}_{33})$ \\\hline
$\Lambda^+_c\to n\pi^+$ &  $\frac{1}{\sqrt{6}}\lambda_d(-a^A_{1}+a^A_{3}-a^A_{4}-a^A_{5}-a^A_{7}-a^A_{8}+a^A_{14}
+2a^A_{28}
-J^a_{18}+J^a_{20}+J^a_{21}+2J^a_{30})$\\
 & $+\frac{1}{\sqrt{6}}\lambda_s(-a^A_{18}+a^A_{20}+a^A_{21}+2a^A_{30})-\frac{1}{\sqrt{6}}\lambda_b(-a^{AP}_2+a^{AP}_4-a^{AP}_7-a^{AP}_{18}-3a^{AP}_{23}
 +a^{AP}_1-a^{AP}_5+a^{AP}_8+a^{AP}_{19}  $\\&$ +3a^{AP}_{24}
 -a^{AP}_{11}+a^{AP}_{12}+a^{AP}_{13}+a^{AP}_{21}+3a^{AP}_{26}-2a^{AP}_{10}+2a^{AP}_{29}+2a^{AP}_{30}
 +6a^{AP}_{32})$ \\\hline
$\Xi^0_c\to \Sigma^-\pi^+$ &  $\frac{1}{\sqrt{2}}\lambda_d(-a^A_{7}-a^A_{8}+a^A_{14}+2a^A_{28}+a^A_{21}+2a^A_{30})
+\frac{1}{\sqrt{2}}\lambda_s(-a^A_{7}+a^A_{9}+a^A_{21}+2a^A_{30})$\\
 & $-\frac{1}{\sqrt{2}}\lambda_b(
 -a^{AP}_{11}+a^{AP}_{12}+a^{AP}_{13}+a^{AP}_{21}+3a^{AP}_{26}-2a^{AP}_{10}
 +2a^{AP}_{29}+2a^{AP}_{30}+6a^{AP}_{32})$ \\\hline
$\Xi^0_c\to \Lambda^0\pi^0$ &  $-\frac{1}{2\sqrt{6}}\lambda_d(a^A_{1}-a^A_{2}-2a^A_{3}+a^A_{7}-a^A_{8}-2a^A_{9}
+a^A_{10}-a^A_{11}-2a^A_{12}+a^A_{13}+2a^A_{29}+a^A_{18}
+a^A_{19}-2a^A_{20}-a^A_{21}-2a^A_{30})$\\&$
-\frac{1}{2\sqrt{6}}\lambda_s(a^A_{4}+2a^A_{5}+a^A_{6}+a^A_{7}+2a^A_{8}+a^A_{9}
+a^A_{18}
-a^A_{19}-2a^A_{20}-a^A_{21}-2a^A_{30})$\\
 & $+\frac{1}{2\sqrt{6}}\lambda_b(a^{AP}_2-a^{AP}_4+a^{AP}_7+a^{AP}_{18}+3a^{AP}_{23}
 -a^{AP}_1+a^{AP}_5-a^{AP}_8-a^{AP}_{19}-3a^{AP}_{24}
 +2a^{AP}_3+2a^{AP}_6-2a^{AP}_9$\\&$ -2a^{AP}_{20}-6a^{AP}_{25} +a^{AP}_{11}-a^{AP}_{12}-a^{AP}_{13}-a^{AP}_{21}-3a^{AP}_{26}
 +2a^{AP}_{10}-2a^{AP}_{29}-2a^{AP}_{30}-6a^{AP}_{32})$ \\\hline
$\Xi^0_c\to \Lambda^0\eta_8$ &  $-\frac{1}{6\sqrt{2}}\lambda_d(-a^A_{1}+a^A_{2}+2a^A_{3}-4a^A_{4}-2a^A_{5}+2a^A_{6}
+a^A_{7}-a^A_{8}-2a^A_{9}-a^A_{10}+a^A_{11}+2a^A_{12}
-a^A_{13}-2a^A_{29}$\\&$-5a^A_{18}-a^A_{19}+4a^A_{20}-a^A_{21}-2a^A_{30})$\\&$
-\frac{1}{6\sqrt{2}}\lambda_s(-4a^A_{1}-2a^A_{2}+2a^A_{3}-a^A_{4}-2a^A_{5}-a^A_{6}
+a^A_{7}
+2a^A_{8}+a^A_{9}-4a^A_{10}-2a^A_{11}+2a^A_{12}+2a^A_{13}+4a^A_{29}$\\&$-5a^A_{18}
-a^A_{19}+4a^A_{20}-a^A_{21}-2a^A_{30})$\\
 & $+\frac{1}{6\sqrt{2}}\lambda_b(-5a^{AP}_2+5a^{AP}_4-5a^{AP}_7-5a^{AP}_{18}
 -15a^{AP}_{23}-a^{AP}_1+a^{AP}_5-a^{AP}_8-a^{AP}_{19}-3a^{AP}_{24}-4a^{AP}_3-4a^{AP}_6
   $\\&$ +4a^{AP}_9+4a^{AP}_{20}+12a^{AP}_{25} +a^{AP}_{11}-a^{AP}_{12}-a^{AP}_{13}-a^{AP}_{21}-3a^{AP}_{26}  +2a^{AP}_{10}-2a^{AP}_{29}-2a^{AP}_{30}-6a^{AP}_{32})$ \\\hline
$\Xi^0_c\to \Lambda^0\eta_1$ &
$-\frac{1}{6}\lambda_d(-a^A_{1}+a^A_{2}+2a^A_{3}+2a^A_{4}+a^A_{5}-a^A_{6}+a^A_{7}
-a^A_{8}-2a^A_{9}-a^A_{10}+a^A_{11}+2a^A_{12}
-a^A_{13}-2a^A_{29}-3a^A_{15}$\\&$+3a^A_{16}+6a^A_{17}+a^A_{18}+2a^A_{19}+a^A_{20}
-a^A_{21}-2a^A_{30}-3a^A_{22}-6a^A_{31})$\\&$
-\frac{1}{6}\lambda_s(2a^A_{1}+a^A_{2}-a^A_{3}-a^A_{4}-2a^A_{5}-a^A_{6}+a^A_{7}
+2a^A_{8}+a^A_{9}+2a^A_{10}+a^A_{11}-a^A_{12}-a^A_{13}-2a^A_{29}+6a^A_{15}
$\\&$+3a^A_{16}
-3a^A_{17}+a^A_{18}+2a^A_{19}+a^A_{20}-a^A_{21}-2a^A_{30}
-3a^A_{22}-6a^A_{31})$\\
 & $+\frac{1}{6}\lambda_b(-a^{AP}_2+a^{AP}_4-a^{AP}_7-a^{AP}_{18}-3a^{AP}_{23}  -2a^{AP}_1+2a^{AP}_5+2a^{AP}_8+2a^{AP}_{19}+6a^{AP}_{24}
 -a^{AP}_3-a^{AP}_6+a^{AP}_9$\\&$+a^{AP}_{20}+3a^{AP}_{25}+a^{AP}_{11}
 -a^{AP}_{12}
 -a^{AP}_{13}-a^{AP}_{21}-3a^{AP}_{26}
 -3a^{AP}_{14}+3a^{AP}_{16}-3a^{AP}_{17}-3a^{AP}_{22}-9a^{AP}_{27}$\\&$+2a^{AP}_{10}
 -2a^{AP}_{29}-2a^{AP}_{30}-6a^{AP}_{32}
 +6a^{AP}_{15}-6a^{AP}_{28}-6a^{AP}_{31}-18a^{AP}_{33})$ \\\hline
$\Xi^0_c\to \Sigma^0\pi^0$ &  $\frac{1}{2\sqrt{2}}\lambda_d(a^A_{1}+a^A_{2}-a^A_{7}-a^A_{8}+a^A_{10}+a^A_{11}
-a^A_{13}-2a^A_{29}+a^A_{18}+a^A_{19}
+a^A_{21}+2a^A_{30})$\\&$
+\frac{1}{2\sqrt{2}}\lambda_s(a^A_{4}-a^A_{6}-a^A_{7}+a^A_{9}+a^A_{18}+a^A_{19}
+a^A_{21}+2a^A_{30})$\\
 & $-\frac{1}{2\sqrt{2}}\lambda_b(a^{AP}_2-a^{AP}_4+a^{AP}_7+a^{AP}_{18}
 +3a^{AP}_{23} +a^{AP}_1-a^{AP}_5+a^{AP}_8+a^{AP}_{19}+3a^{AP}_{24}
 -a^{AP}_{11} $\\
 & $ +a^{AP}_{12}+a^{AP}_{13}+a^{AP}_{21}+3a^{AP}_{26}-2a^{AP}_{10}+2a^{AP}_{29}+2a^{AP}_{30}+6a^{AP}_{32})$ \\\hline
$\Xi^0_c\to \Sigma^0\eta_8$ &  $\frac{1}{2\sqrt{6}}\lambda_d(-a^A_{1}-a^A_{2}-2a^A_{5}-2a^A_{6}-a^A_{7}-a^A_{8}
-a^A_{10}-a^A_{11}+a^A_{13}+2a^A_{29}-a^A_{18}+a^A_{19}
+2a^A_{20}+a^A_{21}+2a^A_{30})$\\&$
+\frac{1}{2\sqrt{2}}\lambda_s(2a^A_{2}+2a^A_{3}-a^A_{4}+a^A_{6}
-a^A_{7}+a^A_{9} +2a^A_{11} +2a^A_{12} -2a^A_{13}-4a^A_{29} -a^A_{18}+a^A_{19}
+2a^A_{20}+a^A_{21}+2a^A_{30})$\\
 & $-\frac{1}{2\sqrt{6}}\lambda_b(-a^{AP}_2+a^{AP}_4-a^{AP}_7-a^{AP}_{18}-3a^{AP}_{23}
 +a^{AP}_1-a^{AP}_5+a^{AP}_8+a^{AP}_{19}+3a^{AP}_{24}
 -2a^{AP}_3-2a^{AP}_6+2a^{AP}_9$\\
 & $ +2a^{AP}_{20}+6a^{AP}_{25}-a^{AP}_{11}+a^{AP}_{12}+a^{AP}_{13}+a^{AP}_{21}+3a^{AP}_{26} -2a^{AP}_{10}+2a^{AP}_{29}+2a^{AP}_{30}+6a^{AP}_{32})$ \\\hline
$\Xi^0_c\to \Sigma^0\eta_1$ &  $\frac{1}{2\sqrt{3}}\lambda_d(-a^A_{1}-a^A_{2}+a^A_{5}+a^A_{6}-a^A_{7}-a^A_{8}
-a^A_{10}-a^A_{11}+a^A_{13}+2a^A_{29}-3a^A_{15}-3a^A_{16}$\\&$-a^A_{18}-2a^A_{19}
-a^A_{20}+a^A_{21}+2a^A_{30} +3a^A_{22}+6a^A_{31} )$\\&$
+\frac{1}{2\sqrt{3}}\lambda_s(-a^A_{2}-a^A_{3}-a^A_{4}+a^A_{6}
-a^A_{7}+a^A_{9} -a^A_{11} -a^A_{12} +a^A_{13}+2a^A_{29} -3a^A_{16}-3a^A_{17}$\\&$
-a^A_{18}-2a^A_{19}
-a^A_{20}+a^A_{21}+2a^A_{30} +3a^A_{22}+6a^A_{31} )$\\
 & $-\frac{1}{2\sqrt{3}}\lambda_b(-a^{AP}_2+a^{AP}_4-a^{AP}_7-a^{AP}_{18}-3a^{AP}_{23}  -2a^{AP}_1+2a^{AP}_5-2a^{AP}_8-2a^{AP}_{19}-6a^{AP}_{24}
 +a^{AP}_3+a^{AP}_6-a^{AP}_9$\\
 & $-a^{AP}_{20}-3a^{AP}_{25}-a^{AP}_{11}+a^{AP}_{12}+a^{AP}_{13}+a^{AP}_{21}+3a^{AP}_{26}  -2a^{AP}_{10}+2a^{AP}_{29}+2a^{AP}_{30}+6a^{AP}_{32}
 +3a^{AP}_{14}-3a^{AP}_{16} $\\&$ +3a^{AP}_{17}+3a^{AP}_{22}+9a^{AP}_{27} -6a^{AP}_{15}+6a^{AP}_{28}+6a^{AP}_{31}+18a^{AP}_{33})$ \\\hline
$\Xi^0_c\to n\overline K^0$ &  $\frac{1}{\sqrt{2}}\lambda_d(a^A_{1}-a^A_{3}+a^A_{4}+a^A_{5}+a^A_{18}-a^A_{20})
+\frac{1}{\sqrt{2}}\lambda_s(a^A_{10}-a^A_{12}+a^A_{18}-a^A_{20})$\\
 & $-\frac{1}{\sqrt{2}}\lambda_b(
 a^{AP}_2-a^{AP}_4+a^{AP}_7+a^{AP}_{18}+3a^{AP}_{23}  +a^{AP}_3+a^{AP}_6-a^{AP}_9-a^{AP}_{20}-3a^{AP}_{25})$ \\\hline\hline
\end{tabular}
\end{table*}
\begin{table*}
\caption{Topological amplitudes of the singly Cabibbo-favored $\mathcal{B}_{c\overline 3}\to \mathcal{B}_8^AM$ decays (2).}\label{ta6}
 \scriptsize
\begin{tabular}{|c|c|}
\hline\hline
 Channel & Topological amplitude \\\hline
 $\Xi^0_c\to \Xi^-K^+$ &  $\frac{1}{2}\lambda_d(-a^A_{7}+a^A_{9}+a^A_{21}+2a^A_{30})
+\frac{1}{2}\lambda_s(-a^A_{7}-a^A_{8}+a^A_{14}+2a^A_{28}+a^A_{21}+2a^A_{30})$\\
 & $-\frac{1}{2}\lambda_b(-a^{AP}_{11}+a^{AP}_{12}+a^{AP}_{13}+a^{AP}_{21}+3a^{AP}_{26}
 -2a^{AP}_{10}+2a^{AP}_{29}+2a^{AP}_{30}+6a^{AP}_{32})$ \\\hline
 $\Xi^0_c\to \Xi^0K^0$ &
$\frac{1}{\sqrt{2}}\lambda_d(a^A_{10}-a^A_{12}+a^A_{18}-a^A_{20})
+\frac{1}{\sqrt{2}}\lambda_s(a^A_{1}-a^A_{3}+a^A_{4}+a^A_{5}+a^A_{18}-a^A_{20})$\\
 & $-\frac{1}{\sqrt{2}}\lambda_b(a^{AP}_2-a^{AP}_4+a^{AP}_7+a^{AP}_{18}+3a^{AP}_{23}  +a^{AP}_3+a^{AP}_6-a^{AP}_9-a^{AP}_{20}-3a^{AP}_{25})$ \\\hline
$\Xi^0_c\to \Sigma^+\pi^-$ &  $\frac{1}{\sqrt{2}}\lambda_d(a^A_{10}+a^A_{11}+a^A_{18}+a^A_{19})
+\frac{1}{\sqrt{2}}\lambda_s(a^A_{4}-a^A_{6}+a^A_{18}+a^A_{19})$\\
 & $-\frac{1}{\sqrt{2}}\lambda_b(a^{AP}_2-a^{AP}_4+a^{AP}_7+a^{AP}_{18}+3a^{AP}_{23}
 +a^{AP}_1-a^{AP}_5+a^{AP}_8+a^{AP}_{19}+3a^{AP}_{24})$ \\\hline
$\Xi^0_c\to pK^-$ &
$\frac{1}{\sqrt{2}}\lambda_d(a^A_{4}-a^A_{6}+a^A_{18}+a^A_{19})
+\frac{1}{\sqrt{2}}\lambda_s(a^A_{10}+a^A_{11}+a^A_{18}+a^A_{19})$\\
 & $-\frac{1}{\sqrt{2}}\lambda_b(a^{AP}_2-a^{AP}_4+a^{AP}_7+a^{AP}_{18}+3a^{AP}_{23}
 +a^{AP}_1-a^{AP}_5+a^{AP}_8+a^{AP}_{19}+3a^{AP}_{24})$ \\\hline
$\Xi^+_c\to \Lambda^0\pi^+$ &  $\frac{1}{2\sqrt{3}}\lambda_d(a^A_{1}-a^A_{2}-2a^A_{3}-a^A_{14}-2a^A_{28}+a^A_{18}
-a^A_{19}-2a^A_{20} -a^A_{21}-2a^A_{30} )
$\\&$+\frac{1}{2\sqrt{3}}\lambda_s(a^A_{4}+2a^A_{5}+a^A_{6}+a^A_{7}+2a^A_{8}+a^A_{9}
+a^A_{18}-a^A_{19}-2a^A_{20} -a^A_{21}-2a^A_{30})$\\
 & $-\frac{1}{2\sqrt{3}}\lambda_b(a^{AP}_2-a^{AP}_4+a^{AP}_7+a^{AP}_{18}+3a^{AP}_{23} -a^{AP}_1+a^{AP}_5-a^{AP}_8-a^{AP}_{19}-3a^{AP}_{24}
 +2a^{AP}_3+2a^{AP}_6-2a^{AP}_9$\\&$-2a^{AP}_{20}-6a^{AP}_{25}+a^{AP}_{11}-a^{AP}_{12}
 -a^{AP}_{13}-a^{AP}_{21}-3a^{AP}_{26} +2a^{AP}_{10}-2a^{AP}_{29}-2a^{AP}_{30}-6a^{AP}_{32})$ \\\hline
$\Xi^+_c\to \Sigma^0\pi^+$ &  $\frac{1}{2}\lambda_d(-a^A_{1}-a^A_{2}+a^A_{14}+2a^A_{28} -a^A_{18}-a^A_{19}+a^A_{21}+2a^A_{30})
+\frac{1}{2}\lambda_s(-a^A_{4}+a^A_{6}-a^A_{7}+a^A_{9}-a^A_{18}-a^A_{19} +a^A_{21}+2a^A_{30})$\\
 & $-\frac{1}{2}\lambda_b(-a^{AP}_2+a^{AP}_4-a^{AP}_7-a^{AP}_{18}-3a^{AP}_{23} - a^{AP}_1+a^{AP}_5-a^{AP}_8-a^{AP}_{19}-3a^{AP}_{24}$\\&$
 -a^{AP}_{11}+a^{AP}_{12}+a^{AP}_{13}+a^{AP}_{21}+3a^{AP}_{26} -2a^{AP}_{10}+2a^{AP}_{29}+2a^{AP}_{30}+6a^{AP}_{32})$ \\\hline
$\Xi^+_c\to \Sigma^+\pi^0$ &  $\frac{1}{2}\lambda_d(a^A_{13}+2a^A_{29}+a^A_{18}+a^A_{19} -a^A_{21}-2a^A_{30}  )
+\frac{1}{2}\lambda_s(a^A_{4}-a^A_{6}+a^A_{7}-a^A_{9} +a^A_{18}+a^A_{19} -a^A_{21}-2a^A_{30} )$\\
 & $-\frac{1}{2}\lambda_b( a^{AP}_2-a^{AP}_4+a^{AP}_7+a^{AP}_{18}+3a^{AP}_{23} + a^{AP}_1-a^{AP}_5+a^{AP}_8+a^{AP}_{19}+3a^{AP}_{24} +a^{AP}_{11}-a^{AP}_{12}-a^{AP}_{13}$\\&$-a^{AP}_{21}-3a^{AP}_{26}+2a^{AP}_{10}
 -2a^{AP}_{29}-2a^{AP}_{30}-6a^{AP}_{32})$\\\hline
$\Xi^+_c\to \Sigma^+\eta_8$ &
$\frac{1}{2\sqrt{3}}\lambda_d(-a^A_{13}-2a^A_{29}+a^A_{18}-a^A_{19} -2a^A_{20}-a^A_{21} -2a^A_{30}  )$\\&$
+\frac{1}{2\sqrt{3}}\lambda_s(-2a^A_{2}-2a^A_{3}+a^A_{4}-a^A_{6} +a^A_{7}-a^A_{9} -2a^A_{11}-2a^A_{12} +2a^A_{13} +4a^A_{29} +a^A_{18}-a^A_{19} -2a^A_{20}-a^A_{21} -2a^A_{30} )$\\
 & $-\frac{1}{2\sqrt{3}}\lambda_b(a^{AP}_2-a^{AP}_4+a^{AP}_7+a^{AP}_{18}+3a^{AP}_{23} -a^{AP}_1+a^{AP}_5-a^{AP}_8-a^{AP}_{19}-3a^{AP}_{24}  +2a^{AP}_3+2a^{AP}_6-2a^{AP}_9$\\&$-2a^{AP}_{20}-6a^{AP}_{25}+a^{AP}_{11}-a^{AP}_{12}
 -a^{AP}_{13}-a^{AP}_{21}-3a^{AP}_{26}  +2a^{AP}_{10}-2a^{AP}_{29}-2a^{AP}_{30}-6a^{AP}_{32})$\\\hline
$\Xi^+_c\to \Sigma^+\eta_1$ &
$\frac{1}{\sqrt{6}}\lambda_d(-a^A_{13}-2a^A_{29}+a^A_{18}+2a^A_{19} +a^A_{20}-a^A_{21} -2a^A_{30} -3a^A_{22} -6a^A_{31}  )+\frac{1}{\sqrt{6}}\lambda_s(a^A_{2}+a^A_{3}+a^A_{4}-a^A_{6}$\\&$
 +a^A_{7}-a^A_{9} +a^A_{11}+a^A_{12} -a^A_{13} -2a^A_{29}+3a^A_{16} +3a^A_{17}+a^A_{18}+2a^A_{19} +a^A_{20}-a^A_{21} -2a^A_{30} -3a^A_{22} -6a^A_{31} )$\\
 & $-\frac{1}{\sqrt{6}}\lambda_b(a^{AP}_2-a^{AP}_4+a^{AP}_7+a^{AP}_{18}+3a^{AP}_{23}  +2a^{AP}_1-2a^{AP}_5+2a^{AP}_8+2a^{AP}_{19}+6a^{AP}_{24}-a^{AP}_3
 -a^{AP}_6$\\
 &$+a^{AP}_9
 +a^{AP}_{20}+3a^{AP}_{25}+a^{AP}_{11}-a^{AP}_{12}-a^{AP}_{13}-a^{AP}_{21}-3a^{AP}_{26} +2a^{AP}_{10}-2a^{AP}_{29}-2a^{AP}_{30}-6a^{AP}_{32}$\\&
 $-3a^{AP}_{14}+3a^{AP}_{16}-3a^{AP}_{17}-3a^{AP}_{22}+9a^{AP}_{27}  +6a^{AP}_{15}-6a^{AP}_{28}-6a^{AP}_{31}-18a^{AP}_{33})$\\\hline
$\Xi^+_c\to p\overline K^0$ & $\frac{1}{\sqrt{2}}\lambda_d(-a^A_{2}-a^A_{3}-a^A_{19}-a^A_{20} )
+\frac{1}{\sqrt{2}}\lambda_s(-a^A_{11}-a^A_{12}-a^A_{19}-a^A_{20})$\\
 & $-\frac{1}{\sqrt{2}}\lambda_b(-a^{AP}_1+a^{AP}_5-a^{AP}_8-a^{AP}_{19}-3a^{AP}_{24} +a^{AP}_3+a^{AP}_6-a^{AP}_9-a^{AP}_{20}-3a^{AP}_{25} )$ \\\hline
$\Xi^+_c\to \Xi^0K^+$ &  $\frac{1}{\sqrt{2}}\lambda_d(a^A_{18}-a^A_{20}-a^A_{21}-2a^A_{30} )
+\frac{1}{\sqrt{2}}\lambda_s(a^A_{1}-a^A_{3}+a^A_{4}+a^A_{5} +a^A_{7}+a^A_{8}-a^A_{14}-2a^A_{28} +a^A_{18}-a^A_{20}-a^A_{21}-2a^A_{30} )$\\
 & $-\frac{1}{\sqrt{2}}\lambda_b(-a^{AP}_1+a^{AP}_5-a^{AP}_8-a^{AP}_{19}-3a^{AP}_{24} +a^{AP}_3+a^{AP}_6-a^{AP}_9-a^{AP}_{20}-3a^{AP}_{25} )$ \\\hline
  \hline
\end{tabular}
\end{table*}

\section{Topological amplitudes in the Standard Model}\label{sm}
\begin{table*}
\caption{Decay amplitudes of the Cabibo-favored and doubly Cabibbo-suppressed $\mathcal{B}_{c\overline{3}}\to \mathcal{B}_{8}M$ decays, in which baryon octet is expressed as second-rank tensor.}\label{ta7}
 \scriptsize
\begin{tabular}{|c|c|c|c|}
\hline\hline
 Channel & Amplitude & Channel &Amplitude\\\hline
 $\Lambda_{c}^{+}\to \Lambda^0\pi^+$ & $\frac{1}{\sqrt{6}}\lambda_1(-2A_2+A_7+A_8)$  &    $\Lambda_{c}^{+}\to pK^0$ & $\lambda_2(A_1+A_4)$ \\\hline
 $\Lambda^{+}_c\to \Sigma^0\pi^+$ & $\frac{1}{\sqrt{2}}\lambda_1(A_7-A_8)$ &   $\Lambda_{c}^{+}\to nK^+$ & $\lambda_2(A_2+A_3)$  \\\hline
  $\Lambda_{c}^{+}\to \Sigma^+\pi^0$ & $\frac{1}{\sqrt{2}}\lambda_1(-A_7+A_8)$  &    $\Xi_{c}^{0}\to \Sigma^-K^+$ & $\lambda_2(A_2+A_5)$ \\\hline
 $\Lambda_{c}^{+}\to \Sigma^+\eta_8$ & $\frac{1}{\sqrt{6}}\lambda_1(-2A_1+A_7+A_8)$  &    $\Xi_{c}^{0}\to \Lambda^0K^0$ & $\frac{1}{\sqrt{6}}\lambda_2(A_4+A_5-2A_6)$
  \\\hline
$\Lambda_{c}^{+}\to \Sigma^+\eta_1$ & $\frac{1}{\sqrt{3}}\lambda_1(A_1+3A_{10}+A_7+A_8)$   &  $\Xi_{c}^{0}\to \Sigma^0K^0$ & $\frac{1}{\sqrt{2}}\lambda_2(A_4-A_5)$\\\hline
  $\Lambda_{c}^{+}\to p\overline{K}^0$ & $\lambda_1(A_4+A_7)$  & $\Xi_{c}^{0}\to p\pi^-$&$\lambda_2(A_1+A_6)$
  \\\hline
  $\Lambda_{c}^{+}\to \Xi^0K^+$ & $\lambda_1(A_3+A_8)$   &    $\Xi_{c}^{0}\to n\pi^0$ & $\frac{1}{\sqrt{2}}\lambda_2(A_3-A_6)$\\\hline
  $\Xi_{c}^{0}\to \Xi^-\pi^+$ & $\lambda_1(A_2+A_5)$ &  $\Xi_{c}^{0}\to n\eta_8$ & $\frac{1}{\sqrt{6}}\lambda_2(A_3-2A_5+A_6)$
 \\\hline
  $\Xi_{c}^{0}\to \Xi^0\pi^0$ & $\frac{1}{\sqrt{2}}\lambda_1(A_3-A_5)$&  $\Xi_{c}^{0}\to n\eta_1$ & $\frac{1}{\sqrt{3}}\lambda_2(A_3+A_5+A_6+3A_9)$\\\hline
 $\Xi_{c}^{0}\to \Xi^0\eta_8$ & $\frac{1}{\sqrt{6}}\lambda_1(A_3+A_5-2A_6)$ &  $\Xi_{c}^+\to \Lambda^0K^+$ & $-\frac{1}{\sqrt{6}}\lambda_2(A_2+A_7-2A_8)$\\\hline
  $\Xi_{c}^{0}\to \Xi^0\eta_1$ & $\frac{1}{\sqrt{3}}\lambda_1(A_3+A_5+A_6+3A_9)$  &  $\Xi_{c}^{+}\to \Sigma^0K^+$ & $\frac{1}{\sqrt{2}}\lambda_2(A_2-A_7)$
  \\\hline
 $\Xi_{c}^{0}\to \Lambda^0\overline{K}^0$ & $\frac{1}{\sqrt{6}}\lambda_1(A_4-2A_5+A_6)$ &  $\Xi_{c}^{+}\to \Sigma^+K^0$ & $-\lambda_2(A_4+A_7)$
  \\\hline
  $\Xi_{c}^{0}\to \Sigma^+K^-$ & $\lambda_1(A_1+A_6)$  &  $\Xi_{c}^+\to p\pi^0$ & $\frac{1}{\sqrt{2}}\lambda_2(A_1-A_8)$\\\hline
 $\Xi_{c}^{0}\to \Sigma^0\overline{K}^0$ & $\frac{1}{\sqrt{2}}\lambda_1(A_4-A_6)$ &  $\Xi_{c}^{+}\to p\eta_8$ & $-\frac{1}{\sqrt{6}}\lambda_2(A_1-2A_7+A_8)$
  \\\hline
  $\Xi_{c}^{+}\to \Xi^0\pi^+$ & $-\lambda_1(A_2+A_3)$ & $\Xi_{c}^{+}\to p\eta_1$ & $-\frac{1}{\sqrt{3}}\lambda_2(A_1+A_7+A_8+3A_{10})$
  \\\hline
  $\Xi_{c}^{+}\to \Sigma^+\overline{K}^0$ & $-\lambda_1(A_1+A_4)$   &  $\Xi_{c}^+\to n\pi^+$ & $-\lambda_2(A_3+A_8)$\\\hline
  \hline
\end{tabular}
\end{table*}
\begin{table*}
\caption{Decay amplitudes of the singly Cabibbo-suppressed $\mathcal{B}_{c\overline{3}}\to \mathcal{B}_{8}M$ decays, in which baryon octet is expressed as second-rank tensor.}\label{ta8}
 \scriptsize
\begin{tabular}{|c|c|}
\hline\hline
 Channel & Amplitude \\\hline
$\Lambda_{c}^{+}\to \Lambda^0K^+$&  $-\frac{1}{\sqrt{6}}\lambda_b(A^P_3-2A^P_4+A^P_5-2A^P_6+A^P_{11}-2A^P_{12})
$\\ &$+\frac{1}{\sqrt{6}}\lambda_d(A_3+A_{11}-2A_{12})
+\frac{1}{\sqrt{6}}\lambda_s(-2A_{2}-2A_{3}+A_{7}-2A_{8}+A_{11}-2A_{12})$
\\\hline
$\Lambda_{c}^{+}\to \Sigma^0K^+$&  $-\frac{1}{\sqrt{2}}\lambda_b(A^P_3+A^P_5+A^P_{11}
+A^P_{17})+\frac{1}{\sqrt{2}}\lambda_d(-A_3+A_{11})
+\frac{1}{\sqrt{2}}\lambda_s(A_{7}+A_{11})$
\\\hline
$\Lambda_{c}^{+}\to \Sigma^+K^0$ & $-\lambda_b(A^P_3+A^P_5+A^P_{11}+3A^P_{17})+\lambda_d(A_1+A_{11}) +\lambda_s(A_{7}+A_{11})$
\\\hline
$\Lambda_{c}^{+}\to p\pi^0$ & $-\frac{1}{\sqrt{2}}\lambda_b(A^P_4+A^P_6+A^P_{12}+3A^P_{15})
+\frac{1}{\sqrt{2}}\lambda_d(-A_4+A_{12})+\frac{1}{\sqrt{2}}\lambda_s(A_{8}+A_{12})$
 \\\hline
$\Lambda_{c}^{+}\to p\eta_8$&  $\frac{1}{\sqrt{6}}\lambda_b(2A^P_3-A^P_4+2A^P_5-A^P_6+2A^P_{11}-A^P_{12}
-3A^P_{15}+6A^P_{17})$\\ &$+\frac{1}{\sqrt{6}}\lambda_d(A_4-2A_{11}+A_{12})
+\frac{1}{\sqrt{6}}\lambda_s(-2A_{1}-2A_{4}-2A_{7}+A_8-2A_{11}+A_{12})$
\\\hline
$\Lambda_{c}^{+}\to p\eta_1$&  $-\frac{1}{\sqrt{3}}\lambda_b(3A^P_2+A^P_3+A^P_4+A^P_5+A^P_6+3A^P_9
+A^P_{11}+A^P_{12}+3A^P_{13}+3A^P_{15}+9A^P_{16}+3A^P_{17})$\\ &$+\frac{1}{\sqrt{3}}\lambda_d(A_4+A_{11}+A_{12}+3A_{13})
+\frac{1}{\sqrt{3}}\lambda_s(A_{1}+A_{4}+A_{7}+A_{8}+3A_{10}+A_{11}+A_{12})$
 \\\hline
$\Lambda_{c}^{+}\to n\pi^+$ & $-\lambda_b(A^P_4+A^P_6+A^P_{12}+3A^P_{15})+\lambda_d(A_2+A_{12})$ $+\lambda_s(A_{8}+A_{12})$
\\\hline
 $\Xi_{c}^{0}\to \Sigma^-\pi^+$ & $-\lambda_b(A^P_4+A^P_6+A^P_7+A^P_8
 +A^P_{12}+A^P_{14}+3A^P_{15}+3A^P_{18})
 $\\ &$+\lambda_d(A_2+A_5+A_{12}+A_{14})+\lambda_s(A_{12}+A_{14})$
\\\hline
$\Xi_{c}^{0}\to \Lambda^0\pi^0$ & $-\frac{1}{2\sqrt{3}}\lambda_b(A^P_3+A^P_4+A^P_5+A^P_6
+A^P_{11}+A^P_{12}+3A^P_{15}+3A^P_{17})$\\ &$+\frac{1}{2\sqrt{3}}\lambda_d(A_3-A_4-A_5-A_6+A_{11}+A_{12})
+\frac{1}{2\sqrt{3}}\lambda_s(A_{11}+A_{12}-2A_{3})$
\\\hline
$\Xi_{c}^{0}\to \Lambda^0\eta_8$ & $-\frac{1}{6}\lambda_b(A^P_3+A^P_4+A^P_5+A^P_6+6A^P_7+6A^P_8+A^P_{11}
+A^P_{12}+6A^P_{14}+3A^P_{15}+3A^P_{17}+18A^P_{18})$\\ &$+\frac{1}{6}\lambda_d(A_3+A_4+A_5+A_6+A_{11}+A_{12}+6A_{14})$\\ &$+\frac{1}{6}\lambda_s(-2A_3-2A_4+4A_5+4A_6+A_{11}
+A_{12}+6A_{14})$
\\\hline
$\Xi_{c}^{0}\to \Lambda^0\eta_1$ & $-\frac{1}{3\sqrt{2}}\lambda_b(3A^P_2+A^P_3+A^P_4+A^P_5+A^P_6+3A^P_9
+A^P_{11}+A^P_{12}+3A^P_{13}+3A^P_{15}+9A^P_{16}+3A^P_{17})$\\ &$+\frac{1}{3\sqrt{2}}\lambda_d(A_3+A_4+A_5+A_6+3A_9+A_{11}+A_{12}+3A_{13})$\\ &$+\frac{1}{3\sqrt{2}}\lambda_s(-2A_3+A_4-2A_5-2A_6-6A_9+A_{11}+A_{12}+3A_{13})$
\\\hline
$\Xi_{c}^{0}\to \Sigma^0\pi^0$ & $-\frac{1}{2}\lambda_b(A^P_3+A^P_4+A^P_5+A^P_6+2A^P_7+2A^P_8
+A^P_{11}+A^P_{12}+2A^P_{14}+3A^P_{15}+3A^P_{17}+6A^P_{18})$\\ &$+\frac{1}{2}\lambda_d(-A_3-A_4+A_5+A_6+A_{11}+A_{12}+2A_{14})
+\frac{1}{2}\lambda_s(A_{11}+A_{12}+2A_{14})$
\\\hline
$\Xi_{c}^{0}\to \Sigma^0\eta_8$ & $-\frac{1}{2\sqrt{3}}\lambda_b(A^P_3+A^P_4+A^P_5+A^P_6+A^P_{11}+A^P_{12}
+3A^P_{15}+3A^P_{17})$\\ &$+\frac{1}{2\sqrt{3}}\lambda_d(-A_3+A_4-A_5-A_6+A_{11}+A_{12})
+\frac{1}{2\sqrt{3}}\lambda_s(A_{11}+A_{12}-2A_{4})$
\\\hline
$\Xi_{c}^{0}\to \Sigma^0\eta_1$ & $-\frac{1}{\sqrt{6}}\lambda_b(3A^P_2+A^P_3+A^P_4+A^P_5+A^P_6+3A^P_9
+A^P_{11}+A^P_{12}+3A^P_{13}+3A^P_{15}+9A^P_{16}+3A^P_{17})$\\ &$+\frac{1}{\sqrt{6}}\lambda_d(-A_3+A_4-A_5-A_6+A_{11}
+A_{12}+3A_{13})+\frac{1}{\sqrt{6}}\lambda_s(A_4+A_{11}+A_{12}+3A_{13})$
\\\hline
$\Xi_{c}^{0}\to n\overline{K}^0$ & $-\lambda_b(A^P_7+A^P_8+A^P_{14}+3A^P_{18})+\lambda_d(A_5+A_{14}) +\lambda_s(A_{6}+A_{14})$
\\\hline
$\Xi_{c}^{0}\to \Xi^-K^+$ & $-\lambda_b(A^P_4+A^P_6+A^P_7+A^P_8+A^P_{12}+A^P_{14}+3A^P_{15}+3A^P_{18})
$\\ &$+\lambda_d(A_{12}+A_{14})+\lambda_s(A_2+A_5+A_{12}+A_{14})$
\\\hline
$\Xi_{c}^{0}\to \Xi^0K^0$ & $-\lambda_b(A^P_7+A^P_8+A^P_{14}+3A^P_{18})+\lambda_d(A_6+A_{14}) +\lambda_s(A_{5}+A_{14})$
\\\hline
$\Xi_{c}^{0}\to \Sigma^+\pi^-$ & $-\lambda_b(A^P_3+A^P_5+A^P_7+A^P_8+A^P_{11}
+A^P_{14}+3A^P_{17}+3A^P_{18})$\\ &$+\lambda_d(A_1+A_6+A_{11}+A_{14})+\lambda_s(A_{11}+A_{14})$
\\\hline
$\Xi_{c}^{0}\to pK^-$ & $-\lambda_b(A^P_3+A^P_5+A^P_7+A^P_8+A^P_{11}+A^P_{14}+3A^P_{17}+3A^P_{18})
$\\ &$+\lambda_d(A_{11}+A_{14})+\lambda_s(A_1+A_6+A_{11}+A_{14})$
\\\hline
$\Xi_{c}^{+}\to \Lambda^0\pi^+$ & $\frac{1}{\sqrt{6}}\lambda_b(A^P_3+A^P_4+A^P_5+A^P_6+A^P_{11}+A^P_{12}
+3A^P_{15}+3A^P_{17})$\\ &$-\frac{1}{\sqrt{6}}\lambda_d(A_2+A_3+A_7+A_8+A_{11}+A_{12})
-\frac{1}{\sqrt{6}}\lambda_s(-2A_3+A_{11}+A_{12})$
\\\hline
$\Xi_{c}^{+}\to \Sigma^0\pi^+$ & $\frac{1}{\sqrt{2}}\lambda_b(A^P_3-A^P_4+A^P_5-A^P_6+A^P_{11}-A^P_{12}
-3A^P_{15}+3A^P_{17})$\\ &$+\frac{1}{\sqrt{2}}\lambda_d(A_2+A_3-A_7+A_8-A_{11}+A_{12})
+\frac{1}{\sqrt{2}}\lambda_s(-A_{11}+A_{12})$
\\\hline
$\Xi_{c}^{+}\to \Sigma^+\pi^0$ & $-\frac{1}{\sqrt{2}}\lambda_b(-A^P_3+A^P_4-A^P_5+A^P_6-A^P_{11}
+A^P_{12}+3A^P_{15}-3A^P_{17})$\\ &$+\frac{1}{\sqrt{2}}\lambda_d(A_1+A_4+A_7-A_8+A_{11}-A_{12})
+\frac{1}{\sqrt{2}}\lambda_s(A_{11}-A_{12})$
\\\hline
$\Xi_{c}^{+}\to \Sigma^+\eta_8$ & $\frac{1}{\sqrt{6}}\lambda_b(A^P_3+A^P_4+A^P_5+A^P_6+A^P_{11}+A^P_{12}
+3A^P_{15}+3A^P_{17})$\\ &$-\frac{1}{\sqrt{6}}\lambda_d(A_1+A_4+A_7+A_8+A_{11}+A_{12})
-\frac{1}{\sqrt{6}}\lambda_s(-2A_4+A_{11}+A_{12})$
\\\hline
$\Xi_{c}^{+}\to \Sigma^+\eta_1$ & $\frac{1}{\sqrt{3}}\lambda_b(3A^P_2+A^P_3+A^P_4+A^P_5
+A^P_6+3A^P_9+A^P_{11}+A^P_{12}+3A^P_{13}+3A^P_{15}+9A^P_{16}+3A^P_{17})$\\ &$-\frac{1}{\sqrt{3}}\lambda_d(A_1+A_4+A_7+A_8+3A_{10}+A_{11}+A_{12}+3A_{13})
-\frac{1}{\sqrt{3}}\lambda_s(A_4+A_{11}+A_{12}+3A_{13})$
\\\hline
$\Xi_{c}^{+}\to p\overline{K}^0$ & $\lambda_b(A^P_3+A^P_5+A^P_{11}+3A^P_{17})-\lambda_d(A_7+A_{11}) -\lambda_s(A_{1}+A_{11})$
\\\hline
$\Xi_{c}^{+}\to \Xi^0K^+$ & $\lambda_b(A^P_4+A^P_6+A^P_{12}+3A^P_{15})-\lambda_d(A_8+A_{12}) -\lambda_s(A_{2}+A_{12})$
\\\hline
\hline
\end{tabular}
\end{table*}

The effective Hamiltonian in charm quark weak decay in the SM is \cite{Buchalla:1995vs}
 \begin{equation}\label{hsm}
 \mathcal H_{\rm eff}=\frac{G_F}{\sqrt 2}
 \left[\sum_{q=d,s}V_{cq_1}^*V_{uq_2}\left(\sum_{q=1}^2C_i
 (\mu)\mathcal{O}_i(\mu)\right)
 -V_{cb}^*V_{ub}\left(\sum_{i=3}^6C_i(\mu)\mathcal{O}_i(\mu)
 +C_{8g}(\mu)\mathcal{O}_{8g}(\mu)\right)\right].
 \end{equation}
The tree operators are
\begin{eqnarray}
\mathcal{O}_1=(\bar{u}_{\alpha}q_{2\beta})_{V-A}
(\bar{q}_{1\beta}c_{\alpha})_{V-A},\qquad
\mathcal{O}_2=(\bar{u}_{\alpha}q_{2\alpha})_{V-A}
(\bar{q}_{1\beta}c_{\beta})_{V-A},
\end{eqnarray}
in which $\alpha,\beta$ are color indices, $q_{1,2}$ are $d/s$
quarks.
The QCD penguin operators are
 \begin{align}
 \mathcal{O}_3&=\sum_{q'=u,d,s}(\bar u_\alpha c_\alpha)_{V-A}(\bar q'_\beta
 q'_\beta)_{V-A},~~~
 \mathcal{O}_4=\sum_{q'=u,d,s}(\bar u_\alpha c_\beta)_{V-A}(\bar q'_\beta q'_\alpha)_{V-A},
 \nonumber\\
 \mathcal{O}_5&=\sum_{q'=u,d,s}(\bar u_\alpha c_\alpha)_{V-A}(\bar q'_\beta
 q'_\beta)_{V+A},~~~
 \mathcal{O}_6=\sum_{q'=u,d,s}(\bar u_\alpha c_\beta)_{V-A}(\bar q'_\beta
 q'_\alpha)_{V+A},
 \end{align}
and the chromomagnetic penguin operator is
\begin{eqnarray}
\mathcal{O}_{8g}=\frac{g}{8\pi^2}m_c{\bar
u}\sigma_{\mu\nu}(1+\gamma_5)T^aG^{a\mu\nu}c.
\end{eqnarray}
The magnetic-penguin contributions can be included into the Wilson coefficients for the penguin operators following the substitutions
\cite{Beneke:2003zv,Beneke:2000ry,Beneke:1999br}
\begin{eqnarray}
C_{3,5}(\mu)\to& C_{3,5}(\mu) + \frac{\alpha_s(\mu)}{8\pi N_c}
\frac{2m_c^2}{\langle l^2\rangle}C_{8g}^{\rm eff}(\mu),\qquad
C_{4,6}(\mu)\to& C_{4,6}(\mu) - \frac{\alpha_s(\mu)}{8\pi }
\frac{2m_c^2}{\langle l^2\rangle}C_{8g}^{\rm eff}(\mu),\label{mag}
\end{eqnarray}
with the effective Wilson coefficient $C_{8g}^{\rm eff}=C_{8g}+C_5$ and $\langle l^2\rangle$ being the averaged invariant mass squared of the virtual gluon emitted from the magnetic penguin operator.
In this work, we classify the topological diagrams into tree- and penguin-induced diagrams according to which type of four-quark operators-tree or penguin- inserted into the effective vertexes.
The advantage of this classification is that we can identify all topologies systematically by writing invariant tensors \cite{Wang:2020gmn}.

According to Eq.~\eqref{hsm}, $H_{ij}^k$ can be obtained from the maps $(\bar uq_1)(\bar q_2c)\rightarrow V^*_{cq_2}V_{uq_1}$ in current-current operators and $(\bar qq)(\bar uc)\rightarrow -V^*_{cb}V_{ub}$ in penguin operators.
The nonzero $H_{ij}^k$ induced by tree operators include
\begin{align}\label{ckm1}
 &(H^{(0)})_{13}^2 = V_{cs}^*V_{ud},  \qquad (H^{(0)})^{2}_{12}=V_{cd}^*V_{ud},\qquad (H^{(0)})^{3}_{13}= V_{cs}^*V_{us}, \qquad (H^{(0)})^{3}_{12}=V_{cd}^*V_{us}.
\end{align}
 The nonzero $H_{ij}^k$ induced by penguin operators include
\begin{align}\label{ckm2}
 &(H^{(1)})_{11}^1 = (H^{(1)})_{21}^2= (H^{(1)})_{31}^3=-V_{cb}^*V_{ub}.
\end{align}
The superscripts (0) and (1) differentiate between tree and penguin operator contributions.
The tree- and penguin-induced topological amplitudes of the $\mathcal{B}_{c\overline 3}\to \mathcal{B}^S_8M$ and $\mathcal{B}_{c\overline 3}\to \mathcal{B}^A_8M$ decays are listed in Tables~\ref{ta1} to \ref{ta6}.
And the decay amplitudes constructed by $(\mathcal{B}_8)^i_j$ are listed in Tables~\ref{ta7} and \ref{ta8}.
We use superscript $P$ to distinguish the penguin induced amplitudes from the tree induced amplitudes.
One can check the topological amplitudes constructed by third- and second-rank octet tensors follow Eq.~\eqref{sol}.
For example, the topological amplitude of the $\Lambda^+_c\to p\overline K^0$ decay is
\begin{align}
  \mathcal{A}(\Lambda^+_c\to p\overline K^0) &=  \frac{1}{\sqrt{2}}\big[\mathcal{A}^S(\Lambda^+_c\to p\overline K^0)+\mathcal{A}^A(\Lambda^+_c\to p\overline K^0)\big]
  \nonumber \\ & =  \frac{1}{2\sqrt{3}}\lambda_1(2a^S_{10}-a^S_{11}-a^S_{12}-3a^S_{13})+ \frac{1}{2}\lambda_1(-a^A_{11}-a^A_{12}+a^A_{13}+2a^A_{29})
 \nonumber \\ & = \lambda_1(A_4+A_7).
\end{align}
Note that the tree-induced amplitudes $A_{15}$ $\sim$ $A_{18}$ and penguin-induced amplitudes $A_{1}^P$ and $A_{10}^P$ do not contribute to the charmed baryon decays.
It is because $H^j_{jk}$ in $A_{15}$ $\sim$ $A_{18}$ terms does not match with the tree operators in the SM.
Additionally, the property of $j\neq k$ in $(\mathcal{B}_{8})_j^k$ conflicts with $j=k$ in $H^j_{kl}$ when the penguin operators are inserted into $A_1$ and $A_{10}$ terms.
The decay amplitudes listed in Tables~\ref{ta1} to \ref{ta8} could help us to test the $SU(3)$ and isospin sum rules given by Refs.~\cite{Jia:2019zxi,Wang:2022kwe,Luo:2023vbx}, and the $U$-spin $CP$ violating sum rules given by Ref.~\cite{Wang:2019dls}.

The linear correlation of decay amplitudes contributing to the $\mathcal{B}_{c\overline 3}\to \mathcal{B}_8M$ modes in the Standard Model is beyond the model-independent analysis.
The nonzero coefficients induced by tree operators in the $SU(3)$ irreducible representations are
\begin{align}\label{ckm3}
 &  H^{(0)}( \overline6)^{22}=-2\,V_{cs}^*V_{ud},\qquad H^{(0)}( \overline 6)^{23}=(V_{cd}^*V_{ud}-V_{cs}^*V_{us}),  \qquad H^{(0)}( \overline 6)^{33}=  2\,V_{cd}^*V_{us},\nonumber \\
   &  H^{(0)}(15)^{1}_{11}=-2\,(V_{cd}^*V_{ud}+V_{cs}^*V_{us}), \qquad H^{(0)}(15)^{2}_{13}= 4\,V_{cs}^*V_{ud},  \qquad  H^{(0)}(15)^{3}_{12}=4\,V_{cd}^*V_{us},\nonumber \\
 &  H^{(0)}(15)^{2}_{12}= 3\,V_{cd}^*V_{ud}-V_{cs}^*V_{us},\qquad H^{(0)}(15)^{3}_{13}=3\,V_{cs}^*V_{us}-V_{cd}^*V_{ud}, \nonumber \\ & H^{(0)}( 3_t)_1=V_{cd}^*V_{ud}+V_{cs}^*V_{us},
\end{align}
and the nonzero coefficients induced by penguin operators in the $SU(3)$ irreducible representations are
\begin{align}\label{ckm4}
 H^{(1)}( 3_t)_1=-V_{cb}^*V_{ub}, \qquad H^{(1)}( 3_p)_1=-3V_{cb}^*V_{ub}.
\end{align}
The $SU(3)$ irreducible amplitudes of $\mathcal{B}_{c\overline 3}\to \mathcal{B}_8M$ modes can be obtained from Tables~\ref{ta7} and \ref{ta8} by replacing $A_{1}\sim A_{18}$ with $b_1\sim b_{18}$ according to Eq.~\eqref{sol3}.
From Eqs.~\eqref{ckm3} and \eqref{ckm4}, it is found the nonzero coefficients of the $3$-dimensional representations, including $H^{(0)}(3_t)_1$,  $H^{(1)}(3_t)_1$ and $H^{(1)}(3_p)_1$, only contain the first component.
There are no penguin induced amplitudes in the $15$- and $6$-dimensional irreducible representations.
Because of the unitarity of the CKM matrix, we have $H^{(0)}(3_t)_1 = V_{cd}^*V_{ud}+V_{cs}^*V_{us} = -V_{cb}^*V_{ub}$, and then
$H^{(0)}(3_t):H^{(1)}(3_t):H^{(1)}(3_p) = -V_{cb}^*V_{ub}:-V_{cb}^*V_{ub}:-3V_{cb}^*V_{ub}$.
Therefore, the amplitudes induced by the $3$-dimensional representations always appear simultaneously and can be absorbed into four parameters,
\begin{align}\label{c5}
 & b^\prime_{11}  = b_{15}+3b^P_{11} + b^P_{15},  \qquad   b^\prime_{12} =b_{16}+3b^P_{12} + b^P_{12}, \qquad   b^\prime_{13} = b_{17}+3b^P_{13} + b^P_{17}, \qquad   b^\prime_{14} = b_{18}+3 b^P_{14} + b^P_{18}.
\end{align}
According to Eq.~\eqref{c5} and Eq.~\eqref{sol2}, the tree induced amplitudes $A_{11} \sim A_{14}$ and all the penguin induced amplitudes can be absorbed into four parameters in the $SU(3)_F$ limit,
\begin{align}\label{c11}
 & A^{\prime}_{11}  = A_{11}+A^P_{3}+A^P_{5}+A^P_{11}+3A^P_{17},\qquad A^{\prime}_{12}  = A_{12}+A^P_{4}+A^P_{6}+A^P_{12}+3A^P_{15},\nonumber\\
& A^{\prime}_{13}  = A_{13}+A^P_{2}+A^P_{9}+A^P_{13}+3A^P_{16},\qquad A^{\prime}_{14}  = A_{14}+A^P_{7}+A^P_{8}+A^P_{14}+3A^P_{18}.
\end{align}
$A^\prime_{11}\sim A^\prime_{14}$ are proportional to $\lambda_b=V_{cb}^*V_{ub}$ since all the penguin induced amplitudes are proportional to $-V_{cb}^*V_{ub}$ and the tree induced amplitudes $A_{11}\sim A_{14}$ always appear simultaneously as $V_{cd}^*V_{ud}A_{i}+V_{cs}^*V_{us}A_{i}=-V_{cb}^*V_{ub}A_{i}$ in the SM.
According to Eq.~\eqref{c11}, all the penguin induced amplitudes are determined once the tree induced amplitudes with quark loops are known.
There is no degree of freedom of penguin induced diagrams in the SM.
Due to the small $\lambda_b$, amplitudes $A^\prime_{11}$ $\sim$ $ A^\prime_{14}$ are negligible in the branching fractions of charmed baryon decays.
But they are dominant in $CP$ asymmetries because of the large weak phase in $\lambda_b$.

According to the K\"orner-Pati-Woo (KPW) theorem \cite{Pati:1970fg,Korner:1970xq}, if two quarks produced by weak operators enter baryon, they are antisymmetric in flavor.
The two quark indices of $\mathcal{O}(15)$ and $\mathcal{O}(\overline 6)$ operators are symmetric and antisymmetric respectively.
The Fierz transformations of $\mathcal{O}(\overline 6)$ operators are opposite to the original ones.
Thereby, if two topological diagrams that the two quarks produced by weak operators enter final state baryon are connected by the Fierz transformation, they are opposite according to the KPW theorem.
This leads to the following amplitude relations,
\begin{align}
  a^{S,A}_1 &= -a^{S,A}_2,\qquad   a^{S,A}_5 = -a^{S,A}_6, \qquad a^{S,A}_7 = -a^{S,A}_{10}, \qquad a^{S,A}_8 = -a^{S,A}_{11}, \nonumber \\ a^{S,A}_9 & = -a^{S,A}_{12},\qquad
 a^{S,A}_{15}  = -a^{S,A}_{16},  \qquad a^{S}_{3} =a^{S}_{4}= a^{S}_{17} = 0.
\end{align}
By substituting these equations into Eq.~\eqref{sol11}, we find only six amplitudes in $b_1$ $\sim$ $b_{10}$ are not suppressed by the KPW theorem,
\begin{align}
b_1&= -(a^A_4 + a^A_7 -a^A_9) + \sqrt{3}(2a^S_5 - a^S_7 +2a^S_8 - a^S_9),\qquad
b_2= (a^A_1 -a^A_3 +a^A_5) + \sqrt{3}(3a^S_1 - a^S_5),\nonumber\\
b_3&= -(a^A_5 - a^A_{15} +a^A_{17}) + \sqrt{3}(a^S_5 + 3a^S_{15}),\qquad
b_4= (a^A_5 + a^A_8 +a^A_9) -\sqrt{3}(a^S_5 +2 a^S_7 -a^S_8 - a^S_9),\nonumber\\
b_5&= \frac{1}{2}(2a^A_1 -2 a^A_3 +2a^A_5+2a^A_8 +2a^A_9+a^A_{13}-a^A_{14}-2a^A_{28}+2a^A_{29}) \nonumber\\ &~~~~~~~~+\frac{\sqrt{3}}{2}(6a^S_1 - 2a^S_5 -4a^S_7 +2 a^S_8+2 a^S_{9}- 3a^S_{13} +3a^S_{14} ),\nonumber\\
b_{10}&= \frac{1}{4}(a^A_{13}+a^A_{14}+2a^A_{28}+2a^A_{29}) -\frac{\sqrt{3}}{4}(3a^S_{13}+3a^S_{14} ).
\end{align}
Considering that one of amplitudes in $b_1\sim b_5$ is not independent according to Eq.~\eqref{c9}, there are five dominant $SU(3)$ irreducible amplitudes contributing to the charmed baryon decays,
\begin{align}\label{c1}
b_1^\prime& =b_1-b_4= -(a^A_4 +a^A_5 + a^A_7 +a^A_8) + \sqrt{3}(3a^S_5 + a^S_7 +a^S_8 - 2a^S_9),\nonumber\\
b_2^\prime&=b_2-b_5= -\frac{1}{2}(2a^A_8 +2a^A_9+a^A_{13}-a^A_{14}-2a^A_{28}+2a^A_{29}) +\frac{\sqrt{3}}{2}(4a^S_7 -2 a^S_8-2 a^S_{9}+ 3a^S_{13} -3a^S_{14} ),\nonumber\\
b_3^\prime&=b_3+b_4=  ( a^A_8 +a^A_9+a^A_{15} -a^A_{17}) -\sqrt{3}(2 a^S_7 -a^S_8 - a^S_9-3a^S_{15}),\nonumber\\
b_4^\prime&=b_4-b_5= -\frac{1}{2}(2a^A_1 -2 a^A_3 +a^A_{13}-a^A_{14}-2a^A_{28}+2a^A_{29})-\frac{\sqrt{3}}{2}(6a^S_1 - 3a^S_{13} +3a^S_{14} ),\nonumber\\
b_{5}^\prime&= b_{10} =\frac{1}{4}(a^A_{13}+a^A_{14}+2a^A_{28}+2a^A_{29}) -\frac{\sqrt{3}}{4}(3a^S_{13}+3a^S_{14} ).
\end{align}
The dominant $SU(3)$ irreducible amplitudes in the charmed baryon decays are consistent with Ref.~\cite{Geng:2023pkr}.
Comparing the decay amplitudes defined in Ref.~\cite{Geng:2023pkr} with this work, we have
\begin{align}
  b_1^\prime =-\frac{1}{2}\tilde{f}^c,\qquad b_2^\prime = -\frac{1}{2}\tilde{f}^b,\qquad b_3^\prime = -\frac{1}{2}\tilde{f}^a,\qquad b_4^\prime = -\frac{1}{2}\tilde{f}^d,\qquad b_5^\prime =\frac{1}{4}\tilde{f}^e.
\end{align}

Taking the $\Xi_{c}^{0}\to \Sigma^+\pi^-$ and $\Xi_{c}^{0}\to pK^-$ modes as examples, we discuss the $U$-spin breaking in the charmed baryon decays.
In the $SU(3)_F$ limit, the amplitudes of $\Xi_{c}^{0}\to \Sigma^+\pi^-$ and $\Xi_{c}^{0}\to pK^-$ decays are given by
\begin{align}\label{a3}
 \mathcal{A}(\Xi_{c}^{0}\to \Sigma^+\pi^-) & =  \lambda_d(A_1+A_6+A_{11}+A_{14})+\lambda_s(A_{11}+A_{14}),\nonumber\\
\mathcal{A}(\Xi_{c}^{0}\to pK^-) & = \lambda_d(A_{11}+A_{14})+\lambda_s(A_1+A_6+A_{11}+A_{14}).
\end{align}
In the effective Hamiltonian \eqref{hsm}, the Wilson coefficients $C_{3-6}$ are much smaller than $C_{1,2}$ \cite{Buchalla:1995vs}.
The penguin induced amplitudes are smaller than the tree induced ones, so we neglect the penguin induced amplitudes in Eq.~\eqref{a3}.
Considering the $U$-spin breaking, the amplitude of $\Xi_{c}^{0}\to \Sigma^+\pi^-$ and $\Xi_{c}^{0}\to pK^-$ decays can be written as
\begin{align}\label{a4}
 \mathcal{A}(\Xi_{c}^{0}\to \Sigma^+\pi^-) & =  \lambda_d(A_1^d+A_6^d+A_{11}^d+A_{14}^d)+\lambda_s(A_{11}^s+A_{14}^s)\nonumber\\
 &~~~~= -\lambda_s(A_1^d+A_6^d-A_{11}^{\Delta U}-A_{14}^{\Delta U})-\lambda_b(A_{11}+A_{14}),\nonumber\\
\mathcal{A}(\Xi_{c}^{0}\to pK^-) & = \lambda_d(A_{11}^d+A_{14}^d)+\lambda_s(A_1^s+A_6^s+A_{11}^s+A_{14}^s)\nonumber\\
&~~~~=\lambda_s(A_1^s+A_6^s+A_{11}^{\Delta U}+A_{14}^{\Delta U})-\lambda_b(A_{11}+A_{14}),
\end{align}
where $A^{d,s}_{11}$ and $A^{d,s}_{14}$ are the decay amplitudes with $d$ or $s$ quarks in the quark-loop, and $A_{11,14}^{\Delta U}= A_{11,14}^{s}-A_{11,14}^{d}$.
Since $\lambda_b/\lambda_{s}$ is very small, we can drop the $SU(3)$ symmetric parts of quark-loop diagrams in the branching fractions,
\begin{align}
 \mathcal{A}(\Xi_{c}^{0}\to \Sigma^+\pi^-) \simeq  -\lambda_s(A_1^d+A_6^d-A_{11}^{\Delta U}-A_{14}^{\Delta U}),\qquad
\mathcal{A}(\Xi_{c}^{0}\to pK^-) \simeq  \lambda_s(A_1^s+A_6^s+A_{11}^{\Delta U}+A_{14}^{\Delta U}).
\end{align}
According to above formulas, if the decay amplitudes of $A_1^d+A_6^d$ and $A_{11}^{\Delta U}+A_{14}^{\Delta U}$ are constructive in $\Xi_{c}^{0}\to \Sigma^+\pi^-$ mode, the decay amplitudes of $A_1^s+A_6^s$ and $A_{11}^{\Delta U}+A_{14}^{\Delta U}$ will be destructive in $\Xi_{c}^{0}\to pK^-$ mode, and vice versa.
In other words, amplitudes $A^{\Delta U}_{11}$ and $A^{\Delta U}_{14}$ could enhance the $U$-spin breaking effects induced by amplitudes $A^{d,s}_{1}$ and $A^{d,s}_6$.
In general, the $U$-spin breaking effect is expected to be around $30\%$.
Thereby, if the branching fraction difference of $\Xi_{c}^{0}\to \Sigma^+\pi^-$ and $\Xi_{c}^{0}\to pK^-$ modes is far from the $30\%$ $U$-spin breaking, it would indicate large quark-loop diagrams.
Consequently, $CP$ asymmetries in these two decay modes might be observable.
The similar analysis has been proposed in charm meson decays \cite{Grossman:2019xcj,Brod:2012ud,Bhattacharya:2012ah,Cheng:2012xb,Muller:2015lua,Muller:2015rna}
and verified by LHCb measurements \cite{Aaij:2019kcg,LHCb:2022lry}.
Even if the $U$-spin breaking in charmed baryon decays is of the order of $\mathcal{O}(1\%)$, the $U$-spin broken parts are the dominant quark-loop contributions in the branching fractions since $\lambda_b/\lambda_{s}$ is at order of $\mathcal{O}(10^{-3})$.
It is impossible to extract the $U$-spin symmetric parts of quark-loop amplitudes, which dominate in the $CP$ asymmetries, by global fitting of branching fractions.
Above discussions can be generalized to other decay modes connected by $U$-spin, such as $\Xi_{c}^{0}\to \Sigma^{-}\pi^+$ and $\Xi_{c}^{0}\to \Xi^{-}K^+$, $\Lambda^+_c\to \Sigma^+K^{(*)0}$ and $\Xi^+_c\to p\overline K^{(*)0}$, and so on.
We suggest measuring the branching fractions the singly Cabibbo-suppressed charmed baryon modes systematically.
It will help us to identify which modes are promising for searching for $CP$ asymmetries in the charmed baryon decays.

\section{Summary}\label{summary}
In summary, we first perform a model independent study of the topological amplitudes of $\mathcal{B}_{c\overline 3}\to \mathcal{B}_8M$ decays in the $SU(3)_F$ limit.
The topological diagrams of the $\mathcal{B}_{c\overline 3}\to \mathcal{B}_8^S M$ decays, along with diagrams involving quark loops, are presented comprehensively for the first time.
The count of the possible contributing topologies to the $\mathcal{B}_{c\overline 3}\to \mathcal{B}_8 M$ decay is done by permutation.
There are twenty-seven and thirty-three topological diagrams in the $\mathcal{B}_{c\overline 3}\to \mathcal{B}_8^SM$ and $\mathcal{B}_{c\overline 3}\to \mathcal{B}_8^AM$ modes, respectively.
To investigate the linear correlation of these topologies, we derive the relation between the topological amplitudes constructed by the third-rank and second-rank octet tensors.
It is found that if the topologies in the $\mathcal{B}_{c\overline 3}\to \mathcal{B}_8^S M$ decays are known, the topologies in the $\mathcal{B}_{c\overline 3}\to \mathcal{B}_8^A M$ decays are also determined, and vice versa.
The equations of $SU(3)$ irreducible amplitudes decomposed by topologies are derived through two different intermediate amplitudes.
However, the inverse solution is not viable because of the larger number of topologies compared to the $SU(3)$ irreducible amplitudes.

Applying our framework to the $\mathcal{B}_{c\overline 3}\to \mathcal{B}_8M$ decays in the Standard Model, we identify thirteen independent tree induced $SU(3)$ irreducible amplitudes that contribute to the $\mathcal{B}_{c\overline 3}\to \mathcal{B}_8 M$ decays. Among these,  four amplitudes associated with three-dimensional operators are important in the $CP$ asymmetries.
The penguin induced amplitudes are determined by the tree induced amplitudes with quark loops.
Considering the suppression from small CKM matrix elements and the K\"orner-Pati-Woo theorem, the dominant $SU(3)$ irreducible amplitudes contributing to the branching fractions of charmed baryon decays are reduced to five.
The quark-loop diagrams could enhance the $U$-spin breaking effects, thereby increasing the branching fraction difference between two decay channels.
An systematic measurement of branching fractions of the singly Cabibbo-suppressed charmed baryon modes could help us to find which modes are promising in searching for $CP$ asymmetries.

\begin{acknowledgements}

We are grateful to Hai-Yang Cheng for useful discussions.
This work was supported in part by the National Natural Science Foundation of China under Grants No. 12105099.

\end{acknowledgements}

\begin{appendix}

\section{Third-rank $SU(3)$ irreducible amplitudes}\label{su3}

The $SU(3)$ irreducible amplitude of $\mathcal{B}_{c\overline 3}\to \mathcal{B}_8^SM$ decay constructed by third-rank octet tensor is
\begin{align}\label{IS}
 \mathcal{A}^S(\mathcal{B}_{c\overline 3}\to \mathcal{B}_8^SM) & = c^S_1(\mathcal{B}_{c\overline 3})_{ij} H(15)^m_{kl}M^i_m (\mathcal{B}_8^S)^{jkl} + c^S_2(\mathcal{B}_{c\overline 3})_{ij}H(\overline 6)^m_{kl}M^i_m (\mathcal{B}_8^S)^{jkl}+ c^S_3(\mathcal{B}_{c\overline 3})_{ij}H(15)^m_{kl}M^i_m (\mathcal{B}_8^S)^{klj} \nonumber\\
   & + c^S_4(\mathcal{B}_{c\overline 3})_{ij} H(15)^i_{kl}M^j_m (\mathcal{B}_8^S)^{klm}  + c^S_5(\mathcal{B}_{c\overline 3})_{ij} H(15)^i_{kl}M^j_m (\mathcal{B}_8^S)^{kml}  + c^S_6(\mathcal{B}_{c\overline 3})_{ij} H(\overline 6)^i_{kl}M^j_m (\mathcal{B}_8^S)^{kml} \nonumber\\
   &+ c^S_7(\mathcal{B}_{c\overline 3})_{ij} H(15)^i_{kl}M^k_m (\mathcal{B}_8^S)^{jlm}+ c^S_8(\mathcal{B}_{c\overline 3})_{ij} H(15)^i_{kl}M^k_m (\mathcal{B}_8^S)^{jml} + c^S_9(\mathcal{B}_{c\overline 3})_{ij} H(15)^i_{kl}M^k_m (\mathcal{B}_8^S)^{lmj} \nonumber\\
 &+ c^S_{10}(\mathcal{B}_{c\overline 3})_{ij} H(\overline 6)^i_{kl}M^k_m (\mathcal{B}_8^S)^{jlm}  + c^S_{11}(\mathcal{B}_{c\overline 3})_{ij} H(\overline 6)^i_{kl}M^k_m (\mathcal{B}_8^S)^{jml} + c^S_{12}(\mathcal{B}_{c\overline 3})_{ij} H(\overline 6)^i_{kl}M^k_m (\mathcal{B}_8^S)^{lmj} \nonumber\\
&+ c^S_{13}(\mathcal{B}_{c\overline 3})_{ij} H(15)^m_{kl}M^l_m (\mathcal{B}_8^S)^{ikj}+ c^S_{14}(\mathcal{B}_{c\overline 3})_{ij} H(\overline 6)^m_{kl}M^l_m (\mathcal{B}_8^S)^{ikj} + c^S_{15}(\mathcal{B}_{c\overline 3})_{ij} H(15)^i_{kl}M^m_m (\mathcal{B}_8^S)^{jkl} \nonumber\\
 &+ c^S_{16}(\mathcal{B}_{c\overline 3})_{ij} H(\overline 6)^i_{kl}M^m_m (\mathcal{B}_8^S)^{jkl} + c^S_{17}(\mathcal{B}_{c\overline 3})_{ij} H(15)^i_{kl}M^m_m (\mathcal{B}_8^S)^{klj}  + c^S_{18}(\mathcal{B}_{c\overline 3})_{ij} H(3_t)_{k}M^i_m (\mathcal{B}_8^S)^{jkm}\nonumber\\
 & + c^S_{19}(\mathcal{B}_{c\overline 3})_{ij} H(3_t)_{k}M^i_m (\mathcal{B}_8^S)^{jmk} + c^S_{20}(\mathcal{B}_{c\overline 3})_{ij} H(3_t)_{k}M^i_m (\mathcal{B}_8^S)^{kmj} + c^S_{21}(\mathcal{B}_{c\overline 3})_{ij} H(3_t)_{k}M^k_m (\mathcal{B}_8^S)^{imj}\nonumber\\
& + c^S_{22}(\mathcal{B}_{c\overline 3})_{ij} H(3_t)_{k}M^m_m (\mathcal{B}_8^S)^{ikj} + c^S_{23}(\mathcal{B}_{c\overline 3})_{ij} H(3_p)_{k}M^i_m (\mathcal{B}_8^S)^{jkm}+ c^S_{24}(\mathcal{B}_{c\overline 3})_{ij} H(3_p)_{k}M^i_m (\mathcal{B}_8^S)^{jmk}\nonumber\\
& + c^S_{25}(\mathcal{B}_{c\overline 3})_{ij} H(3_p)_{k}M^i_m (\mathcal{B}_8^S)^{kmj} + c^S_{26}(\mathcal{B}_{c\overline 3})_{ij} H(3_p)_{k}M^k_m (\mathcal{B}_8^S)^{imj}\nonumber\\&+ c^S_{27}(\mathcal{B}_{c\overline 3})_{ij} H(3_p)_{k}M^m_m (\mathcal{B}_8^S)^{ikj}.
\end{align}
The equivalence relations of the third-rank $SU(3)$ irreducible amplitudes and the topological amplitudes are derived to be
\begin{align}\label{sol6}
   & c^S_1 = \frac{a^S_1+a^S_2}{8},\quad c^S_2 = \frac{a^S_1-a^S_2}{4},  \quad c^S_3 = \frac{a^S_3}{8},  \quad c^S_4 = \frac{a^S_4}{8}, \quad c^S_5 = \frac{a^S_5+a^S_6}{8}, \quad  c^S_6 = \frac{a^S_5-a^S_6}{4},\nonumber\\
   & c^S_7 = \frac{a^S_7+a^S_{10}}{8},\quad c^S_{10} = \frac{a^S_7-a^S_{10}}{4},  \quad c^S_8 = \frac{a^S_8 + a^S_{11}}{8},  \quad c^S_{11} = \frac{a^S_8 - a^S_{11}}{4}, \quad c^S_9 = \frac{a^S_9+a^S_{12}}{8}, \quad  c^S_{12} = \frac{a^S_9-a^S_{12}}{4},\nonumber\\
   & c^S_{13} = \frac{a^S_{13}+a^S_{14}}{8},\quad c^S_{14} = \frac{a^S_{13}-a^S_{14}}{4},  \quad c^S_{15} = \frac{a^S_{15} + a^S_{16}}{8},  \quad c^S_{16} = \frac{a^S_{15} - a^S_{16}}{4}, \quad c^S_{17} = \frac{a^S_{17}}{8},\nonumber\\
   &  c^S_{18} = \frac{3}{8}a^S_1- \frac{1}{8}a^S_2 - \frac{1}{4}a^S_4- \frac{1}{8}a^S_7 +\frac{3}{8}a^S_{10}+ a^S_{18},\quad c^S_{23} = -\frac{1}{8}a^S_1+ \frac{3}{8}a^S_2 - \frac{1}{4}a^S_4+ \frac{3}{8}a^S_7 -\frac{1}{8}a^S_{10}+ a^S_{23},\nonumber\\
   &  c^S_{19} = -\frac{1}{8}a^S_1+ \frac{3}{8}a^S_2 + \frac{1}{8}a^S_{5}- \frac{3}{8}a^S_{6} -\frac{1}{8}a^S_{8}+ \frac{3}{8}a^S_{11}+ a^S_{19},\quad c^S_{24} = \frac{3}{8}a^S_1- \frac{1}{8}a^S_2 - \frac{3}{8}a^S_{5}+ \frac{1}{8}a^S_{6} +\frac{3}{8}a^S_{8}- \frac{1}{8}a^S_{11}+ a^S_{24},\nonumber\\
   &  c^S_{20} = \frac{1}{4}a^S_3- \frac{3}{8}a^S_5 + \frac{1}{8}a^S_{6}- \frac{1}{8}a^S_{9} +\frac{3}{8}a^S_{12}+ a^S_{20},\quad c^S_{25} = \frac{1}{4}a^S_3+ \frac{1}{8}a^S_5 - \frac{3}{8}a^S_{6}+ \frac{3}{8}a^S_{9} -\frac{1}{8}a^S_{12}+ a^S_{25},
\nonumber\\
   &  c^S_{21} = -\frac{3}{8}a^S_8+ \frac{1}{8}a^S_{11} + \frac{3}{8}a^S_{9}- \frac{1}{8}a^S_{12} -\frac{1}{8}a^S_{13} + \frac{3}{8}a^S_{14}+a^S_{21},\quad  c^S_{26} = \frac{1}{8}a^S_8- \frac{3}{8}a^S_{11} - \frac{1}{8}a^S_{9}+\frac{3}{8}a^S_{12} +\frac{3}{8}a^S_{13} - \frac{1}{8}a^S_{14}+a^S_{26},
\nonumber\\
   &  c^S_{22} = \frac{3}{8}a^S_{13}- \frac{1}{8}a^S_{14} -\frac{3}{8}a^S_{15}+ \frac{1}{8}a^S_{16} +\frac{1}{4}a^S_{17} +a^S_{22},\quad  c^S_{27} = -\frac{1}{8}a^S_{13}+\frac{3}{8}a^S_{14} +\frac{1}{8}a^S_{15}- \frac{3}{8}a^S_{16} +\frac{1}{4}a^S_{17} +a^S_{27}.
\end{align}
The inverse solution of Eq.~\eqref{sol6} is
\begin{align}\label{sol8}
   & a^S_1 = 4c^S_1+2c^S_2,\quad a^S_2 = 4c^S_1-2c^S_2,  \quad a^S_3 = 8c^S_3,  \quad a^S_4 = 8c^S_4, \quad a^S_5 = 4c^S_5+2c^S_6, \quad  a^S_6 = 4c^S_5-2c^S_6,\nonumber\\
   & a^S_7 = 4c^S_7+2c^S_{10},\quad a^S_{10} = 4c^S_7-2c^S_{10},  \quad a^S_8 = 4c^S_8+2c^S_{11},  \quad a^S_{11} = 4c^S_8-2c^S_{11}, \quad a^S_9 = 4c^S_9+2c^S_{12}, \nonumber\\
   & a^S_{12} = 4c^S_9-2c^S_{12}, \quad a^S_{13} = 4c^S_{13}+2c^S_{14},\quad a^S_{14} = 4c^S_{13}-2c^S_{14},  \quad a^S_{15} = 4c^S_{15}+2c^S_{16},  \quad a^S_{16} = 4c^S_{15}-2c^S_{16}, \quad a^S_{17} = 8c^S_{17},\nonumber\\
   &  a^S_{18} = c^S_{18}-c^S_1- c^S_2 +2 c^S_4- c^S_7 +c^S_{10},\quad a^S_{23} = c^S_{23}-c^S_1+ c^S_2 +2 c^S_4- c^S_7 -c^S_{10},\nonumber\\
   &  a^S_{19} = c^S_{19}-c^S_1+ c^S_2 + c^S_5- c^S_6 -c^S_{8}+c^S_{11},\quad a^S_{24} = c^S_{24}-c^S_1- c^S_2 + c^S_5+ c^S_6 -c^S_{8}-c^S_{11},\nonumber\\
   &  a^S_{20} = c^S_{20}-2c^S_3+c^S_5 + c^S_6- c^S_9 +c^S_{12},\quad a^S_{25} = c^S_{25}-2c^S_3+ c^S_5 - c^S_6-c^S_9 -c^S_{12},\nonumber\\
   &  a^S_{21} = c^S_{21}+c^S_8+ c^S_{11} - c^S_9- c^S_{12} -c^S_{13}+c^S_{14},\quad a^S_{26} = c^S_{26}+c^S_8- c^S_{11} - c^S_9+ c^S_{12} -c^S_{13}-c^S_{14},\nonumber\\
   &  a^S_{22} = c^S_{22}-c^S_{13}- c^S_{14} + c^S_{15}+ c^S_{16} -2c^S_{17},\quad a^S_{27} = c^S_{27}-c^S_{13}+ c^S_{14} + c^S_{15}- c^S_{16} -2c^S_{17}.
\end{align}
The $SU(3)$ irreducible amplitude of $\mathcal{B}_{c\overline 3}\to \mathcal{B}_8^AM$ decay constructed by third-rank octet tensor is
\begin{align}\label{IA}
 \mathcal{A}^A(\mathcal{B}_{c\overline 3}\to \mathcal{B}_8^AM) & = c^A_1(\mathcal{B}_{c\overline 3})_{ij} H(15)^m_{kl}M^i_m (\mathcal{B}_8^A)^{jkl} + c^A_2(\mathcal{B}_{c\overline 3})_{ij}H(\overline 6)^m_{kl}M^i_m (\mathcal{B}_8^A)^{jkl}+ c^A_3(\mathcal{B}_{c\overline 3})_{ij}H(\overline 6)^m_{kl}M^i_m (\mathcal{B}_8^A)^{klj} \nonumber\\
   & + c^A_4(\mathcal{B}_{c\overline 3})_{ij} H(\overline 6)^i_{kl}M^j_m (\mathcal{B}_8^A)^{klm}  + c^A_5(\mathcal{B}_{c\overline 3})_{ij} H(15)^i_{kl}M^j_m (\mathcal{B}_8^A)^{kml}  + c^A_6(\mathcal{B}_{c\overline 3})_{ij} H(\overline 6)^i_{kl}M^j_m (\mathcal{B}_8^A)^{kml} \nonumber\\
   &+ c^A_7(\mathcal{B}_{c\overline 3})_{ij} H(15)^i_{kl}M^k_m (\mathcal{B}_8^A)^{jlm}+ c^A_8(\mathcal{B}_{c\overline 3})_{ij} H(15)^i_{kl}M^k_m (\mathcal{B}_8^A)^{jml} + c^A_9(\mathcal{B}_{c\overline 3})_{ij} H(15)^i_{kl}M^k_m (\mathcal{B}_8^A)^{lmj} \nonumber\\
 &+ c^A_{10}(\mathcal{B}_{c\overline 3})_{ij} H(\overline 6)^i_{kl}M^k_m (\mathcal{B}_8^A)^{jlm}  + c^A_{11}(\mathcal{B}_{c\overline 3})_{ij} H(\overline 6)^i_{kl}M^k_m (\mathcal{B}_8^A)^{jml} + c^A_{12}(\mathcal{B}_{c\overline 3})_{ij} H(\overline 6)^i_{kl}M^k_m (\mathcal{B}_8^A)^{lmj} \nonumber\\
&+ c^A_{13}(\mathcal{B}_{c\overline 3})_{ij} H(15)^m_{kl}M^l_m (\mathcal{B}_8^A)^{ikj}+ c^A_{14}(\mathcal{B}_{c\overline 3})_{ij} H(\overline 6)^m_{kl}M^l_m (\mathcal{B}_8^A)^{ikj} + c^A_{15}(\mathcal{B}_{c\overline 3})_{ij} H(15)^i_{kl}M^m_m (\mathcal{B}_8^A)^{jkl} \nonumber\\
 &+ c^A_{16}(\mathcal{B}_{c\overline 3})_{ij} H(\overline 6)^i_{kl}M^m_m (\mathcal{B}_8^A)^{jkl} + c^A_{17}(\mathcal{B}_{c\overline 3})_{ij} H(\overline 6)^i_{kl}M^m_m (\mathcal{B}_8^A)^{klj}  + c^A_{18}(\mathcal{B}_{c\overline 3})_{ij} H(3_t)_{k}M^i_m (\mathcal{B}_8^A)^{jkm}\nonumber\\
 & + c^A_{19}(\mathcal{B}_{c\overline 3})_{ij} H(3_t)_{k}M^i_m (\mathcal{B}_8^A)^{jmk} + c^A_{20}(\mathcal{B}_{c\overline 3})_{ij} H(3_t)_{k}M^i_m (\mathcal{B}_8^A)^{kmj} + c^A_{21}(\mathcal{B}_{c\overline 3})_{ij} H(3_t)_{k}M^k_m (\mathcal{B}_8^A)^{imj}\nonumber\\
& + c^A_{22}(\mathcal{B}_{c\overline 3})_{ij} H(3_t)_{k}M^m_m (\mathcal{B}_8^A)^{ikj} + c^A_{23}(\mathcal{B}_{c\overline 3})_{ij} H( 3_p)_{k}M^i_m (\mathcal{B}_8^A)^{jkm}+ c^A_{24}(\mathcal{B}_{c\overline 3})_{ij} H( 3_p)_{k}M^i_m (\mathcal{B}_8^A)^{jmk}\nonumber\\
& + c^A_{25}(\mathcal{B}_{c\overline 3})_{ij} H( 3_p)_{k}M^i_m (\mathcal{B}_8^A)^{kmj} + c^A_{26}(\mathcal{B}_{c\overline 3})_{ij} H( 3_p)_{k}M^k_m (\mathcal{B}_8^A)^{imj}+ c^A_{27}(\mathcal{B}_{c\overline 3})_{ij} H( 3_p)_{k}M^m_m (\mathcal{B}_8^A)^{ikj}\nonumber\\
&+ c^A_{28}(\mathcal{B}_{c\overline 3})_{ij} H(15)^m_{kl}M^k_m (\mathcal{B}_8^A)^{ijl} + c^A_{29}(\mathcal{B}_{c\overline 3})_{ij} H(\overline 6)^m_{kl}M^k_m (\mathcal{B}_8^A)^{ijl} + c^A_{30}(\mathcal{B}_{c\overline 3})_{ij} H(3_t)_{k}M^k_m (\mathcal{B}_8^A)^{ijm} \nonumber\\
&+ c^A_{31}(\mathcal{B}_{c\overline 3})_{ij} H(3_t)_{k}M^m_m (\mathcal{B}_8^A)^{ijk}+ c^A_{32}(\mathcal{B}_{c\overline 3})_{ij} H(3_p)_{k}M^k_m (\mathcal{B}_8^A)^{ijm}\nonumber\\&+ c^A_{33}(\mathcal{B}_{c\overline 3})_{ij} H(3_p)_{k}M^m_m (\mathcal{B}_8^A)^{ijk}.
\end{align}
The equivalence relations of the third-rank $SU(3)$ irreducible amplitudes and the topological amplitudes are derived to be
\begin{align}\label{sol7}
   & c^A_1 = \frac{a^A_1+a^A_2}{8},\quad c^A_2 = \frac{a^A_1-a^A_2}{4},  \quad c^A_3 = \frac{a^A_3}{4},  \quad c^A_4 = \frac{a^A_4}{4}, \quad c^A_5 = \frac{a^A_5+a^A_6}{8}, \quad  c^A_6 = \frac{a^A_5-a^A_6}{4},\quad c^A_7 = \frac{a^A_7+a^A_{10}}{8},\nonumber\\
   &  c^A_{10} = \frac{a^A_7-a^A_{10}}{4},  \quad c^A_8 = \frac{a^A_8 + a^A_{11}}{8},  \quad c^A_{11} = \frac{a^A_8 - a^A_{11}}{4}, \quad c^A_9 = \frac{a^A_9+a^A_{12}}{8}, \quad  c^A_{12} = \frac{a^A_9-a^A_{12}}{4},\quad c^A_{13} = \frac{a^A_{13}+a^A_{14}}{8},\nonumber\\
   &  c^A_{14} = \frac{a^A_{13}-a^A_{14}}{4},  \quad c^A_{15} = \frac{a^A_{15} + a^A_{16}}{8},  \quad c^A_{16} = \frac{a^A_{15} - a^A_{16}}{4}, \quad c^A_{17} = \frac{a^A_{17}}{4},\quad c^A_{28} = \frac{a^A_{28} + a^A_{29}}{8},  \quad c^A_{29} = \frac{a^A_{28} - a^A_{29}}{4},\nonumber\\
   &  c^A_{18} = \frac{3}{8}a^A_1- \frac{1}{8}a^A_2 + \frac{1}{2}a^A_4- \frac{1}{8}a^A_7 +\frac{3}{8}a^A_{10}+ J^S_{18},\quad c^A_{23} = -\frac{1}{8}a^A_1+ \frac{3}{8}a^A_2 - \frac{1}{2}a^A_4+ \frac{3}{8}a^A_7 -\frac{1}{8}a^A_{10}+ a^A_{23},\nonumber\\
   &  c^A_{19} = -\frac{1}{8}a^A_1+ \frac{3}{8}a^A_2 + \frac{1}{8}a^A_{5}- \frac{3}{8}a^A_{6} -\frac{1}{8}a^A_{8}+ \frac{3}{8}a^A_{11}+ a^A_{19},\quad c^A_{24} = \frac{3}{8}a^A_1- \frac{1}{8}a^A_2 - \frac{3}{8}a^A_{5}+ \frac{1}{8}a^A_{6} +\frac{3}{8}a^A_{8}- \frac{1}{8}a^A_{11}+ a^A_{24},\nonumber\\
   &  c^A_{20} = \frac{1}{2}a^A_3- \frac{3}{8}a^A_5 + \frac{1}{8}a^A_{6}- \frac{1}{8}a^A_{9} +\frac{3}{8}a^A_{12}+ a^A_{20},\quad c^A_{25} = -\frac{1}{2}a^A_3+ \frac{1}{8}a^A_5 - \frac{3}{8}a^A_{6}+ \frac{3}{8}a^A_{9} -\frac{1}{8}a^A_{12}+ a^S_{25},
\nonumber\\
   &  c^A_{21} = -\frac{3}{8}a^A_8+ \frac{1}{8}a^A_{11} + \frac{3}{8}a^A_{9}- \frac{1}{8}a^A_{12} -\frac{1}{8}a^A_{13} + \frac{3}{8}a^A_{14}+a^A_{21},\quad  c^A_{26} = \frac{1}{8}a^A_8- \frac{3}{8}a^A_{11} - \frac{1}{8}a^A_{9}+\frac{3}{8}a^A_{12} +\frac{3}{8}a^A_{13} - \frac{1}{8}a^A_{14}+a^A_{26},
\nonumber\\
   &  c^A_{22} = \frac{3}{8}a^A_{13}- \frac{1}{8}a^A_{14} -\frac{3}{8}a^A_{15}+ \frac{1}{8}a^A_{16} -\frac{1}{2}a^A_{17} +a^A_{22},\quad  c^A_{27} = -\frac{1}{8}a^A_{13}+\frac{3}{8}a^A_{14} +\frac{1}{8}a^A_{15}- \frac{3}{8}a^A_{16} +\frac{1}{2}a^A_{17} +a^A_{27},
   \nonumber\\
   &  c^A_{30} = -\frac{3}{8}a^A_{7}+ \frac{1}{8}a^A_{10} +\frac{3}{8}a^A_{28}- \frac{1}{8}a^A_{29} +a^A_{30},\quad  c^A_{32} = \frac{1}{8}a^A_{7}- \frac{3}{8}a^A_{10} -\frac{1}{8}a^A_{28}+ \frac{3}{8}a^A_{29} +a^A_{32},   \nonumber\\
   &  c^A_{31} = \frac{1}{8}a^A_{15}- \frac{3}{8}a^A_{16} -\frac{1}{8}a^A_{28}+ \frac{3}{8}a^A_{29} +a^A_{31},\quad  c^A_{33} = -\frac{3}{8}a^A_{15}+ \frac{1}{8}a^A_{16} +\frac{3}{8}a^A_{28}- \frac{1}{8}a^A_{29} +a^A_{33}.
\end{align}
The inverse solution of Eq.~\eqref{sol7} is
\begin{align}\label{sol9}
   & a^A_1 = 4c^A_1+2c^A_2,\quad a^A_2 = 4c^A_1-2c^A_2,  \quad a^A_3 = 4c^A_3,  \quad a^A_4 = 4c^A_4, \quad a^A_5 = 4c^A_5+2c^A_6, \quad  a^A_6 = 4c^A_5-2c^A_6,\nonumber\\
   & a^A_7 = 4c^A_7+2c^A_{10}, \quad a^A_{10} = 4c^A_7-2c^A_{10},  \quad a^A_8 = 4c^A_8+2c^A_{11},  \quad a^A_{11} = 4c^A_8-2c^A_{11}, \quad a^A_9 = 4c^A_9+2c^A_{12}, \nonumber\\
   & a^A_{12} = 4c^A_9-2c^A_{12}, \quad a^A_{13} = 4c^A_{13}+2c^A_{14},\quad a^A_{14} = 4c^A_{13}-2c^A_{14},  \quad a^A_{15} = 4c^A_{15}+2c^A_{16},  \quad a^A_{16} = 4c^A_{15}-2c^A_{16}, \nonumber\\
   & a^A_{17} = 4c^A_{17},\quad a^A_{28} = 4c^A_{28}+2c^A_{29},\quad a^A_{29} = 4c^A_{28}-2c^A_{29},\nonumber\\
   &  a^A_{18} = c^A_{18}-c^A_1- c^A_2 -2 c^A_4- c^A_7 +c^A_{10},\quad a^A_{23} = c^A_{23}-c^A_1+ c^A_2 +2 c^A_4- c^A_7 -c^A_{10},\nonumber\\
   &  a^A_{19} = c^A_{19}-c^A_1+ c^A_2 + c^A_5- c^A_6 -c^A_{8}+c^A_{11},\quad a^A_{24} = c^A_{24}-c^A_1- c^A_2 + c^A_5+ c^A_6 -c^A_{8}-c^A_{11},\nonumber\\
   &  a^A_{20} = c^A_{20}-2c^A_3+c^A_5 + c^A_6- c^A_9 +c^A_{12},\quad a^A_{25} = c^A_{25}+2c^A_3+ c^A_5 - c^A_6-c^A_9 -c^A_{12},\nonumber\\
   &  a^A_{21} = c^A_{21}+c^A_8+ c^A_{11} - c^A_9- c^A_{12} -c^A_{13}+c^A_{14},\quad a^A_{26} = c^A_{26}+c^A_8- c^A_{11} - c^A_9+ c^A_{12} -c^A_{13}-c^A_{14},\nonumber\\
   &  a^A_{22} = c^A_{22}-c^A_{13}- c^A_{14} + c^A_{15}+ c^A_{16} +2c^A_{17},\quad a^A_{27} = c^A_{27}-c^A_{13}+ c^A_{14} + c^A_{15}- c^A_{16} -2c^A_{17},\nonumber\\
   &  a^A_{30} = c^A_{30}+c^A_{7}+ c^A_{10} - c^A_{28}- c^A_{29},\quad a^A_{32} = c^A_{32}+c^A_{7}- c^A_{10} - c^A_{28}+ c^A_{29},\nonumber\\
    &  a^A_{31} = c^A_{31}+c^A_{15}- c^A_{16} - c^A_{28}+ c^A_{29},\quad a^A_{33} = c^A_{33}+c^A_{15}+ c^A_{16} - c^A_{28}- c^A_{29}.
\end{align}

The relations between the third- and second-rank $SU(3)$ irreducible amplitudes can be derived by inserting Eqs.~\eqref{sy1} and \eqref{sy2} into each term of Eq.~\eqref{IS} and Eq.~\eqref{IA}.
For example, the first term of Eq.~\eqref{IS} is simplified to be
\begin{align}
c^S_1(\mathcal{B}_{c\overline 3})_{ij} H(15)^m_{kl}M^i_m ( \mathcal{B}_8^S)^{jkl} & = c^S_1\epsilon_{ijp}(\mathcal{B}_{c\overline 3})^{p} H(15)^m_{kl}M^i_m \epsilon^{ljq}(\mathcal{B}_8)^k_q+c^S_1\epsilon_{ijp}(\mathcal{B}_{c\overline 3})^{p} H(15)^m_{kl}M^i_m\epsilon^{lkq}(\mathcal{B}_8)^j_q\nonumber\\
& = c^S_1\,(\mathcal{B}_{c\overline 3})^{i} H(15)^j_{kl}M^l_j( \mathcal{B}_8)^k_i-c^S_1\,(\mathcal{B}_{c\overline 3})^{i} H(15)^j_{ki}M^l_j( \mathcal{B}_8)^k_l.
\end{align}
So $c^S_1$ contributes to Eq.~\eqref{amp4} as
\begin{align}
&b_7 = -2\sqrt{3}\,c^S_1+..., \qquad b_{10} = 2\sqrt{3}\,c^S_1+...\,\,.
\end{align}
The relations between third- and second-rank $SU(3)$ irreducible amplitudes are derived to be
\begin{align}\label{sol10}
& b_1 = -2(2c^A_4+c^A_{10}-c^A_{12})
   +2\sqrt{3}(2c^S_6-c^S_{10}+2c^S_{11}-c^S_{12}),\nonumber\\
& b_2 = 2(2c^A_2-2c^A_{3}+c^A_{6})
   +2\sqrt{3}(3c^S_2-c^S_{6}),\nonumber\\
& b_3 = -2(c^A_6-c^A_{16}+2c^A_{17})
   +2\sqrt{3}(c^S_6+3c^S_{16}),\nonumber\\
& b_4 = 2(c^A_6+c^A_{11}+c^A_{12})
   -2\sqrt{3}(c^S_6+2c^S_{10}-c^S_{11}-c^S_{12}),\nonumber\\
& b_5 = 2(c^A_2-2c^A_{3}+c^A_{6}+c^A_{11}+c^A_{12}+c^A_{14}-2c^A_{29})
   +2\sqrt{3}(3c^S_2-c^S_{6}-2c^S_{10}+c^S_{11}+c^S_{12}-3c^S_{14}),\nonumber\\
& b_6 = 2(c^A_7-c^A_{9})
   +2\sqrt{3}(c^S_7-2c^S_{8}+c^S_{9}),\nonumber\\
& b_7 = 2(c^A_1+c^A_{5})
   -2\sqrt{3}(c^S_1-2c^S_{3}+2c^S_4-c^S_{5}),\nonumber\\
& b_8 = -2(c^A_5-c^A_{15})
   +2\sqrt{3}(2c^S_4-c^S_{5}-c^S_{15}+2c^S_{17}),\nonumber\\
& b_9 = 2(c^A_5+c^A_{8}+c^A_{9})
   -2\sqrt{3}(2c^S_4-c^S_{5}+2c^S_7-c^S_{8}-c^S_{9}),\nonumber\\
& b_{10} = -2(c^A_1+c^A_{5}+c^A_{8}+c^A_{9}-c^A_{13}-2c^A_{28})
   +2\sqrt{3}(c^S_1-2c^S_{3}+2c^S_{4}-c^S_{5}+2c^S_{7}
   -c^S_{8}-c^S_{9}-3c^S_{13}),\nonumber\\
& b_{11} = 2(c^A_{23}-c^A_{25})
   +2\sqrt{3}(c^S_{23}-2c^S_{24}+c^S_{25}),\nonumber\\
& b_{12} = -2(c^A_{24}+c^A_{25}-c^A_{27}-2c^A_{33})
   +2\sqrt{3}(2c^S_{23}-c^S_{24}-c^S_{25}-3c^S_{27}),\nonumber\\
& b_{13} = 2(c^A_{24}+c^A_{25})
   -2\sqrt{3}(2c^S_{23}-c^S_{24}-c^S_{25}),\nonumber\\
& b_{14} = -2(c^A_{23}-c^A_{25}-c^A_{26}-2c^A_{32})
   -2\sqrt{3}(c^S_{23}-2c^S_{24}+c^S_{25}+3c^S_{26}),\nonumber\\
& b_{15} = 2(c^A_{18}-c^A_{20})
   +2\sqrt{3}(c^S_{18}-2c^S_{19}+c^S_{20}),\nonumber\\
& b_{16} = -2(c^A_{19}+c^A_{20}-c^A_{22}-2c^A_{31})
   +2\sqrt{3}(2c^S_{18}-c^S_{19}-c^S_{20}-3c^S_{22}),\nonumber\\
& b_{17} = 2(c^A_{19}+c^A_{20})
   -2\sqrt{3}(2c^S_{18}-c^S_{19}-c^S_{20}),\nonumber\\
& b_{18} = -2(c^A_{18}-c^A_{20}-c^A_{21}-2c^A_{30})
   -2\sqrt{3}(c^S_{18}-2c^S_{19}+c^S_{20}+3c^S_{21}).
\end{align}
From Eq.~\eqref{sol10}, it is found the 3-rank $SU(3)$ irreducible amplitudes contribute to the 2-rank $SU(3)$ irreducible amplitudes without the mixing of $15$-, $\overline 6$, $3_{t}$ and $3_{p}$ operators.
By substituting Eqs.~\eqref{sol6} and \eqref{sol7} into Eq.~\eqref{sol10}, we rederive Eq.~\eqref{sol11} within different approach.

\end{appendix}

\end{document}